\documentclass[prd,twocolumn,nofootinbib,longbibliography,10pt]{revtex4-1}
\usepackage{textcomp}
\usepackage{amsmath}
\usepackage{amsfonts}
\usepackage{calc}
\usepackage{mathrsfs}
\usepackage{amssymb}
\usepackage{amsmath}
\usepackage{array}
\usepackage{color}
\usepackage{bm}
\usepackage{graphicx}
\usepackage{hyperref}

\newcommand{\beq}{\begin{equation}}
\newcommand{\eeq}{\end{equation}}
\renewcommand{\O}{\mathcal{O}}
\renewcommand{\P}{\mathcal{P}}
\newcommand{\res}{\mathcal{R}}
\renewcommand{\S}{{\rm S}}
\newcommand{\R}{{\rm R}}
\DeclareMathOperator{\STF}{STF}
\newcommand{\e}{\epsilon}
\newcommand{\nhat}{\hat{n}}
\newcommand{\E}{\mathcal{E}}
\newcommand{\B}{\mathcal{B}}
\newcommand{\tail}{h^{\rm tail}}
\newcommand{\exact}[1]{{\sf #1}}
\newcommand{\sft}{{\sf t}}
\newcommand{\sfx}{{\sf x}}
\newcommand{\sfy}{{\sf y}}
\newcommand{\sfz}{{\sf z}}
\newcommand{\sfr}{{\sf r}}
\newcommand{\sfv}{{\sf v}}

\newcommand{\Lie}{{\cal L}}
\newcommand{\ord}[2]{{}^{#1}\hspace{-.6pt}#2}

\begin{document}
\title{Nonlinear gravitational self-force: second-order equation of motion} 
\author{Adam Pound} 
\affiliation{Mathematical Sciences and STAG Research Centre, University of Southampton, Southampton,
United Kingdom, SO17 1BJ}
\date{\today}

\begin{abstract}
When a small, uncharged, compact object is immersed in an external background spacetime, at zeroth order in its mass it moves as a test particle in the background. At linear order, its own gravitational field alters the geometry around it, and it moves instead as a test particle in a certain effective metric satisfying the linearized vacuum Einstein equation. In the letter [Phys. Rev. Lett. 109, 051101 (2012)], using a method of matched asymptotic expansions, I showed that the same statement holds true at second order: if the object's leading-order spin and quadrupole moment vanish, then through second order in its mass it moves on a geodesic of a certain smooth, locally causal vacuum metric defined in its local neighbourhood. Here I present the complete details of the derivation of that result. In addition, I extend the result, which had previously been derived in gauges smoothly related to Lorenz, to a class of highly regular gauges that should be optimal for numerical self-force computations. 
\end{abstract}

\maketitle

\section{Introduction}
Over the last two decades, there has been renewed interest in a fundamental question of general relativity: How does a small object move when immersed in an external spacetime? In other words, how is the object's motion altered from the test-particle description when one accounts for the object's own gravitational field, finite size, and internal composition? This question is now of astrophysical interest, due to the advent of gravitational wave astronomy. Binaries of compact objects with dissimilar masses will directly exhibit corrections to the test-particle approximation. This is true even of the intermediate-mass-ratio binaries that should be detected~\cite{Brown:07} by second-generation ground-based detectors such as Advanced LIGO~\cite{aLIGO} and Virgo~\cite{aVirgo}. It is doubly true of extreme-mass-ratio inspirals (EMRIs), in which a stellar-mass black hole or neutron star spirals into a supermassive black hole in a galactic core; the clear separation of  scales in these systems will allow a precise delineation of the post-test-particle effects in the smaller object's motion. EMRIs, while outside the frequency band of LIGO and Virgo, will be key sources for the planned space-based detector LISA~\cite{eLISA:13}. 

The principal approach to modeling these systems is  self-force theory, which seeks to describe a small object's motion by treating it as a source of perturbation $h_{\mu\nu}$ of an external background spacetime $g_{\mu\nu}$~\cite{Poisson-Pound-Vega:11,Pound:15a,Barack:09}. In this description, the object is accelerated by the self-force, the back-reaction of the object's field on its own motion. The formalism in this approach is now on a sound theoretical basis \cite{Gralla-Wald:08,Pound:10a,Harte:12}, has well-developed computational methods~\cite{Wardell:15, vandeMeent:16}, has yielded a range of physical predictions~\cite{Poisson-Pound-Vega:11,Barack:09, Amaro-Seoane-etal:14}, and has had impact on binary modeling outside the EMRI regime, providing important input for post-Newtonian theory, fully nonlinear numerical relativity, and effective-one-body theory~\cite{Damour:09,Blanchet-etal:10b,LeTiec:14,Damour-etal:16,Zimmerman-etal:16}. 

\subsection{The generalized equivalence principle in self-force theory}
Until recently, self-force theory has focused on linear perturbation theory. 
At that level, the primary result of the self-force program is a \emph{generalized equivalence principle} (a phrase I adopt from Ref.~\cite{Futamase-Itoh:07}). The ordinary equivalence principle dictates that all freely falling test masses, given identical initial conditions, follow the same geodesic trajectory in an external gravitational field, regardless of their inertial mass or internal composition. The generalized equivalence principle extends that statement to gravitating objects: neglecting finite-size effects, isolated small compact objects, be they material bodies or black holes, follow geodesic paths in a certain \emph{effective metric} $g^{\rm eff}_{\mu\nu}=g_{\mu\nu}+h^{\R1}_{\mu\nu}$ that satisfies the vacuum Einstein field equation (EFE), where the \emph{Detweiler-Whiting regular field} $h^{\R1}_{\mu\nu}$ is a certain piece of the perturbation $h_{\mu\nu}$~\cite{Detweiler:01,Detweiler-Whiting:03}. Unlike the ordinary equivalence principle, the generalized principle does \emph{not} suggest that the motion is identical for all bodies. They all move on geodesics, but they move on geodesics of \emph{different}  geometries, because $h^{\R1}_{\mu\nu}$ is proportional to their own gravitational mass and determined by their own past histories~\cite{Detweiler-Whiting:03}. However, the sense of the equivalence principle is preserved, in that each object feels no gravitational force, instead falling freely in what it sees as an ``external'' gravitational field---even though it is responsible for a piece of that field.

On the face of it, the conclusion that an object's worldline is a geodesic in some effective metric might not seem especially meaningful or useful; any equation of motion can be written as the geodesic equation in {\it some} effective metric~\cite{Pound:15a}. However, the statement {\it is} both meaningful and useful if the effective metric satisfies suitable conditions, such as the following:
\begin{enumerate}
\item the effective metric is ``physical'', in the sense that it satisfies the vacuum EFE and on the worldline it (and its derivatives) depend only on the causal past, and
\item there is a practical way to actually calculate the effective metric and solve the self-forced equation of motion.
\end{enumerate}
At first order, these conditions are both met by $g_{\mu\nu}+h^{\R 1}_{\mu\nu}$.

Beyond first order, several foundational analyses have been performed~\cite{Rosenthal:06a,Rosenthal:06b, Detweiler:12,Harte:12, Pound:12a,Gralla:12,Pound:12b,Pound:15a}. Harte has established~\cite{Harte:12} that even in a completely nonperturbative description of a material object, one can construct an effective metric in which the object moves as a test body, subject to forces only due to finite-size effects. However, besides the limitation to material bodies, which excludes black holes, Harte's effective metric is not a solution to the vacuum EFE, and there is no immediate means of calculating it numerically. Fortunately, perturbative approaches have overcome both of these restrictions, at least through second order in the small object's mass. As a practical way of computing the effective metric, all authors, beginning with Rosenthal~\cite{Rosenthal:06a}, have proposed some variant of a ``puncture scheme''~\cite{Barack-Golbourn:07,Vega-Detweiler:07,Wardell:15}, in which a local expansion of the metric near the small object (valid for both black holes and material bodies) is converted into a singular ``puncture''. The curvature of the puncture is then treated as a source for the effective metric. In the letter~\cite{Pound:12a} I presented a definition of an effective metric satisyfing the ``physical'' conditions described above, and I showed that if the object is approximately nonspinning and spherical, then through second order it moves on a geodesic of this effective metric, thereby extending the generalized equivalence principle to second order. I stress that this is a derived result involving no ``regularization'' and no presumed relationship between the motion and the effective metric.

Due to the space constraints of a letter, Ref.~\cite{Pound:12a} necessarily omitted many details. References~\cite{Pound:12b,Pound:15a} filled in some of those details, specifically the explicit form of the puncture and effective metric, the effective metric's causality on the worldline, and the nature of the puncture scheme. The present paper covers the other half of the problem, detailing the derivation of the equation of motion. It also extends the result to a class of {\em highly regular gauges} that should prove useful in numerical implementations.


\subsection{Equations of motion from matched asymptotic expansions}\label{worldline-definition}
In Newtonian mechanics, we typically wish to use as little information as possible to describe extended objects---for example, by treating them as point particles. Similarly, in self-force theory our primary goal is to determine the motion of a small object and obtain the metric {\em outside} of it, without having to concern ourselves with its potentially complicated internal dynamics. In the perturbative context, this is achieved  with the method of matched asymptotic expansions~\cite{Eckhaus:79, Kevorkian-Cole:96} ({e.g.}, in Refs.~\cite{DEath:75, Kates:80,Thorne-Hartle:85,Mino-Sasaki-Tanaka:97,Detweiler:01,Gralla-Wald:08,Pound:10a} and the second-and-higher-order self-force literature). Before ladening the reader with the detailed application of this method in deriving the second-order equation of motion, I  first provide an overview of the derivation strategy I follow. 

We suppose that the small object is in a spacetime with metric ${\sf g}_{\mu\nu}(\e)$, where the parameter $\e$ encodes the dependence on the object's mass $m$ and size $\ell$; we can think of $\e$ being proportional to $m$, though it will be convenient to use it as a formal expansion parameter and set it equal to 1 at the end of the calculation. We take the object to be compact, such that $m\sim \ell$. In the ``self-consistent'' approach~\cite{Pound:10a}, the metric outside the object is expanded as
\beq\label{outer}
{\sf g}_{\mu\nu} = g_{\mu\nu} + \e h^1_{\mu\nu}[\gamma] + \e^2 h^2_{\mu\nu}[\gamma]+\ldots
\eeq
The object creates perturbations $h^n_{\mu\nu}$ of the external background metric $g_{\mu\nu}$, and those perturbations are functionals of the object's motion, as represented by an $\e$-dependent worldline $\gamma$ in the background manifold. For simplicity, I take the object to be in a vacuum region, such that $g_{\mu\nu}$ is a vacuum metric.

Near $\gamma$, at distances $r\sim\e$, the gravity of the small object begins to dominate over the background, and the expansion~\eqref{outer} ceases to be accurate. Hence, we introduce a different expansion in this region. We first rescale the distance $r$ to $\tilde r:=r/\e $, such that  $\tilde r\sim 1$ when $r\sim \e$. We then expand in the limit $\e\to0$ at fixed $\tilde r$:
\beq\label{inner}
{\sf g}_{\mu\nu}(r,\e) = g^{\rm obj}_{\mu\nu}(\tilde r) + \e H^1_{\mu\nu}(\tilde r)+\e^2 H^2_{\mu\nu}(\tilde r) + \ldots
\eeq
Here the background metric becomes $g^{\rm obj}_{\mu\nu}$, the metric of the object if it were isolated, and the perturbations are produced by the external background field (and the object's interactions with that field). While the expansion \eqref{outer} lets the object shrink to zero mass and size while holding external distances fixed, the expansion at fixed $\tilde r$ zooms in on the object, keeping its mass and size fixed while other distances are blown up. I refer to Eq.~\eqref{outer} as the ``outer expansion'' and to Eq.~\eqref{inner} as the ``inner expansion''.

Since both are expansions of the same metric, they must agree (given a suitably well-behaved ${\sf g}_{\mu\nu}$~\cite{Eckhaus:79}). More precisely, if we re-expand the outer expansion for small $r$---{i.e.}, near the worldline---then we obtain a double expansion in powers of $\e$ and $r$. If we re-express the inner expansion in terms of $r=\e\tilde r$ and then re-expand for small $\e$ at fixed $r$---{i.e.}, for distances $\tilde r\gg 1$, relatively far from the small object---then we obtain another such double expansion. These two double expansions, which  can be expected to be accurate in a ``buffer region'' $\e\ll r\ll 1$, must agree order by order in $\e$ and $r$.

The existence of a well-behaved inner expansion constrains $h^n_{\mu\nu}$ to have the form  
\beq\label{hnform}
h^n_{\mu\nu}= \sum_{p\geq-n}r^p h^{n,p}_{\mu\nu}
\eeq
when expanded for small $r$ (allowing for logarithms of $r$ in the coefficients). In other words, $h^n_{\mu\nu}\sim\frac{1}{r^n}$. Any more negative power of $r$ would lead to a poorly behaved inner expansion with negative powers of $\e$; for example, if $h^n_{\mu\nu}\sim \frac{1}{r^{n+1}}$, then a term like $\e^n h^n_{\mu\nu}\sim \frac{\e^n}{r^{n+1}}$ in Eq.~\eqref{outer} would correspond to a term $\sim \frac{1}{\e \tilde r^{n+1}}$ in the inner expansion. $g_{\mu\nu}$ is likewise constrained to take the form 
\beq
g_{\mu\nu}=\sum_{p\geq0}r^p g^p_{\mu\nu}, 
\eeq
which also follows from $g_{\mu\nu}$ being a smooth vacuum metric.

Analogously, the existence of a well-behaved outer expansion constrains $g^{\rm obj}_{\mu\nu}$ and $H^n_{\mu\nu}$ to have the forms 
\beq
g^{\rm obj}_{\mu\nu}=\sum_{p\geq0} \frac{1}{\tilde r^p}g^{\rm obj, p}_{\mu\nu} 
\eeq
and  
\beq\label{Hnform}
H^n_{\mu\nu}=\sum_{p\geq-n} \frac{1}{\tilde r^p}H^{n p}_{\mu\nu}
\eeq
when expanded for large $\tilde r$ (allowing for logarithms of $\tilde r$ in the coefficients). This implies that $g^{\rm obj}_{\mu\nu}$ is asymptotically flat. It is also quasistationary (see Sec.~\ref{inner_expansion}). Hence, in the buffer region it can be expressed in terms of its multipole moments. If we introduce a Cartesian coordinate system $(t,x^i)$ centered on $\gamma$, where $x^i=rn^i$, with $n^i=(n^x,n^y,n^z)$ being orthogonal unit vectors, then the expansion in the buffer region looks schematically like
\begin{align}
g^{\rm obj}_{\mu\nu}&\sim 1+\frac{m}{\tilde r} + \frac{M_i n^i+\epsilon_{ijk}S^j n^k}{\tilde r^2} \nonumber\\
						&\quad + \frac{M_{ij} n^in^j+\epsilon_{ijk}S^j{}_q n^kn^q}{\tilde r^3}+\ldots \label{gobj}.
\end{align}
Here $m$ is $g^{\rm obj}_{\mu\nu}$'s Arnowitt-Deser-Misner (ADM) mass, $M_i$ its mass dipole moment, $S^i$ its ADM angular momentum, and $M_{ij}$ and $S_{ij}$ its mass and current quadrupole moments. (I omit terms like $m^2/\tilde r^2$ and $mM_i/\tilde r^3$ for visual clarity.) 

$g^{\rm obj}_{\mu\nu}$'s moments encode the internal composition of the object, and they determine the most negative powers of $r$ in the outer perturbations $h^n_{\mu\nu}$. For example, when rewritten in terms of $r$, the $\frac{m}{\tilde r}$ term in Eq.~\eqref{gobj} becomes $\frac{\e m}{r}$, which fixes the $1/r$ term in $h^1_{\mu\nu}$. Hence, $h^1_{\mu\nu}$ has the form 
\beq\label{h1form}
h^1_{\mu\nu} \sim \frac{m}{r}+\O(r^0).
\eeq
Similarly, 
\beq\label{h2form}
h^2_{\mu\nu} \sim \frac{m^2+M_in^i+\epsilon_{ijk}S^j n^k}{r^2}+\O(r^{-1}),
\eeq
(now keeping the $m^2/r^2$ term for completeness) and so on for the higher-order perturbations $h^n_{\mu\nu}$. The fact that the $n$th moments scale as $\e^n$, and hence first appear in the $n$th-order perturbation, is a consequence of the object's assumed compactness. 

In short, the perturbations $h^n_{\mu\nu}$ are locally determined by the object's first $n$ multipole moments. This means that rather than requiring a full model of the object's  internal dynamics, to obtain a finite order of approximation we merely need to specify a finite number of  moments. This simplification is closely tied to the point-particle approximation: as first shown by D'Eath~\cite{DEath:75} (see also Refs.~\cite{Gralla-Wald:08,Pound:10a}), the more explicit form of Eq.~\eqref{h1form} suffices to show that $h^1_{\mu\nu}$ is identical to the linear perturbation produced by a point mass $m$ moving on $\gamma$.

So far in this description, I have said nothing of the object's motion. All we know is that it lies somewhere near some worldline $\gamma$, in the region $r\sim \e$.\footnote{Note that $\gamma$ need not be ``inside'' the object, which would not be sensible for a black hole. In general, $\gamma$ exists in the background manifold on which $g_{\mu\nu}$ is the metric, not in the manifold on which $g^{\rm obj}_{\mu\nu}$ is the metric. $\gamma$ is then approximately associated with the object's ``position'' through the existence of the inner expansion, and more closely associated with its center-of-mass ``position'' through  the properties of the metric in the buffer region.} To fix $\gamma$ to be a good representative of the object's position, we recall that a mass dipole moment can be interpreted as a displacement $\delta z^i = M^i/m$ of the center of mass from the origin of the coordinates; equivalently, it is generated if we begin in a mass-centered coordinate system and perform a small coordinate transformation $x^i\to x^i +M^i/m$. Since our coordinates are centered on $\gamma$, a nonzero $M^i$ would indicate that $\gamma$ does not represent the object's center of mass. Hence, we set $M^i$ to zero. This ensures that $\gamma$ is at the center of mass of the leading-order metric $g^{\rm obj}_{\mu\nu}$. To constrain $\gamma$ at higher orders, similar conditions must also be imposed on the perturbations $H^n_{\mu\nu}$; these will be discussed momentarily.

With this minimal setup in place, there are two ways  to determine the  equation of motion governing $\gamma$. One route, detailed in Ref.~\cite{Pound:12b} (following Ref.~\cite{Pound:10a}), is to solve the vacuum EFE (outside the object) for the perturbations $h^n_{\mu\nu}$ order by order in $\e$ and $r$, beginning with expansions of the form~\eqref{hnform}. Solving the EFE in this way, combined with a center-of-mass condition, determines the acceleration of $\gamma$. (It also provides a local expansion of $h^n_{\mu\nu}$ near $\gamma$, written in terms of the object's multipole moments, which can be used to define a puncture for use in practical computations.) More concretely, if we expand $\gamma$'s acceleration as $a^\mu = f_0^\mu+\e f_1^\mu+\ldots$, then an equation for the $n$th-order force (per unit mass) $f_n^\mu$ follows from the field equation for $h^{n+1}_{\mu\nu}$. The fact that $f_0^\mu=0$ ({i.e.}, that the motion is approximately geodesic in $g_{\mu\nu}$) follows from the equations for $h^1_{\mu\nu}$; the standard result for the first-order self-force, $f_1^\mu$, follows from the equations for $h^2_{\mu\nu}$; and the second-order self-force, $f^\mu_2$, which is the order of interest in this paper, would follow from the equations for $h^3_{\mu\nu}$. However, solving the third-order field equations in the outer expansion is quite burdensome. 

Fortunately, there is a second, easier path. First, calculate the metric in the inner expansion in a ``rest gauge'', in which the object is manifestly at rest  on $\gamma$. (More generally, if the object is spinning and nonspherical, it should move as a test body on $\gamma$ in this gauge, but I will nevertheless refer to it as a rest gauge.) Next, re-express this expansion as an outer expansion in the buffer region, and then transform to whichever gauge is desired for the outer expansion---call it the ``practical gauge''. The existence of a rest gauge is intimately related to the fact that there is {\em some} effective metric in which $\gamma$ is a geodesic; the transformation identifies {\em which} effective metric that is. A key to this approach is that the gauge transformation must be constrained to preserve the location of the center of mass on $\gamma$.

To apply this procedure to determine $f^\mu_1$, we must specify that $M^i$ vanishes in the rest gauge, as described above. To apply it to determine $f^\mu_2$, we require the next-order extension of this condition. The natural choice is to impose that the mass dipole moment in $H^1_{\mu\nu}$ vanishes. In an appropriate gauge, $H^1_{\mu\nu}$ can be written in the form 
\begin{align}
H^1_{\mu\nu}&\sim \tilde r a_i n^i + \tilde r^0 \delta u^t  + \frac{\delta m}{\tilde r} \nonumber\\
&\quad+ \frac{\delta M_i n^i + \epsilon_{ijk}\delta S^jn^k}{\tilde r^2} + \ldots,\label{H1form}
\end{align}
where $a_i$ can be interpreted as an acceleration of the $\tilde r\to\infty$ asymptotic frame relative to the rest frame of the object, $\delta u^t$ as a mismatch between the proper times of the two frames, and $\delta m$, $\delta M^i$, and $\delta S^i$ as corrections to the mass, mass dipole moment, and spin. We can then naturally set $\delta M^i=0$ as our center-of-mass condition. Because the metric $g^{\rm obj}_{\mu\nu}+\e H^1_{\mu\nu}$ appears not to be asymptotically flat, this $\delta M^i$ does not strictly correspond to the standard multipole moments defined for stationary, asymptotically flat spacetimes~\cite{Geroch:70,Hansen:74}. Nevertheless, $\delta M^i=0$ is a natural center-of-mass condition: $\delta M^i$ appears as part of a gauge perturbation that is easily isolated from the rest of $H^1_{\mu\nu}$, and it would arise from a small coordinate translation in precisely the same manner as $M^i$. However, though that argument would become important at higher orders, where it becomes impossible to write the metric in asymptotically flat form, it need not be stressed at this order, because the first two terms in Eq.~\eqref{H1form} are pure gauge, meaning $g^{\rm obj}_{\mu\nu}+\e H^1_{\mu\nu}$ {\em is} asymptotically flat. In fact, the only gauge-invariant content in $H^1_{\mu\nu}$ corresponds to trivial corrections to the multipole moments of $g^{\rm obj}_{\mu\nu}$, meaning the entirety of $H^1_{\mu\nu}$ can be set to zero by absorbing those trivial corrections into the background.

When the metric of the inner expansion is re-expanded in the buffer region, it yields an outer expansion in the rest gauge:
\beq\label{re-expanded inner}
{\sf g}_{\mu\nu} = g_{\mu\nu} + \e h^{1'}_{\mu\nu}[\gamma] + \e^2 h^{2'}_{\mu\nu}[\gamma]+\ldots,
\eeq
where each of the terms is expressed as an expansion for small $r$. If the rest gauge and the gauge of Eq.~\eqref{outer} are fully fixed, then there must exist a unique gauge transformation between them, subject to the crucial condition that the transformation does not induce a nonzero $M^i$ or $\delta M^i$. This condition ensures that the object remains centered on the same worldline $\gamma$ after the transformation. I will refer to a transformation satisfying it as {\em worldline-preserving}. For the calculations in this paper, the condition reduces to a simple form: a transformation with the coordinate representation
\begin{align}
x'^\mu &= x^\mu-\epsilon \xi^\mu_1-\epsilon^2\!\! \left(\xi^\mu_2-\frac{1}{2}\xi^\nu_1\partial_\nu\xi_1^\mu\right) +\O(\e^3)\label{coord_transformation}
\end{align}
is worldline-preserving if and only if
\begin{equation}\label{worldline-preserving}
\lim_{r\to0}\oint_S \frac{dS}{r^2}\, \xi^a_n = 0,
\end{equation}
at all times $t$, where $S$ is a small sphere of radius $r$ around $\gamma(t)$. In other words, $\xi^\mu_n$ must have no net direction on the worldline.

The condition~\eqref{worldline-preserving} is intuitively meaningful, as a gauge transformation violating it would manifestly move the origin of the coordinate system. Hence, we could impose this condition without making any reference to the mass dipole moment. However, tying it to the mass dipole moment helps to clarify what may appear to be a mysterious elimination of one whole order of calculation. The first-order equation of motion can be derived as a consequence of the second-order EFE, and the second-order equation of motion as a consequence of the third-order EFE. Yet by referring to the transformation from a rest gauge, we seem able to derive the first-order equation of motion solely from the first-order metric, and the second-order equation from the second-order metric, effectively saving an order. We can understand this by noting that what the second-order EFE actually determines is an evolution equation for $M^i$ relative to any given worldline~\cite{Pound:10a}; setting $M^i=0$ for all time then picks out the first-order acceleration of the correct, center-of-mass worldline. Analogously, the third-order EFE determines the evolution of $\delta M^i$, and setting $\delta M^i=0$ picks out the second-order acceleration of this worldline. But we also have that the mass dipole moment $M^i$, which appears in the second-order metric perturbation, is fully determined by the {\em first-order} gauge: if we transform away from a rest gauge, then $M^i=-m\xi_1^i$. The first-order gauge transformation therefore determines the same information as the second-order field equation. In the same way, $\delta M^i=-m\xi_2^i$ (if $\xi^i_1=0$). This is an illustration of the deep connection between gauge and motion in perturbation theory~\cite{Barack-Ori:01,Gralla:11,Gralla:12,Pound:15b}.

This strategy of transforming from a rest gauge to a practical gauge, although not as intuitively clear as the direct derivation of the equation of motion from the EFE for $h^n_{\mu\nu}$, underlies many derivations in the literature.  Mino, Sasaki, and Tanaka~\cite{Mino-Sasaki-Tanaka:97} used it in essentially the same way as I do here in their original derivation of the first-order equation of motion. Rosenthal~\cite{Rosenthal:06b} used similar ideas in his derivation of a second-order equation of motion. However, he relied on an axiomatic list of possible ingredients in the self-force, rather than the first-principles approach I take here, and his formulation ended with an equation of motion in an impractical form in which the first-order perturbation is required to remain in a rest gauge. Detweiler~\cite{Detweiler:12} argued that, given the form of the metric in a rest gauge, the motion must be geodesic in some suitable effective metric, though he did not consider the problem of transforming to a practical gauge and identifying the effective metric in it. Most recently, shortly after my letter~\cite{Pound:12a}, Gralla~\cite{Gralla:12} used a closely related method in his derivation of a second-order equation of motion. His formulation was somewhat different in that he did not seek an effective metric in which the motion is geodesic. But a more important distinction is that in both his and Rosenthal's approaches, their rest gauges refer to a different representative worldline. Gralla explicitly uses a perturbative description, in which the worldline is expanded as $\gamma=\gamma_0+\e \gamma_1+\e^2\gamma_2+\ldots$, with $\gamma_0$ being a background geodesic and $\gamma_1$ and $\gamma_2$ being small deviation vectors defined on that geodesic.\footnote{See Refs.~\cite{Pound:10a,Pound:15a,Pound:15b} for in-depth discussions of the relationship between this approximation and the self-consistent one I use throughout this paper.} This description is sensible on timescales of order $\e^0$ because if the acceleration is of order $\e$, then the deviation of the accelerated object from a background geodesic is also of order $\e$. In this treatment, the ``rest gauge'' puts the center of mass at rest on a background geodesic $\gamma_0$, such that $\gamma_1=\gamma_2=0$. Rather than being worldline-preserving, the  transformation to a practical gauge is then allowed to be arbitrary, and the evolution equations for $\gamma_1$ and $\gamma_2$ (or, equivalently, $M^i$ and $\delta M^i$) are derived from the evolution of $\xi^\mu_1$ and $\xi^\mu_2$ along $\gamma_0$. Although Rosenthal does not use this type of description, he likewise uses a rest gauge in which the object moves on a geodesic of $g_{\mu\nu}$ and a transformation that translates the object onto an accelerated path. In both cases, these approaches are restricted to timescales of order $\e^0$, meaning they cannot accurately describe effects such as the inspiral of an EMRI, which occurs on the timescale $1/\e$. The treatment here avoids that restriction.

\subsection{Outline of this paper}
In the bulk of the paper, I work through each step of the derivation outlined above, specializing to an object with vanishing spin and quadrupole moments at leading order. Sections~\ref{outer_expansion} and \ref{inner_expansion} present the form of the metric perturbation through second order, with Sec.~\ref{outer_expansion} summarizing the calculation in the Lorenz gauge, and Sec.~\ref{inner_expansion} in a rest gauge. Section~\ref{matching} presents the gauge transformation between the two solutions, which leads to the equation of motion. This stage of the derivation also illustrates an ambiguity in the definition of the self-induced tidal moments acting on the body, as first computed by Dolan {et al.}~\cite{Dolan-etal:14} and Bini and Damour~\cite{Bini-Damour:14}. 

In Sec.~\ref{geodesic_motion}, I show that the derived equation of motion is equivalent to geodesic motion in the effective metric $g_{\mu\nu}+h^{\R}_{\mu\nu}$ defined in Refs.~\cite{Pound:12a,Pound:12b}. 

Section~\ref{gauge} extends this result to non-Lorenz gauges. After a brief review of the extension to gauges smoothly related to Lorenz, the bulk of this section is devoted to a derivation in a class of highly regular gauges. These gauges remove the dominant, $m^2/r^2$ part of the metric at second order,  circumventing many challenges that generically arise in second-order numerical schemes. 

The final section of the paper summarizes my results and discusses the prospects for  numerical computations in these highly regular gauges.

Throughout, I work in units with $G=c=1$. Greek indices range from 0 to 3 and are raised and lowered with the background metric $g_{\mu\nu}$. Lowercase Latin indices refer to spatial coordinates $x^a=(x,y,z)$ in the outer expansion. Lowercase sans-serif indices refer to spatial coordinates $\sfx^{\sf a}=(\sfx,\sfy,\sfz)$ in the inner expansion. Both are raised and lowered with the flat-space Euclidean metric $\delta_{ij}$. Uppercase Latin indices denote multi-indices, as in $L:= i_1\cdots i_\ell$, and an expression such as $A_{|L|-1}$ denotes $A_{i_1\cdots i_{|l|-1}}$. Parentheses around indices indicate symmetrization; square brackets, antisymmetrization. Angular parentheses, as in $\langle abc\rangle$, indicate the symmetric trace-free (STF) combination of the enclosed indices, where the trace is taken with $\delta_{ab}$. A hat over a tensor, as in $\hat T_{abc}$, likewise indicates that the tensor is STF with respect to $\delta_{ab}$. $\nabla$ and a semicolon both denote covariant derivatives compatible with $g_{\mu\nu}$. An overbar, as in $\bar h_{\mu\nu}$, denotes trace reversal with $g_{\mu\nu}$, as in $\bar h_{\mu\nu}=h_{\mu\nu}-\tfrac{1}{2}g_{\mu\nu}g^{\rho\sigma}h_{\rho\sigma}.$


\section{Outer expansion in the Lorenz gauge}\label{outer_expansion}
This section reviews the outer expansion through second order. Further details can be found in Refs.~\cite{Pound:10a,Pound:12b,Pound:15a}.


\subsection{Form of the expansion}
To find the outer expansion, I utilize the self-consistent framework developed in Ref.~\cite{Pound:10a}. The metric is written as
\begin{equation}\label{external ansatz}
{\sf g}_{\mu\nu}(x,\e) =g_{\mu\nu}(x)+h_{\mu\nu}(x,\e;\gamma),
\end{equation}
where $x$ denotes any suitable set of coordinates that do not depend on $\gamma$, and the semicolon is used as a compact alternative to $h_{\mu\nu}[\gamma](x)$. The metric perturbation is expanded {\em while holding the $\e$-dependent worldline $\gamma$ fixed}:
\begin{equation}
h_{\mu\nu}(x,\e;\gamma)=\sum_{n\geq1}\e^n h^n_{\mu\nu}(x;\gamma).
\end{equation}
This expansion self-consistently incorporates the metric's dependence on the long-term evolution of the worldline such as, for example, the inspiral in an EMRI.

By imposing the gauge condition $\nabla^\nu\bar h_{\mu\nu}=0$, I reduce the vacuum EFE $R_{\mu\nu}[{\sf g}]=0$ outside the object to the weakly nonlinear wave equation 
\beq\label{wave equation}
E_{\mu\nu}[h]=2\delta^2R_{\mu\nu}[h]+\O(h^3), 
\eeq
where $E_{\mu\nu}$ is the relativistic wave operator
\begin{equation}
E_{\mu\nu}[h] = \left(g^\rho_\mu g^\sigma_\nu\nabla^\gamma\nabla_{\gamma} +2R_\mu{}^\rho{}_\nu{}^\sigma\right)\!h_{\rho\sigma},
\end{equation}
with $R_\mu{}^\rho{}_\nu{}^\sigma$ the Riemann tensor of $g_{\mu\nu}$, and $\delta^2 R_{\mu\nu}$ is the second-order Ricci tensor, given by
\begin{align}
\delta^2R_{\alpha\beta}[h] &=-\tfrac{1}{2}\bar h^{\mu\nu}{}_{;\nu}\left(2h_{\mu(\alpha;\beta)}-h_{\alpha\beta;\mu}\right) 
					+\tfrac{1}{4}h^{\mu\nu}{}_{;\alpha}h_{\mu\nu;\beta}\nonumber\\
					&\quad +\tfrac{1}{2}h^{\mu}{}_{\beta}{}^{;\nu}\left(h_{\mu\alpha;\nu} -h_{\nu\alpha;\mu}\right)\nonumber\\
					&\quad-\tfrac{1}{2}h^{\mu\nu}\left(2h_{\mu(\alpha;\beta)\nu}-h_{\alpha\beta;\mu\nu}-h_{\mu\nu;\alpha\beta}\right).\label{d2R}
\end{align}
Equation~\eqref{wave equation} can be expanded and solved order by order in $\e$ while holding $\gamma$ fixed, leading to a sequence of wave equations beginning with
\begin{align}
E_{\mu\nu}[h^1] &=0, \label{EFE1}\\
E_{\mu\nu}[h^2] &=2\delta^2 R_{\mu\nu}[h^1]. \label{EFE2}
\end{align}
Each of these equations can be solved for an arbitrary $\gamma$, and for an arbitrary set of multipole moments defined on $\gamma$. The evolution equations governing $\gamma$ and the moments are then found by imposing the gauge condition in the buffer region. Because the wave equation is constraint-preserving, these evolution equations (together with suitable initial data) suffice to enforce the gauge condition globally~\cite{Pound:12b}.

\subsection{General solution in the buffer region}
Here we are only interested in the form of the solution near $\gamma$. I transform away from the global coordinates $x$ to local coordinates that {\em are} dependent on the worldline: Fermi-Walker coordinates $(t, x^a)$ centered on $\gamma$. These are defined such that $x^i=r n^i$, where $r$ is the proper distance from $\gamma$ along a spatial geodesic that is sent out from $\gamma$ perpendicularly, and $n^i$ is a unit radial vector that labels the direction along which the geodesic is sent out. For a given point $z$ on $\gamma$, the set of all such geodesics form a three-dimensional spatial surface. Each such surface is labelled with a coordinate time $t$, equal to the proper time on $\gamma$ at the point $z$. Reference~\cite{Poisson-Pound-Vega:11} contains a pedagogical introduction.

Because the self-force will naturally involve a derivative of $h^2_{\mu\nu}$ (thinking naively of $h_{\mu\nu}$ as a potential and its derivative as a force), to derive the equation of motion we will  require $h^2_{\mu\nu}$ through order $r$ in these coordinates. Since $h^2_{\mu\nu}$ begins at order $r^{-2}$ according to Eq.~\eqref{h2form}, this implies that we need a total of four orders in $r$: that is, $h^1_{\mu\nu}$ through order $r^2$ because it begins at order $1/r$ according to Eq.~\eqref{h1form}, and $g_{\mu\nu}$ through order $r^3$ because it begins at order $r^0$. Through that order, the background metric is given by
\begin{subequations}\label{background}%
\begin{align}
g_{tt} &= -1-2a_ix^i-\left(R_{titj}+a_ia_j\right)x^ix^j\nonumber\\
&\quad-\frac{1}{3}\left(4R_{titj}a_k+R_{titj;k}\right)x^ix^jx^k+\O(r^4),\\
g_{ta} &= -\frac{2}{3}R_{tiaj}x^ix^j-\frac{1}{3}R_{tiaj}a_kx^ix^jx^k\nonumber\\
&\quad -\frac{1}{4}R_{tiaj;k}x^ix^jx^k+\O(r^4),\\
g_{ab} &= \delta_{ab}-\frac{1}{3}R_{aibj}x^ix^j-\frac{1}{6}R_{aibj;k}x^ix^jx^k\nonumber\\&\quad+\O(r^4),
\end{align}
\end{subequations}
where $a^\mu:=\frac{Du^\mu}{dt}=(0,a^i)$ is the acceleration of the worldline [with $u^\mu=(1,0,0,0)$ the normalized four-velocity], the Riemann tensor and its derivatives are evaluated on the worldline, and a quantity such as $R_{aibj;k}$ denotes a component of a covariant derivative rather than a derivative of a component. The metric takes the form of Minkowski along the worldline, plus corrections away from the worldline due to acceleration and curvature. 

Because $g_{\mu\nu}$ is Ricci-flat, the components of the Riemann tensor and its first derivatives can be written in terms of  STF tensors $\E_{ab}$, $\B_{ab}$, $\E_{abc}$, and $\B_{abc}$, which I define as
\begin{align}
\E_{ab} &:= R_{tatb}, \label{Eab}\\
\B_{ab} &:= \frac{1}{2}\epsilon^{pq}{}_{(a}R_{b)tpq}, \label{Bab}\\
\E_{abc} &:= \mathop{\rm STF}_{abc}R_{tatb;c}, \label{Eabc}\\
\B_{abc} &:= \frac{3}{8}\mathop{\rm STF}_{abc}\epsilon^{pq}{}_{a}R_{btpq;c},
\end{align}
where ``STF'' indicates the STF combination of the specified indices. $\E_{ab}$ and $\B_{ab}$ are the electric- and magnetic-type tidal quadrupole moments of the background spacetime, and $\E_{abc}$ and $\B_{abc}$ are the electric- and magnetic-type tidal octupole moments. Appendix D3 of Ref.~\cite{Poisson-Vlasov:09} provides identities for decomposing each component of the Riemann tensor and its derivatives in terms of these tidal moments.

In these coordinates, the fields $h^n_{\mu\nu}$ near $\gamma$ can be found by substituting Eq.~\eqref{hnform} into the EFE. Because spatial derivatives reduce the power of $r$ by one while temporal derivatives do not, the wave operator becomes $E_{\mu\nu}[h]=\partial^i\partial_i h_{\mu\nu}+\O(h/r)$, and solving order by order in $r$  reduces to solving a sequence of flat-space Poisson equations. 

This procedure is facilitated by adopting the more specific expansion
\begin{equation}\label{hn mode expansion}
h^n_{\mu\nu}=\sum_{p\geq-n,q\geq0,l\geq0}r^p(\ln r)^qh^{(n,p,q)}_{\mu\nu L}(t)\nhat^L,
\end{equation}
where $\nhat^L:=n^{\langle L\rangle}:=n^{\langle i_1}\cdots n^{i_l\rangle}$, and for brevity, in later expressions I will write $h^{(n,p,0)}_{\mu\nu L}=h^{(n,p)}_{\mu\nu L}$. The quantity $\nhat^L$ plays the same role as a scalar spherical harmonic: it satisfies the eigenvalue equation $r^2\partial_i\partial^i \nhat^L=-l(l+1)\nhat^L$, thereby reducing the Poisson equations to algebraic equations. It also satisfies the useful identities $\partial_a r=n_a$, $n^a\partial_a\nhat^L=0$, and $\partial^a\partial_a (r^p\nhat^L)=(p-l)(p+l+1)r^{p-2}\nhat^L$. I refer to Ref.~\cite{Blanchet-Damour:86} for others.

At each order in $r$, a new homogeneous solution arises, corresponding to one of the standard solutions $r^l$ or $1/r^{l+1}$ to the Laplace equation. The solution to all orders in $\e$ and $r$ can then be written in terms of the coefficients of these solutions, $h^{(n,p)}_{\mu\nu P}$ for $p\geq0$ and $h^{(n,p)}_{\mu\nu |P|-1}$ for $p<0$~\cite{Pound:12b}. For $p=-n$, these coefficients correspond to the multipole moments of $g^{\rm obj}_{\mu\nu}$; for $-n<p<0$, they correspond to (potentially gauge) corrections to those moments; for $p\geq0$ they together make up a smooth solution to the vacuum field equation even at $r=0$. If no additional boundary conditions are specified, then the nonnegative-$p$ coefficients $h^{(n,p)}_{\mu\nu P}$ remain entirely arbitrary (up to relationships imposed by the gauge condition). They become determined when global ({e.g.}, retarded) boundary conditions are imposed.

Motivated by this division of terms, I split $h^n_{\mu\nu}$ into a ``self-field'' $h^{\S n}_{\mu\nu}$ and an ``effective field'' $h^{\R n}_{\mu\nu}$, as defined in Ref.~\cite{Pound:12a}. The effective field I define to be the piece of the total solution containing none of the negative-$p$ coefficients $h^{(n,p)}_{\mu\nu |P|-1}$; this makes $g_{\mu\nu}+\sum_n \e^n h^{\R n}_{\mu\nu}$ a smooth vacuum metric at $r=0$, determined by global boundary conditions. The self-field I define to be the remainder of the full field; it carries the local information about the object, and at $r=0$ it diverges as $r^{-n}$. Due to their behaviors at the origin, I refer to $h^{\S n}_{\mu\nu}$ and $h^{\R n}_{\mu\nu}$ as the singular and regular fields, respectively. While $r=0$ is outside the domain of validity of the outer expansion, this extension of the fields to all points $r>0$ (and to $r=0$ in the case of the regular field) has no impact on the field in the region $r\gg\e$, and it is essential in practice: at first order it is used to show that $h^1_{\mu\nu}$ is identical to the field of a point mass; and at higher orders it allows us to define practical puncture schemes that can compute the metric outside the object while bypassing its internal dynamics~\cite{Pound:12b,Pound:15a}.
 
At first order, the above definitions  lead to a singular field in which each term is explicitly proportional to the object's mass $m$. It is given by
\begin{subequations}\label{hS1}
\allowdisplaybreaks
\begin{align}
h^{\S1}_{tt} &= \frac{2m}{r}+3ma_i n^i+\tfrac{5}{3} mr\mathcal{E}_{ab} 
								\hat{n}^{ab}\nonumber\\
			 &\quad +\tfrac{7}{12}mr^2\mathcal{E}_{abc} \hat{n}^{abc}+\O(r^3,r^2a,ra^2),\\
h^{\S1}_{ta} &= mr\left(\tfrac{2}{3} \mathcal{B}^{bc} \epsilon_{acd} \hat{n}_{b}{}^{d} - 2 \dot a_{a}\right)
								-mr^2\bigl(\tfrac{19}{30}  \dot{\mathcal{E}}_{ab} \hat{n}^{b}\nonumber\\   
			&\quad + \tfrac{1}{18}  \dot{\mathcal{E}}^{bc} \hat{n}_{abc}  
								+ \tfrac{2}{9} \mathcal{B}^{bcd} \epsilon_{ab}{}^{i} \hat{n}_{cdi}\bigr)\nonumber\\
			&\quad +\O(r^3,r^2a,ra^2),\\
h^{\S1}_{ab} &= \frac{2m\delta_{ab}}{r}-m\delta_{ab}a_in^i-mr\bigl(\tfrac{38}{9} \mathcal{E}_{ab} 
								- \tfrac{4}{3} \mathcal{E}_{(a}{}^{c} \hat{n}_{b)c} \nonumber\\
				&\quad + \delta_{ab}\mathcal{E}_{cd} \hat{n}^{cd}\bigr) - mr^2\Bigl[ 
								 \left(\tfrac{31}{15}  \mathcal{E}_{abc} 
								+  \tfrac{68}{45}  \dot{\mathcal{B}}_{(a}{}^{d} \epsilon_{b)cd} 
								\right)\hat{n}^{c}\nonumber\\
							&\quad	- \tfrac{2}{3}\mathcal{E}_{(a}{}^{cd} 
								 \hat{n}_{b)cd}  - \tfrac{2}{9}  \dot{\mathcal{B}}^{cd} \epsilon^i{}_{c(a} \hat{n}_{b)di} 
								+\tfrac{5}{12} \delta_{ab}\mathcal{E}^{cdi}  \hat{n}_{cdi} \Big]\nonumber\\ 				&\quad +\O(r^3,r^2a,ra^2),
\end{align}
\end{subequations}
where an overdot denotes differentiation with respect to $t$, as in $\dot a^i = \frac{da^i}{dt}$. The gauge condition determines the evolution equations $\dot m =0$ and $a^\mu=\O(\e)$. Because $a^\mu=\O(\e)$, I have omitted terms that will be unnecessary for the matching procedure; the complete expression is given in the supplemental material~\cite{supplemental}. 

The first-order regular field is given by 
\begin{align}
 h^{\R1}_{\mu\nu} &= h^{(1,0)}_{\mu\nu} + r h^{(1,1)}_{\mu\nu i}n^i\nonumber\\
&\quad +r^2\left(h^{(1,2)}_{\mu\nu ij}\nhat^{ij} +  h^{(1,2)}_{\mu\nu}\right) + \O(r^3).\label{hR1}
\end{align}
In the order-by-order solution described above, the wave equation leads to $h^{(1,2)}_{\mu\nu}=\tfrac{1}{6}(h^{(1,0)}_{\mu\nu,tt}-2R_{\mu}{}^\alpha{}_\nu{}^\beta\big|_\gamma h^{(1,0)}_{\alpha\beta})$, such that all the coefficients in Eq.~\eqref{hR1} are written entirely in terms of the coefficients $h^{(n,p)}_{\mu\nu P}$. We can also write the regular field as the Taylor series
\begin{align}
h^{\R1}_{\mu\nu} &= \left. h^{\R1}_{\mu\nu}\right|_{\gamma} + \left.h^{\R1}_{\mu\nu,i}\right|_{\gamma}x^i\nonumber\\
&\quad +\frac{1}{2}\left. h^{\R1}_{\mu\nu,ij}\right|_{\gamma}x^ix^j + \O(r^3),
\end{align}
where the coefficients are related to the pieces of the full metric as
\begin{align}
\left. h^{\R1}_{\mu\nu}\right|_{\gamma} &= h^{(1,0)}_{\mu\nu},\\
\left.h^{\R1}_{\mu\nu,i}\right|_{\gamma} &= h^{(1,1)}_{\mu\nu i},\\
\left.h^{\R1}_{\mu\nu,\langle ij\rangle}\right|_{\gamma} &= 2h^{(1,2)}_{\mu\nu ij},\\
\left.h^{\R1}_{\mu\nu,i}{}^i\right|_{\gamma} &= 6h^{(1,2)}_{\mu\nu}.
\end{align}

The first-order singular and regular fields defined this way agree with the Detweiler-Whiting definitions~\cite{Detweiler-Whiting:03} at least through the displayed orders in $r$~\cite{Pound-Miller:14}.

At second order in $\e$, the singular field generically involves the object's mass dipole moment and spin, as in Eq.~\eqref{h2form}. With those moments set to zero, the singular field has three pieces:
\beq\label{hS2}
h^{\S2}_{\mu\nu} = h^{\S\S}_{\mu\nu} + h^{\S\R}_{\mu\nu} + h^{\delta m}_{\mu\nu}.  
\eeq

The first piece comprises terms explicitly proportional to $m^2$,
\begin{subequations}\label{hSS}%
\allowdisplaybreaks\begin{align}
h^{\S\S}_{tt} &= -\frac{2m^2}{r^2} - \tfrac{7}{3}m^2\mathcal{E}_{ab} 
							\hat{n}^{ab}-  \tfrac{3}{2} m^2 r \mathcal{E}_{abc} \hat{n}^{abc}\nonumber\\
							&\quad+\O(r^2\ln r,a),\\
h^{\S\S}_{ta} &= - \tfrac{10}{3}m^2\mathcal{B}^{bc} \epsilon_{acd} \hat{n}_{b}{}^{d}+\tfrac{26}{15} m^2r\log r\, \dot{\mathcal{E}}_ {ai}  
\hat{n}^{i}\nonumber\\
&\quad - \tfrac{1}{45} m^2r\bigl(31 \dot{\mathcal{E}}^{bc} \hat{n}_{abc} + 100 
\mathcal{B}^{bcd} \epsilon_{ab}{}^{i} \hat{n}_ {cdi}\bigr)\nonumber\\&\quad+\O(r^2\ln r,a),\\
h^{\S\S}_{ab} &= \frac{\tfrac{8}{3}m^2\delta_{ab} - 7m^2\nhat_{ab}}{r^2}
 + m^2\Big(4 \mathcal{E}_{c(a} \hat{n}_{b)}{}^{c} \nonumber\\
						&\quad - \tfrac{4}{3} \delta_{ab}\mathcal{E}_{cd}  \hat{n}^{cd} + \tfrac{7}{5}\mathcal{E}_{cd} \hat{n}_{ab}{}^{cd}\Big) -\tfrac{16}{15}m^2\mathcal{E}_{ab}\ln r\nonumber\\
						&\quad-m^2r\log r \left( \tfrac{4}{5} \mathcal{E}_{abc} \hat{n}^{c} +  \tfrac{12}{5}  \dot{\mathcal{B}}_{(a}{}^{d} \epsilon_{b)cd} \hat{n}^{c}\right) \nonumber\\
						&\quad + m^2 r \left(\tfrac{493}{180} \mathcal{E}_{(a}{}^{cd} \hat{n}_{b)cd} +  \tfrac{52}{45} \dot{\mathcal{B}}^{cd} \epsilon^i{}_{c(a} \hat{n}_{b)di} \right. \nonumber\\
						&\quad \left.-  \tfrac{131}{108}\delta_{ab} \mathcal{E}^{cdi}  \hat{n}_{cdi} + \tfrac{1}{3} \mathcal{E}^{cdi} \hat{n}_{abcdi}\right)\nonumber\\
						&\quad+\O(r^2\ln r,a).
\end{align}
\end{subequations}
Here I have dropped all acceleration terms, which can be found in the supplemental material. This field satisfies $E_{\mu\nu}[h^{\S\S}]=2\delta^2 R_{\mu\nu}[h^{\S1}]$ pointwise away from $r=0$; because the source is quadratic in $h^{\S1}_{\mu\nu}$, the equation is not distributionally well defined on regions that include $r=0$. 

The second piece comprises terms of the form $mh^{\R1}$. Because the explicit expressions are very lengthy, I give only the leading order:
\begin{subequations}\label{hSR}%
\begin{align}
h^{\S\R}_{tt} &= -\frac{m h^{\R1}_{ab} \hat{n}^{ab}}{r}+\O(r^0),\\
h^{\S\R}_{ta} &= -\frac{m h_{tb}^{\R1}\hat{n}_a{}^b}{r}+\O(r^0),\\
h^{\S\R}_{ab} &= \frac{m}{r}\Big[2h^{\R1}_{c(a}\hat{n}_{b)}{}^c -\delta_{ab} h^{\R1}_{cd} \hat{n}^{cd} \nonumber\\
						&\quad - \left(h^{\R1}_{ij}\delta^{ij}+h_{tt}^{\R1}\right)\hat{n}_{ab}\Big] +\O(r^0),
\end{align}
\end{subequations}
where components of $h^{\R1}_{\mu\nu}$ are evaluated at $r=0$. The subleading terms, which are shown in the supplemental material, are of the form $mr\partial h^{\R1}$ and $mr^2\partial^2 h^{\R1}$. This field satisfies $E_{\mu\nu}[h^{\S\R}]=2\delta^2 R_{\mu\nu}[h^{\S1},h^{\R1}]+2\delta^2 R_{\mu\nu}[h^{\R1},h^{\S1}]$ pointwise away from $r=0$, where I have written $\delta^2 R_{\mu\nu}$ as a bilinear operator in the natural way. Because the source is a linear operator acting on the singular field, this equation is distributionally well defined even if $r=0$ is included; however, I have not confirmed that its two sides are equal as distributions.

The final piece of the singular field is the $\delta m$ term:
\begin{subequations}\label{hdm}
\begin{align}
h^{\delta m}_{tt} &= \frac{\delta m_{tt}}{r} + r\left(\tfrac{5}{6} \delta m_{tt} \mathcal{E}^{ab} \hat{n}_{ab} -  \mathcal{B}^{bc} \delta m_t{}^{a} \epsilon_{acd} \hat{n}_{b}{}^{d} \right.\nonumber\\ &\quad \left.+ \tfrac{1}{2} \ddot{\delta m}_{tt}\right) +\O(r^2,a),\\
h^{\delta m}_{ta} &= \frac{\delta m_{ta}}{r} + r\left(\tfrac{1}{3} \delta m_t{}^{b} \mathcal{E}_{(b}{}^{c} \hat{n}_{a)c}- \tfrac{19}{18} \delta m_t{}^{b} \mathcal{E}_{ab}\right. \nonumber\\
&\quad-  \mathcal{B}^{bc} \delta m_{b}{}^{d} \epsilon_{acd}  + \tfrac{1}{6} \delta m_{ta} \mathcal{E}^{bc} \hat{n}_{bc} + \tfrac{1}{6} \mathcal{B}^{bc} \delta m_{tt} \epsilon_{acd} \hat{n}_{b}{}^{d}\nonumber\\
&\quad \left.+ \tfrac{1}{2} \mathcal{B}^{bc} \delta m_{a}{}^{d} \epsilon_{cdi} \hat{n}_{b}{}^{i} + \tfrac{1}{2}\ddot{\delta m}_{ta}\right)+\O(r^2,a),\\
h^{\delta m}_{ab} &= \frac{\delta m_{ab}}{r}+r\left(\tfrac{17}{9} \delta m_{(a}{}^{c} \mathcal{E}_{b)c}- \delta m^{c}{}_{c} \mathcal{E}_{ab} -  \delta m_{tt} \mathcal{E}_{ab}\right.\nonumber\\
&\quad  +  2\delta m_t{}^{c} \mathcal{B}_{(a}{}^{d}\epsilon_{b)cd}  -  \delta m^{cd} \mathcal{E}_{cd} \delta_{ab} + \tfrac{1}{3} \mathcal{E}_{c}{}^{d} \delta m_{(a}{}^{c} \hat{n}_{b)d}  \nonumber\\
&\quad + \tfrac{1}{3} \delta m_{(a}{}^{c} \mathcal{E}_{b)}{}^{d} \hat{n}_{cd} -  \tfrac{1}{2} \delta m_{ab} \mathcal{E}^{cd} \hat{n}_{cd} \nonumber\\
&\quad\left.+ \tfrac{1}{3} \mathcal{B}^{cd} \delta m_{t(a} \epsilon_{b)di} \hat{n}_{c}{}^{i} + \tfrac{1}{2} \ddot{\delta m}_{ab}\right)+\O(r^2,a),
\end{align}
\end{subequations}
where the acceleration terms are given in the supplemental material. $\delta m_{\mu\nu}$ is a tensor on $\gamma$ that can be thought of as a mass correction  (though only in the loosest  sense),\footnote{Besides being pure gauge, $h^{\delta m}_{\mu\nu}$ corresponds to an $l=0$ perturbation only in terms of {\em scalar} harmonics; Eq.~\eqref{hn mode expansion} is equivalent to a scalar-harmonic expansion of each Cartesian component. In terms of tensor harmonics, $h^{\delta m}_{\mu\nu}$'s $ta$ component is $l=1$, and the trace-free part of its $ab$ component is $l=2$~\cite{Blanchet-Damour:86}. However, it is nevertheless useful to separate $h^{\delta m}_{\mu\nu}$ from $h^{\S\R}_{\mu\nu}$ because it satisfies Eq.~\eqref{Ehdm}. If we view the right-hand side of that equation as a stress-energy tensor, we see that the trace part of $\delta m_{ab}$ can naively be interpreted as a kinetic energy on the worldline, the $ta$, $l=1$ piece as a flux of energy out of the worldline, and the trace-free, $l=2$ part as a flux of momentum.} the same that appears in Eq.~\eqref{H1form}. Its components, as determined by the gauge condition, are
\begin{subequations}\label{dm}
\begin{align}
\delta m_{tt} &= - \tfrac{1}{3} m h^{\R1}_{ab}\delta^{ab} - 2 m h_{tt}^{\R1},\\
\delta m_{ta} &= - \tfrac{4}{3} m h_{ta}^{\R1}, \\
\delta m_{ab} &= \tfrac{2}{3} m h^{\R1}_{ab} + \tfrac{1}{3} m \delta_{ab} h^{\R1}_{cd}\delta^{cd}\nonumber\\
							&\quad + \tfrac{2}{3} m \delta_{ab} h_{tt}^{\R1}.
\end{align}
\end{subequations}
where all components of the regular field are evaluated at $r=0$. This ``mass correction'' is pure gauge, as we will see in the matching procedure. It can be freely altered by adding a term of the form $2\,\delta m\,\delta_{\mu\nu}$ that contains invariant mass, but I absorb that term into $m$.  $h^{\delta m}_{\mu\nu}$ satisfies $E_{\mu\nu}[h^{\delta m}]=0$ at points away from $r=0$, and it satisfies the point-particle-like equation 
\begin{equation}\label{Ehdm}
E_{\mu\nu}[h^{\delta m}] = - 4\pi\delta m_{\mu\nu}(t)\delta^3(x^i)
\end{equation}
on a region that includes $r=0$.

In addition to determining $\delta m_{\mu\nu}$, the gauge condition (together with the center-of-mass condition $M^i=0$) determines that the worldline $\gamma$ has acceleration
\begin{equation}\label{a1_Fermi}
a_a = \e\left[\frac{1}{2}\partial_a h^{\R1}_{tt}-\partial_t h^{\R1}_{ta}\right]+\O(\e^2),
\end{equation}
which can be written in the covariant form
\beq\label{a1-cov}
a^\mu = \frac{\e}{2}P^{\mu\nu}\left(h^{{\rm R}1}_{\sigma\lambda;\nu}-2h^{{\rm R}1}_{\nu\sigma;\lambda}\right)u^\sigma u^\lambda+\O(\e^2),
\eeq
where $P^{\mu\nu}:=g^{\mu\nu}+u^\mu u^\nu$. This is the usual result for the first-order equation of motion. If $g^{\rm obj}_{\mu\nu}$ had spin, then this equation would include the Papapetrou spin force~\cite{Gralla-Wald:08,Pound:10a}.

Finally, the second-order regular field is given by
\begin{align}
 h^{\R2}_{\mu\nu} &= h^{(2,0)}_{\mu\nu} + r h^{(2,1)}_{\mu\nu i}n^i + \O(r^2).\label{hR2}
\end{align}

Note that in this paper, I have followed Ref.~\cite{Pound:12a} (and the earlier Ref.~\cite{Pound:10a}) by defining the second-order singular and regular fields based on a multipole decomposition of the metric perturbation. This differs slightly from the definition in Ref.~\cite{Pound:12b}, which instead defined the singular and regular fields based on the exactly analogous multipole decomposition of the trace-reversed second-order field $\bar h^{\mu\nu}_{2}$. These definitions are not quite equivalent. That is, if $h^{\R2}_{\mu\nu}$ is as defined here and $\bar h_{2}^{\R\rho\sigma}$ is as defined in Ref.~\cite{Pound:12b}, then  $h^{\R2}_{\mu\nu}\neq g_{\mu\alpha} g_{\nu\beta}\left(\bar h_{2}^{\R\alpha\beta}-\frac{1}{2}g^{\alpha\beta}g_{\rho\sigma}\bar h_{2}^{\R\rho\sigma}\right)$. The relationship between the two can be found by calculating the trace reverse of $\bar h_{2}^{\mu\nu}$ and decomposing the result into the form \eqref{hn mode expansion}. The uniqueness of the decomposition allows one to read off the pieces of $\bar h_{2}^{\R\mu\nu}$ and $\bar h_{2}^{\S\mu\nu}$ appearing in $h^{\R2}_{\mu\nu}=h^{(2,0)}_{\mu\nu}+r h^{(2,1)}_{\mu\nu i}n^i+\O(r^2)$. The results of that procedure are shown in Appendix~\ref{hR relations}.

\subsection{Expanding the acceleration}\label{acceleration expansion}
The Fermi-Walker coordinates I use are tethered to an $\e$-dependent worldline. This introduces an $\e$ dependence that would not appear in the original coordinates $x$ of Eq.~\eqref{external ansatz}. Even the background $g_{\mu\nu}$ hence inherits a dependence on $\e$ in this coordinate system. This dependence comes in two forms: implicitly within any function of $t$, since a tensor evaluated at $\gamma(t)$ automatically inherits $\gamma(t)$'s $\e$ dependence; and explicitly in the overt appearance of the acceleration $a^\mu\sim\e$.

Working in  a system that moves with the accelerating worldline necessitates holding the implicit $\e$ dependence unexpanded; expanding it would effectively expand tensors around their values on a nearby, $\e$-independent geodesic. However, it is natural to expand the explicit $\e$ dependence, as locally there is no way to distinguish between a small term that comes from $a^\mu$ and a small term that comes from $h^n_{\mu\nu}$. Indeed, the inner expansion will not  make this distinction.

Hence, prior to matching the metrics, I substitute the expansion $a^\mu=\sum_{n>0}\e^nf_n^\mu$ into $g_{\mu\nu}$ and $h^n_{\mu\nu}$ and regroup terms. I write this re-expansion as, for example, $g_{\mu\nu}=\ord{0}{g}_{\mu\nu}+\e\,\ord{1}{g}_{\mu\nu}+\e^2\,\ord{2}{g}_{\mu\nu}+\O(\e^3)$. Explicitly, the terms in the expansion of $g_{\mu\nu}$ are
\begin{subequations}\label{delta0g}
\begin{align}
\ord{0}{g}_{tt} &= -1 - r^2 \mathcal{E}^{ab} \hat{n}_{ab}  - r^3\frac{1}{3} \mathcal{E}^{abc} \hat{n}_{abc}+\O(r^4),\\
\ord{0}{g}_{ta} &= -\frac{2}{3}r^2 \mathcal{B}^{bc} \epsilon_{acd} \hat{n}_{b}{}^{d} + \frac{1}{20}r^3 \dot{\mathcal{E}}_{ab} \hat{n}^{b}- \frac{1}{12}r^3 \dot{\mathcal{E}}^{bc} \hat{n}_{abc}\nonumber\\
				&\quad  - \frac{1}{3} r^3\mathcal{B}^{bcd} \epsilon_{ab}{}^{i} \hat{n}_{cdi}+\O(r^4),\\
\ord{0}{g}_{ab} &= \delta_{ab} -\frac{1}{9}r^2 (\mathcal{E}_{ab} - 6 \mathcal{E}_{(a}{}^{c} \hat{n}_{b)c} 
				+ 3 \mathcal{E}^{cd} \delta_{ab} \hat{n}_{cd})\nonumber\\
				&\quad + \frac{1}{90}r^3\! \left(30 \mathcal{E}_{(a}{}^{cd} \hat{n}_{b)cd} -3 \mathcal{E}_{abc} \hat{n}^{c} 
				- 8 \dot{\mathcal{B}}_{(a}{}^{d} \epsilon_{b)cd} \hat{n}^{c}\right. \nonumber\\
				&\quad \left.+ 10 \dot{\mathcal{B}}^{cd} \epsilon_{c(a}{}^{i} \hat{n}_{b)di}\! -\! 15 \mathcal{E}^{cdi} \delta_{ab} \hat{n}_{cdi}\right)\!+\O(r^4),
\end{align}
\end{subequations}
and
\begin{align}
\ord{1}{g}_{\mu\nu} &= -2f^1_ix^i\delta_\mu^t\delta_\nu^t+\O(r^3),\label{d1g}\\
\ord{2}{g}_{\mu\nu} &= -2f^2_ix^i\delta_\mu^t\delta_\nu^t+\O(r^2).\label{d2g}
\end{align}
The linear term in the expansion $h^1_{\mu\nu}=\ord{0}{h}^1_{\mu\nu}+\e\, \ord{1}{h}^1_{\mu\nu}+\O(\e^2)$ is
\begin{subequations}\label{d1h1}
\begin{align}
\ord{1}{h}^1_{tt} &= \ord{1}{h}^{(1,0)}_{tt} + \ord{1}{h}^{(1,1)}_{tti}x^i + 3ma^1_in^i +\O(r^2),\\
\ord{1}{h}^1_{ta} &= \ord{1}{h}^{(1,0)}_{ta} +  \ord{1}{h}^{(1,1)}_{tai}x^i - 2mr\dot a^1_a +\O(r^2),\\
\ord{1}{h}^1_{ab} &= \ord{1}{h}^{(1,0)}_{ab} +\ord{1}{h}^{(1,1)}_{abi}x^i - \delta_{ab}ma^1_in^i  +\O(r^2).
\end{align}
\end{subequations}
Since $h^{(1,p)}_{\mu\nu P}$ is undetermined until global boundary conditions are imposed, we cannot always necessarily find exact expressions for the $\ord{1}{h}^{(1,p)}_{\mu\nu P}$ terms in the above equations. However, if we assume retarded boundary conditions, these quantities can be obtained from an analytical expansion of the retarded integral; the results of that expansion are given in Eqs.~\eqref{hR explicit} and \eqref{dhR explicit}. At present, there is no such analytical form at second order, and we cannot provide explicit results for $\ord{k}{h}^{(2,p)}_{\mu\nu P}$. But such expressions will not be necessary.

Given these expansions, the metric can be written as 
\beq\label{gNew}
{\sf g}_{\mu\nu} = \ord{0}{g}_{\mu\nu}+\e h^{1\dagger}_{\mu\nu}+\e^2h^{2\dagger}_{\mu\nu}+\O(\e^3).
\eeq
The first-order perturbation becomes
\beq\label{h1New}
h^{1\dagger}_{\mu\nu} = \ord{0}{h}^{\S1}_{\mu\nu}+\ord{0}{h}^{\R1}_{\mu\nu}+ \ord{1}{g}_{\mu\nu},
\eeq
where $h^{\S1}_{\mu\nu}$ and $h^{\R1}_{\mu\nu}$ are given by Eqs.~\eqref{hS1} and \eqref{hR1}, and $\ord{1}{g}_{\mu\nu}$ by Eq.~\eqref{d1g}. Last, the second subleading perturbation becomes
\beq\label{h2New}
h^{2\dagger}_{\mu\nu} = \ord{0}{h}^{\S\S}_{\mu\nu}+\ord{0}{h}^{\S\R}_{\mu\nu}+\ord{0}{h}^{\delta m}_{\mu\nu}+\ord{0}{h}^{\R2}_{\mu\nu} + \ord{1}{h}^1_{\mu\nu}+\ord{2}{g}_{\mu\nu},
\eeq
where $h^{\S\S}_{\mu\nu}$, $h^{\S\R}_{\mu\nu}$, $h^{\delta m}_{\mu\nu}$, and $h^{\R2}_{\mu\nu}$ are given by Eqs.~\eqref{hSS}, \eqref{hSR}, \eqref{hdm}, and \eqref{hR2}, $\ord{1}{h}^1_{\mu\nu}$ by Eq.~\eqref{d1h1}, and $\ord{2}{g}_{\mu\nu}$ by Eq.~\eqref{d2g}.

To obtain a unique gauge transformation between this expansion and that in the rest gauge, it will be useful to decompose the coefficients that appear in the regular field (at both first and second order) into irreducible form. This decomposition is described in Appendix~\ref{STF decomposition}, and it is given by 
\begin{subequations}\label{hR decomposition}
\begin{align}
h^{(n,p)}_{ttP} &= \hat A^{(n,p)}_P,\\
h^{(n,p)}_{taP} &= \hat C^{(n,p)}_{aP} + \epsilon^j{}_{a\langle i_p}\hat D^{(n,p)}_{P-1\rangle j}
								+ \delta_{a\langle i_p}\hat B^{(n,p)}_{P-1\rangle},\\
h^{(n,p)}_{abP} &= \delta_{ab}\hat K^{(n,p)}_P + \hat H^{(n,p)}_{abP} +\mathop{\STF}_{ab}\Big(\epsilon^c{}_{ai_p}\hat I^{(n,p)}_{bcP-1}\nonumber\\
						&\quad+\delta_{ai_p}\hat F^{(n,p)}_{b P-1} +\delta_{ai_p}\epsilon^c{}_{bi_{p-1}}\hat G^{(n,p)}_{cP-2} \nonumber\\
						&\quad+\delta_{ai_p}\delta_{bi_{p-1}}\hat E^{(n,p)}_{P-2}\Big).
\end{align}
\end{subequations}
For brevity, after expanding the acceleration I combine STF tensors as, for example, $\hat{A}^{(2,p)\dagger}_P := \ord{0}{\hat{A}}^{(2,p)}_P + \ord{1}{\hat{A}}^{(1,p)}_P$. The wave equation leaves each of these STF tensors to be freely specified by boundary conditions. However, the gauge condition imposes the following relationships between them:
\begin{subequations}\label{STF gauge relations 1}
\allowdisplaybreaks
\begin{align}
\ord{0}{\hat B}^{(1,1)} & = \tfrac{1}{6} \partial_t\, \ord{0}{\hat{A}}^{(1,0)} + \tfrac{1}{2} \partial_t \ord{0}{\hat{K}}^{(1,0)},\\
\ord{0}{\hat F}^{(1,1)}_a &= - \tfrac{3}{10} \ord{0}{\hat{A}}^{(1,1)}_{a} + \tfrac{3}{10} \ord{0}{\hat{K}}^{(1,1)}_{a} + \tfrac{3}{5} \partial_t \ord{0}{\hat{C}}^{(1,0)}_{a},\\
\ord{0}{\hat A}^{(1,2)}_{ab} &= \tfrac{2}{3}\, \ord{0}{\hat{A}}^{(1,0)} \mathcal{E}_{ab} -  \tfrac{4}{3} \mathcal{B}_{(a}{}^{d}  \epsilon_{b)}{}^c{}_{d} \ord{0}{\hat{C}}^{(1,0)}_{c}  
						-  \tfrac{7}{6}\,\ord{0}{ \hat{F}}^{(1,2)}_{ab}\nonumber\\
						&\quad - \tfrac{13}{9} \mathcal{E}_{\langle a}{}^{c}\, \ord{0}{\hat{H}}^{(1,0)}_{b\rangle c} + \tfrac{5}{9} \mathcal{E}{}_{ab} \ord{0}{\hat{K}}^{(1,0)} + \ord{0}{\hat{K}}^{(1,2)}_{ab} \nonumber\\
						&\quad + \partial_t \ord{0}{\hat{C}}^{(1,1)}_{ab} -  \tfrac{1}{3} \partial_t\partial_t\, \ord{0}{\hat{H}}^{(1,0)}_{ab},\\
\ord{0}{\hat B}^{(1,2)}_a &= \tfrac{2}{15} \ord{0}{\hat{C}}^{(1,0)}_{b} \mathcal{E}_{a}{}^{b} + \tfrac{1}{5} \mathcal{B}{}^{bc} \epsilon{}_{ac}{}^d \ord{0}{\hat{H}}^{(1,0)}_{bd} 
					+ \tfrac{3}{20} \partial_t \ord{0}{\hat{A}}^{(1,1)}_{a} \nonumber\\
					&\quad + \tfrac{9}{20} \partial_t \ord{0}{\hat{K}}^{(1,1)}_{a} -  \tfrac{1}{10} \partial_t\partial_t\, \ord{0}{\hat{C}}^{(1,0)}_{a},\\
\ord{0}{\hat E}^{(1,2)} &= \tfrac{1}{45} \mathcal{E}{}^{ab}\, \ord{0}{\hat{H}}^{(1,0)}_{ab} + \tfrac{1}{5} \partial_t\partial_t\, \ord{0}{\hat{K}}^{(1,0)},\\
\ord{0}{\hat G}^{(1,2)}_a &= \tfrac{1}{15} \mathcal{E}{}^{b}_{c}\, \epsilon{}_{a}{}^{cd}\, \ord{0}{\hat{H}}^{(1,0)}_{db} + \tfrac{2}{5} \partial_t \ord{0}{\hat{D}}^{(1,1)}_{a},
\end{align}
\end{subequations}
and
\begin{subequations}\label{STF gauge relations 2}
\begin{align}
\hat B^{(2,1)\dagger} &= \tfrac{1}{6} \partial_t \hat{A}^{(2,0)\dagger} + \tfrac{1}{2} \partial_t \hat{K}^{(2,0)\dagger} - \tfrac{1}{3} \ord{0}{\hat{C}}^{(1,0)}_{a} f_1^{a},\\
\hat F^{(2,1)\dagger}_a &= - \tfrac{3}{10} \hat{A}^{(2,1)\dagger}_{a} + \tfrac{3}{10} \hat{K}^{(2,1)\dagger}_{a} + \tfrac{3}{5} \partial_t \hat{C}^{(2,0)\dagger}_{a} \nonumber\\
				&\quad -  \tfrac{3}{5} f_1^{b}\, \ord{0}{\hat{H}}^{(1,0)}_{ab} - \tfrac{3}{5} f^1_{a}\, \ord{0}{\hat{K}}^{(1,0)}. 
\end{align}
\end{subequations}


\section{Inner expansion in a rest gauge}\label{inner_expansion}

With the outer expansion determined in the buffer region, the goal is now to find an inner expansion that is compatible with the outer, that describes the metric in a rest gauge, and that is sufficiently general for the matching calculation. 

\subsection{Form of the expansion}
We could obtain the inner expansion directly in terms of scaled Fermi-Walker coordinates $(t,\tilde x^a)$. However, it is more convenient to work in a coordinate system tailored to the inner expansion. So let $\sfx^\alpha=(\sft,\sfx^a)$ be some quasi-Cartesian coordinate system centered on the object, introduce the scaled coordinates $\tilde\sfx^a=\sfx^a/\e$, and assume the expansion
\begin{align}
\exact{g}_{\mu\nu}(\sfx^\alpha,\e) & = g^{\rm obj}_{\mu\nu}(\sft,\tilde\sfx^a) + \sum_{n\geq 1}\e^n H^n_{\mu\nu}(\sft,\tilde\sfx^a).\label{internal ansatz}
\end{align}
Here all quantities with indices, such as $g^{\rm obj}_{\mu\nu}(\sft,\tilde\sfx^a)$, are the components of tensors, such as $g^{\rm obj}_{\mu\nu}(\sft,\tilde\sfx^a)d\sfx^\mu d\sfx^\nu$, in unscaled coordinates $\sfx^\alpha$, but expressed as functions of the scaled coordinates $(\sft,\tilde\sfx^a)$. One could equivalently write the expansion for the components in the scaled coordinates, in which case overall factors of $\e$ and $\e^2$ would be introduced into time-space and space-space components, respectively.\footnote{These overall factors are not of practical relevance, but they do mean that in the limit $\e\to0$ in these coordinates, the metric becomes one-dimensional, similar to the behavior of the metric in the post-Newtonian limit. If a regular limit is desired, it can be obtained by rescaling time as well, such that $\tilde \sft=(\sft-\sft_0)/\e$, and then introducing a conformally rescaled metric $\tilde {\sf g} _{\mu\nu}=\frac{1}{\e^2}{\sf g}_{\mu\nu}$, as was done by D'Eath~\cite{DEath:75} and later by Gralla and Wald~\cite{Gralla-Wald:08}. In that approach, the inner expansion zooms in not only on a small region around the object, but also on a small interval of time.} 

For this expansion to be appropriately related to the outer one, I will enforce three conditions: (a) there is no mass dipole moment in the metric, such that the object is effectively mass-centered at $\tilde \sfx=0$, (b) the transformation from $\sfx^a$ to Fermi-Walker coordinates $x^a$ does not change the position of the origin, such that the center-of-mass position $\tilde \sfx^a=0$ can be identified with $\gamma$, and (c) the transformation is the form $\sfx^\alpha=\sfx^\alpha_0(x^\beta)+\O(\e)$, with no negative powers of $\e$, such that the inner expansion correctly refers to an expansion at fixed $x^a/\e$ rather than, say, at fixed $x^a/\e^2$. Furthermore, for the expansions to match, it is understood that any dependence on $\sft$ can include some dependence on $\e$ in the same manner as the outer expansion, folding in the $\e$ dependence of $\gamma$.


Although an inner expansion can be used to find an accurate metric even in the object's interior, here I am only interested in the metric in the buffer region. Hence, I seek a solution to the EFE in a vacuum region outside the object.  Substituting Eq.~\eqref{internal ansatz} into the vacuum EFE leads to
\begin{align}
0 &= G_{\mu\nu}[g^{\rm obj}] + \e \delta G_{\mu\nu}[H^1] + \e^2\delta G_{\mu\nu}[H^2] + \e^3\delta G_{\mu\nu}[H^3] \nonumber\\
	&\quad + \e^2\delta^2 G_{\mu\nu}[H^1] + \e^3\delta^2 G_{\mu\nu}[H^1,H^2] \nonumber\\
	&\quad + \e^3\delta^2 G_{\mu\nu}[H^2,H^1] + \e^3\delta^3 G_{\mu\nu}[H^1] \ldots \label{inner_EFE}
\end{align}
where $\delta^n G_{\mu\nu}[H]$ contains $n$ powers of $H_{\mu\nu}$ and its derivatives. Now note that derivatives with respect to $\sft$ are suppressed by a factor of $\e$ compared to derivatives with respect to $\sfx^a$. Hence,
\begin{align}
G_{\mu\nu} &= \e^{-2}\!\left(G^{(0)}_{\mu\nu}+\e G^{(1)}_{\mu\nu}+\e^2 G^{(2)}_{\mu\nu}\right)\!,\\
\delta^k G_{\mu\nu} &= \e^{-2}\!\left(\delta^kG^{(0)}_{\mu\nu}+\e \delta^kG^{(1)}_{\mu\nu}+\e^2 \delta^kG^{(2)}_{\mu\nu}\right)\!,
\end{align}
where the overall factors of $\e^{-2}$ result from the rescaling $\tilde \sfx=\sfx/\e$, and $G_{\mu\nu}^{(n)}$ and $\delta^kG_{\mu\nu}^{(n)}$ contain $n$ derivatives with respect to $\sft$. Picking off coefficients of $\e^n$ in Eq.~\eqref{inner_EFE} therefore leads to a sequence of linear equations for the perturbations $H^n_{\mu\nu}$,
\begin{align}
G^{(0)}_{\mu\nu}[g^{\rm obj}] &= 0,\label{inner_EFE0}\\
\delta G^{(0)}_{\mu\nu}[H^1] &= -G^{(1)}_{\mu\nu}[g^{\rm obj}],\label{inner_EFE1}\\
\delta G^{(0)}_{\mu\nu}[H^2] &= -\delta^2 G^{(0)}_{\mu\nu}[H^1] - \delta G^{(1)}_{\mu\nu}[H^1]  \nonumber\\
								&\quad - G^{(2)}_{\mu\nu}[g^{\rm obj}],\label{inner_EFE2}\\
\delta G^{(0)}_{\mu\nu}[H^3] &= -\delta^3 G^{(0)}_{\mu\nu}[H^1]  -\delta^2 G^{(0)}_{\mu\nu}[H^1,H^2] \nonumber\\
								&\quad -\delta^2 G^{(0)}_{\mu\nu}[H^2,H^1]  -\delta^2 G^{(1)}_{\mu\nu}[H^1]  \nonumber\\
								&\quad -\delta G^{(2)}_{\mu\nu}[H^1] - \delta G^{(1)}_{\mu\nu}[H^2].\label{inner_EFE3}
\end{align}
Equating explicit powers of $\e$ in this way, despite the implicit $\e$ dependence contained in functions of $\sft$, applies the same rules as were used in the re-expansion of the outer expansion in the buffer region: implicit functional dependences on $\gamma$ are held fixed during the expansion procedure, while quantities with explict powers of $\e$, such as the acceleration terms in the outer expansion, are not. The dependence on $\sft$ will be determined by the matching procedure and by the time derivatives in Eqs.~\eqref{inner_EFE0}--\eqref{inner_EFE3}; while each of these is a linear equation for a given $H^n_{\mu\nu}$, it is also an equation for the time-evolution of lower-order terms.

\subsection{General solution in the buffer region}

In most self-force derivations using matched asymptotic expansions, authors take the inner background $g^{\rm obj}_{\mu\nu}$ to be the spacetime of a Schwarzschild black hole, and they construct the perturbations from the tidal moments of the ``external'' gravitational field (which implicitly includes some piece of $h_{\mu\nu}$, to be determined through matching to the outer expansion). Here I will do likewise, but I stress that there is no loss of generality in doing so: I am only interested in the solution in the buffer region, where the tidally perturbed Schwarzschild metric describes the spacetime outside {\em any} nearly spherical, nearly static, compact object.

Before presenting the metric, I review its derivation, with an eye toward its generality. First, I specialize to an object that is approximately spherical. Specifically, I impose that all the $l>0$ moments of the background metric $g^{\rm obj}_{\mu\nu}$ vanish, such that in the exterior of the object, $g^{\rm obj}_{\mu\nu}$ is the Schwarzschild metric in mass-centered coordinates.\footnote{\label{footnote}We could relax this condition to instead only specify that $g^{\rm obj}_{\mu\nu}$'s spin, mass dipole moment, and quadrupole moments vanish. These moments are the only ones that would affect the acceleration at the orders of interest. The spin would couple to the tidal moment $\B_{ab}$ to generate an acceleration term of the form $\sim \e^2 a_{\sf i}\tilde\sfx^{\sf i}$ in $\e^2H^{2}_{\mu\nu}$, corresponding to a first-order acceleration term $\sim \e a_{\sf i}\sfx^{\sf i}$ that would appear in $\e\,\ord{1}{g}_{\mu\nu}$ in the outer expansion. Similarly, the quadrupole moments would couple to the tidal moments $\E_{abc}$ and $\B_{abc}$ to generate an acceleration term  in $\e^3H^{3}_{\mu\nu}$, corresponding to a second-order acceleration term in $\e^2\,\ord{2}{g}_{\mu\nu}$. The spin-induced force is the standard Papapetrou spin force, rederived in self-force contexts in Refs.~\cite{Mino-Sasaki-Tanaka:97b,Gralla-Wald:08,Pound:10a}. The quadrupole-induced forces, although not yet derived consistently within the type of perturbative expansion used here, can be expected to agree with the  test-body-type forces derived in various contexts by, e.g., Dixon~\cite{Dixon:74}, Thorne and Hartle~\cite{Thorne-Hartle:85}, and Harte~\cite{Harte:12}. Any moments in $g^{\rm obj}_{\mu\nu}$ higher than quadrupolar would impact the outer expansion at too high an order to be relevant in the present analysis. Analogously, though $H^1_{\mu\nu}$ can include corrections to the moments, the only relevant one would be the spin; and in $H^{n>1}_{\mu\nu}$, not even a correction to the spin would be relevant.} In principle, because the background metric is only required to satisfy the time-independent equation~\eqref{inner_EFE0}, the mass of $g^{\rm obj}_{\mu\nu}$ could be allowed to depend on $\sft$. However, we already know that this mass is equal to the parameter $m$ in the outer expansion, and so we can appeal to the previous result that $m$ is constant in time; or we can establish that the mass is constant directly from Eq.~\eqref{inner_EFE1}~\cite{Pound:10b}. Intuitively, this follows from the fact that there is nothing to source a growth in the mass.

With $G^{(1)}_{\mu\nu}[g^{\rm obj}]=0$ in Eq.~\eqref{inner_EFE1}, the first-order perturbation $H^1_{\mu\nu}$ is left to satisfy the time-independent linearized vacuum equation $\delta G^{(0)}_{\mu\nu}[H^1]=0$ on a Schwarzschild background. From Eq.~\eqref{Hnform}, we also require $H^1_{\mu\nu}$ to be no more than  linear in $\tilde\sfr$ at large $\tilde\sfr$. A linear term $\sim \e \tilde\sfr$ would have to match an $\e$-independent term $\sim\e^0\sfr$  in the outer expansion; but the only such term is the zeroth-order acceleration term $f^0_ix^i$ in $\ord{0}{g}_{\mu\nu}$, which we know to vanish. As with the time independence of $m$, this is also easily established entirely within the inner expansion: The well-known  linear-in-$\tilde\sfr$ solution~\cite{Zerilli:70,Martel-Poisson:05} is time dependent, with the acceleration coefficient corresponding to the second time derivative of a mass dipole moment. Because of its time dependence, this fails to satisfy Eq.~\eqref{inner_EFE1} (and in any case, we would demand that it vanish because of its inclusion of a mass dipole moment). More generally, for stationary solutions that grow no faster than $\tilde\sfr$, standard results~\cite{Regge-Wheeler:57,Zerilli:70} show that the only invariant content of the perturbation consists of corrections to the background moments. The mass can be straightforwardly found to be constant from Eq.~\eqref{inner_EFE2}, in the same manner as $m$ can be from Eq.~\eqref{inner_EFE1}, and then absorbed into $m$. Again keeping the object spherical, I set all higher moments to zero. At higher order, effects such as tidal heating and torquing~\cite{Poisson:04b} will force the moments to become time dependent, preventing us from making this choice, but that complication does not arise at the orders considered here. (Though we could also straightforwardly relax this choice without affecting our results; see footnote~\ref{footnote}.) We hence have $H^1_{\mu\nu}=0$.

At second order, we again arrive at a time-independent linearized vacuum equation, $\delta G^{(0)}_{\mu\nu}[H^2]=0$. From Eq.~\eqref{Hnform}, $H^2_{\mu\nu}$ can now grow as $\tilde\sfr^2$. Again referring to standard results, we find that the invariant content in a solution with this behavior is purely quadrupolar, with even- and odd-parity pieces. We can write these pieces in terms of two rank-2 STF tensors, and matching to the outer expansion dictates that they be the tidal moments $\E_{ab}$ and $\B_{ab}$; see Ref.~\cite{Pound:10b} for a first-principles construction. In addition, $H^2_{\mu\nu}$ once again contains gauge solutions and corrections to the object's intrinsic moments, and I  again freely set them to zero. 

At third order, the perturbtion $H^3_{\mu\nu}$ must satisfy Eq.~\eqref{inner_EFE3}, which now becomes the time-independent inhomogeneous equation $\delta G^{(0)}_{\mu\nu}[H^3]=-\delta G^{(1)}_{\mu\nu}[H^2]$. From Eq.~\eqref{Hnform}, $H^2_{\mu\nu}$ can now grow as $\tilde\sfr^3$, and standard results show that this behavior corresponds to a purely octupolar perturbation. This can be written in terms of two rank-3 STF tensors, which matching will dictate to be the tidal moments $\E_{abc}$ and $\B_{abc}$. We also have both homogeneous and inhomogeneous quadrupolar solutions. I write the former in terms of STF tensors $\delta\E_{ab}$ and $\delta\B_{ab}$ that represent corrections to  $g_{\mu\nu}$'s tidal moments. The latter will be written in terms of time derivatives $\dot\E_{ab}$ and $\dot\B_{ab}$. I once again freely set all other solutions to zero.

None of the above has any dependence on the nature of the object, except insofar as it is sufficiently spherical. Hence, I can freely take as my solution the metric of a tidally perturbed black hole, which has exactly the form just described. In Ref.~\cite{Poisson:05}, Poisson provides such a metric in a convenient form. It is written in advanced Eddington-Finkelstein coordinates $(\sfv,\tilde \sfx^{\sf a})$, in which the background metric reads
\begin{subequations}\label{gobj-lightcone}
\begin{align}
g^{\rm obj}_{\sf  v v} &= -f, \\
g^{\rm obj}_{\sf v  a} &=  n_{\sf a}, \\
g^{\rm obj}_{\sf a b} &= \delta_{\sf a b}- n_{\sf a b},
\end{align}
\end{subequations}
where $f:=1-2m/\tilde\sfr$. Poisson's metric was originally given in spherical polar coordinates; here I have converted to Cartesian coordinates $\tilde \sfx^{\sf a}$ in the standard Euclidean way. Like in Fermi coordinates, $n^{\sf a}=\frac{\tilde\sfx^{\sf a}}{\tilde\sfr}=\frac{\sfx^{\sf a}}{\sfr}$ and $\delta_{\sf a b}n^{\sf a}n^{\sf b}=1$.

The perturbations are written in a lightcone gauge, which sets $H^n_{\alpha\sfr} = 0$, or in the Cartesian coordinates used here, $H^n_{\alpha \sf a}n^{\sf a} = 0$. This gauge choice preserves the geometrical meaning of the advanced coordinates in the perturbed spacetime: $\sfv$ is constant on each ingoing light cone, $\tilde\sfr$ is an affine parameter on ingoing light rays, and $n^{\sf a}$ labels each  ray's direction. In this gauge, the perturbations are given by
\begin{equation}\label{H1}
H^1_{\mu\nu} =0,
\end{equation}%
\begin{subequations}\label{H2}%
\begin{align}
H^{2}_{\sf vv} &= -\tilde\sfr^2 e_1\E_{\sf ij} n^{\sf ij}, \\
H^{2}_{\sf v a} &= -\tfrac{2}{3} \tilde\sfr^2\!\left[e_4\left(\delta_{\sf a}^{\sf c} - n_{\sf a}{}^{\sf c}\right)\E_{\sf cd}  n^{\sf d} 
						- b_4\epsilon_{\sf apq}\B^{\sf q}_{\ \sf c} n^{\sf pc} \right]\!, \\
H^{2}_{\sf  a  b} &= -\tfrac{1}{3}\tilde\sfr^2\!\left[e_7\big(2\E_{\sf a b}-4\E_{\sf i( a} n_{\sf b)}{}^{\sf i} +\E_{\sf ij} n_{\sf a b}{}^{\sf ij} \right.\nonumber\\
						&\quad \left.  +\delta_{\sf ab}\E_{\sf ij} n^{\sf ij}\big)
						-2b_7\epsilon_{\sf pq( a}\big(\delta^{\sf c}{}_{\sf b)} -  n^{\sf c}{}_{\sf b)}\big) n^{\sf p} \B^{\sf q}_{\ \sf c}\right]\!,
\end{align}
\end{subequations}
and
\begin{subequations}\label{H3}
\begin{align}
H^3_{\sf vv} &= \tfrac{1}{3}\tilde\sfr^3\big( e_2\dot{\E}_{\sf ij} n^{\sf ij}- e_3\E_{\sf ijk} n^{\sf ijk}\big)-\tilde\sfr^2 e_1\delta\E_{\sf ij} n^{ij}, \\
H^3_{\sf v a} &= \tilde\sfr^3\Big\{\tfrac{1}{3}\big[e_5 \left(\delta_{\sf a}^{\sf c} -  n_{\sf a}{}^{\sf c}\right)\dot{\E}_{\sf cd}  n^{\sf d}
						- b_5\epsilon_{\sf apq}\dot{\B}^{\sf q}_{\ \sf c}  n^{\sf pc} \big]\nonumber\\
						&\quad -\!\tfrac{1}{4}\!\left[e_6\! \left(\delta_{\sf a}^{ \sf c} -  n_{\sf a}{}^c\right)\!\E_{\sf cdi} n^{\sf di}\! 
						- \tfrac{4}{3}b_6\epsilon_{\sf a pq}\B^q_{\ \sf cd}  n^{\sf pcd} \right]\!\!\Big\} \nonumber\\
						&\quad -\tfrac{2}{3} \tilde\sfr^2\!\left[e_4\!\left(\delta_{\sf a}^{\sf c} - n_{\sf a}{}^{\sf c}\right)\!\delta\E_{\sf cd}  n^{\sf d} \!
						- b_4\epsilon_{\sf apq}\delta\B^{\sf q}_{\ \sf c} n^{\sf pc} \right]\!,\\
H^3_{\sf  a b} &= \tilde\sfr^3\Big\{\tfrac{5}{18}\big[e_8\big(2\dot{\E}_{\sf a b}-4\dot{\E}_{\sf i( a} n_{\sf b)}{}^{\sf i}
						+\dot{\E}_{\sf ij} n_{\sf a b}{}^{\sf ij}\nonumber\\
						&\quad+\delta_{\sf a b}\dot{\E}_{\sf ij} n^{\sf ij}\big) -2b_8\epsilon_{\sf pq( a}\big(\delta^{\sf c}_{\ \sf b)} -  n^{\sf c}{}_{\sf b)}\big) n^{\sf p} \dot{\B}^{\sf q}_{\ \sf c}\big] \nonumber\\
						&\quad -\tfrac{1}{6}\big[e_9\big(2\E_{\sf abk} -4\E_{\sf ki( a} n_{\sf b)}{}^i+\E_{\sf ijk} n_{\sf a b}{}^{\sf ij}\nonumber\\
						&\quad +\delta_{\sf a b}\E_{\sf ijk} n^{\sf ij}\big) n^{\sf k} -\tfrac{8}{3}b_9\epsilon_{\sf pq( a}\big(\delta^{\sf c}_{\  \sf b)} \nonumber\\
						&\quad -  n^{\sf c}{}_{\sf b)}\big) n^{\sf pj} \B^{\sf q}_{\ \sf cj}\big]\Big\}  -\tfrac{1}{3}\tilde\sfr^2\!\left[e_7\big(2\delta\E_{\sf a b}-4\delta\E_{\sf i( a} n_{\sf b)}{}^i \right.\nonumber\\
						&\quad   +\delta\E_{\sf ij} n_{\sf a b}{}^{\sf ij} +\delta_{\sf ab}\delta\E_{\sf ij} n^{\sf ij}\big)\nonumber\\
						&\quad \left.	-2b_7\epsilon_{\sf pq( a}\big(\delta^{\sf c}{}_{\sf b)} -  n^{\sf c}{}_{\sf b)}\big) n^{\sf p} \delta\B^{\sf q}_{\ \sf c}\right]\! ,
\end{align}
\end{subequations}
where $e_i(\tilde\sfr)$ and $b_i(\tilde\sfr)$ are given in Appendix~\ref{radial_functions}. In addition to rewriting Poisson's metric in Cartesian coordinates, I have added the solution involving $\delta\E_{ab}$ and $\delta\B_{ab}$; these would otherwise be absorbed into the moments in $H^2_{\mu\nu}$, which would then no longer equal the moments $\E_{ab}$ and $\B_{ab}$ of $g_{\mu\nu}$. 

The coefficients $e_i(\tilde\sfr)$ and $b_i(\tilde\sfr)$ all go to 1 at $\tilde\sfr\to\infty$, and the numerical normalizations of the solutions ensure that in that limit, the metric reduces to that of a generic vacuum spacetime in local advanced coordinates centered on some worldline~\cite{Preston-Poisson:06}. However, it reduces to $\ord{0}{g}_{\mu\nu}$, not to $g_{\mu\nu}$; as anticipated in Sec.~\ref{outer_expansion}, the inner expansion automatically expands the acceleration. But our gauge choices have eliminated all acceleration terms from the perturbations, and we can see by inspection that the object is manifestly at rest at the origin of the coordinate system. Hence, we have found a solution in a rest gauge, as desired. This rest-gauge form of the metric makes clear that locally, the object is only perturbed by tidal fields (through order $\e^3$). Matching to the outer expansion in the Lorenz gauge will reveal the origin of these tidal fields. With the chosen normalization of the solutions, $\E_{ab}$, $\B_{ab}$, $\E_{abc}$, and $\B_{abc}$ will trivially agree with the tidal moments of $g_{\mu\nu}$. The subleading moments $\delta\E_{ab}$ and $\delta\B_{ab}$ will be found to be intimately related to the regular field $h^{\R1}_{\mu\nu}$, and $\E_{ab}+\e\delta\E_{ab}$ and $\B_{ab}+\e\delta\B_{ab}$ will be nearly (but not identically) the tidal moments of the ``external'' effective metric $g_{\mu\nu}+\e h^{\R1}_{\mu\nu}$. 

I remind the reader that I {\em do not} take the metric in Eqs.~\eqref{gobj-lightcone}--\eqref{H3} to be valid for all $\tilde \sfr$. I only take it to be valid once it has been re-expanded for large $\tilde \sfr$ (or equivalently, for small $\e$ at fixed $\sfr$); once in that expanded form, it is no longer specific to a black hole, but instead describes the spacetime around any object that is sufficiently spherical in the sense described above. More concretely, the tidally perturbed metric is only specific to a black hole because it contains an event horizon and its construction has imposed regularity on the horizon as a boundary condition. But the horizon is irrelevant in the buffer region, and the horizon-regularity only serves to eliminate higher moments in the perturbations, which in specializing to a spherical object, I have set to zero in any case.


\subsection{Preliminary transformation}\label{preliminary transformation}
At this stage we could rewrite the inner expansion in terms of $\sf r=\e\tilde r$ and re-expand in $\e$ to obtain the outer expansion~\eqref{re-expanded inner}. We could then seek the transformation to the metric in the Lorenz gauge. However, we can also guess part of that transformation in advance.  

First, note that the coordinates of the inner expansion are based on ingoing null geodesics, while the coordinates in the outer expansion are based on spatial geodesics orthogonal to the worldline. Hence, the transformation must account for this difference. This implies that in the $m\to0$ limit, the transformation will have to reduce to the one between advanced local coordinates centered on $\gamma$ and Fermi-Walker coordinates. That transformation, which can be obtained following the method in Sec. 13 of Ref.~\cite{Poisson-Pound-Vega:11}, is given by
\begin{subequations}\label{adv-to-Fermi}
\begin{align}
\sfv &= t +r + \Delta v_0,\\
\sfx^a &= x^a + \Delta x^a_0,
\end{align}
\end{subequations}
where
\beq\label{Dv0}
\Delta v_0  = - \tfrac{1}{6}r^3 \E_{ij} n^{ij} -  \tfrac{1}{24}r^4 (\dot{\E}_{ij} n^{ij} + \E_{ijk}n^{ijk}) +\O(r^5),
\eeq
and
\begin{align}
\Delta x^a_0 & = r^3\big(\tfrac{1}{6} \E^{a}{}_{i} n^{i} -  \tfrac{1}{3} \B_{i}{}^{b} \epsilon^{a}{}_{jb} n^{ij}\big) + r^4\big(\tfrac{1}{18} \dot{\E}^{a}{}_{i} n^{i} \nonumber\\
			&\quad + \tfrac{1}{24} \E^{a}{}_{ij} n^{ij} -  \tfrac{1}{9} \dot{\B}_{i}{}^{k} \epsilon^{a}{}_{jk} n^{ij} + \tfrac{1}{36} \dot{\E}_{ij} n^{aij}  \nonumber\\
			&\quad -  \tfrac{1}{9} \B_{ij}{}^{b} \epsilon^{a}{}_{kb} n^{ijk}\big) +\O(r^5).\label{Dx0}
\end{align}
The radial functions are related as $\sfr = r + \Delta r_0$, with
\beq\label{Dr0}
\Delta r_0 = \tfrac{1}{6}r^3 \E_{ij} n^{ij} + \tfrac{1}{24}r^4 (2\dot{\E}_{ij} n^{ij} + \E_{ijk}n^{ijk}) +\O(r^5).
\eeq
(In these formulas I have omitted acceleration terms.)

If we first re-expand our inner expansion in the buffer region and then apply the above transformation, then the zeroth-order term in the expansion~\eqref{re-expanded inner} will correctly match the background metric $g_{\mu\nu}$ (or more precisely, $\ord{0}{g}_{\mu\nu}$) in Fermi-Walker coordinates. But again, I opt to guess more of the transformation. In the inner expansion, the background metric $g^{\rm obj}_{\mu\nu}$ is expressed in ingoing Eddington-Finkelstein coordinates, which when converted to an outer expansion will yield $h^{1'}_{\mu\nu}=\frac{2m}{\sfr}\delta^v_\mu\delta^v_\nu$. In the Lorenz gauge, on the other hand, we have $h^1_{\mu\nu}=\frac{2m}{\sfr}\delta_{\mu\nu}+\O(r^0)$ in Fermi-Walker coordinates. To match that form, we can change to coordinates in which $g^{\rm obj}_{\mu\nu}=\frac{2m}{\tilde\sfr}\delta_{\mu\nu}+\O(1/\tilde \sfr^2)$. One way of accomplishing that is by transforming to harmonic coordinates, using 
\begin{subequations}\label{adv-to-harmonic}
\begin{align}
\sfv &= t + \e\left[\tilde r +m+ 2m\ln\left(\frac{\tilde r-m}{2m}\right)\right]\!,\\
\tilde \sfx^a &= \tilde x^a + mn^a,
\end{align}
\end{subequations}
with $\tilde \sfr =\tilde r+m$. This transformation would put the inner background in the form
\begin{subequations}\label{gobj harmonic}
\begin{align}
g^{\rm obj}_{tt} &= -f,\\
g^{\rm obj}_{t\tilde a} &= 0,\\
g^{\rm obj}_{\tilde a \tilde b} &= (1+m/\tilde r)^2(\delta_{\tilde a\tilde b}-n_{\tilde a\tilde b})+f^{-1}n_{\tilde a\tilde b},
\end{align}
\end{subequations}
where $f=1-2m/\tilde\sfr=\frac{\tilde r-m}{\tilde r+m}$.

To combine the transformations~\eqref{adv-to-Fermi} and \eqref{adv-to-harmonic}, I consider a small change in Eq.~\eqref{adv-to-harmonic}, leading to
\begin{align}
\Delta \sfv &= \Delta t + \Delta \sfr + 2m f^{-1}\frac{\Delta \sfr}{\sfr},\\
\Delta  x^a &= \Delta \sfx^a - m\frac{\Delta  \sfx^a}{ \sfr} +mn^a\frac{\Delta\sfr}{\sfr}.
\end{align}
I then define a gauge vector with components $\xi^t=\Delta t + 2m f^{-1}\frac{\Delta \sfr}{\sfr}$ and $\xi^{ a}=\Delta  x^a$, with $\Delta t=\Delta v_0-\Delta r_0$, $\Delta\sfx^a=\Delta x^a_0$, and $\Delta\sfr=\Delta r_0$, and with $\Delta v_0$, $\Delta x^a_0$, and $\Delta r_0$  given by Eqs.~\eqref{Dv0}--\eqref{Dr0} with $r$ replaced by $\sfr$; at the end, all quantities are then expressed in terms of the scaled coordinates. This construction may (rightly!) be deemed to be ad hoc, but since any worldline-preserving  transformation can be chosen, the choice is ultimately immaterial; in practice, the results of this choice are marginally simpler than some other alternatives, such as simply using Eqs.~\eqref{Dv0}--\eqref{Dr0} as the gauge transformation. 

Concretely, I perform the background transformation~\eqref{adv-to-harmonic} in conjunction with a gauge transformation $\e^2H^2_{\mu\nu}+\e^3H^3_{\mu\nu}\to \e^2 H^2_{\mu\nu}+\e^3 H^3_{\mu\nu}+\Lie_\xi g^{\rm obj}_{\mu\nu}$, where $\xi^\mu=\e^3\xi^\mu_3+\e^4\xi^\mu_4$ is given by
\begin{align}
\xi^t_3 &= - \tfrac{1}{3} \tilde\sfr^3 \E_{ij} n^{ij} + \tfrac{1}{3} m f^{-1} \tilde\sfr^2\E_{ij} n^{ij},\\
\xi^{ a}_3 &= \tilde\sfr^3(1-m/\tilde\sfr)\left(\tfrac{1}{6} \E^{a}{}_{i} n^{i} - \tfrac{1}{3} \B_{i}{}^{k} \epsilon^{a}{}_{jk} n^{ij}\right)\nonumber\\
			&\quad +\tfrac{1}{6} m\tilde\sfr^2 \E_{ij} n^{aij},
\end{align}
and
\begin{align}
\xi^t_4 &= -  \tfrac{1}{8} \tilde\sfr^4 \dot{\E}_{ij}n^{ij} -  \tfrac{1}{12} \tilde\sfr^4 \E_{ijk} n^{ijk} \nonumber\\
		&\quad + \tfrac{1}{12} m f^{-1} \tilde\sfr^3\bigl(2 \dot{\E}_{ij}n^{ij} + \E_{ijk} n^{ijk}\bigr),\\
\xi^{ a}_4 &= m\tilde\sfr^3\big(\tfrac{1}{12} \dot{\E}_{ij} n^{aij} + \tfrac{1}{24} \E_{ijk} n^{aijk}\big) \nonumber\\
		&\quad + \tilde\sfr^4(1-m/\tilde\sfr)\Big(\tfrac{1}{18} \dot{\E}^{a}{}_{i} n^{i} + \tfrac{1}{24} \E^{a}{}_{ij}n^{ij} -  \tfrac{1}{9} \dot{\B}_{i}{}^{k} \epsilon^{a}{}_{jk} n^{ij} \nonumber\\
		&\quad + \tfrac{1}{36} \dot{\E}_{ij} n^{aij} -  \tfrac{1}{9} \B_{ij}{}^{b} \epsilon^{a}{}_{kb}n^{ijk}\Big),
\end{align}
with $\tilde\sfr=\tilde r+m$. Note that the gauge vectors begin one order higher than their effects because the $x^a$ derivatives in $\Lie_\xi g^{\rm obj}_{\mu\nu}$ reduce the order by one. Also note that after performing the background transformation, functions of $\sfv$ need to be expanded around their values at $t$. Finally, note that the transformation is worldline preserving: because it contains no order-$\e$ or $\e^2$ pieces, it trivially preserves the condition $H^1_{\mu\nu}=0$, and therefore preserves $\delta M_i=0$.

After performing the transformation, I convert to the unscaled coordinate $r=\e\tilde r$ and re-expand for small $\e$. For example, 
\begin{subequations}
\begin{align}
g^{\rm obj}_{tt} &= -\frac{\tilde r-m}{\tilde r+m}\\
						&= -\frac{r-\e m}{r+\e m}\\
						&= -1 + \frac{2 \e m}{r} - \frac{2 \e^2m^2}{r^2} + \O\bigl(\e^3\bigr).
\end{align}
\end{subequations}
The end result is a new expression for the metric in the outer expansion in the buffer region, 
\beq\label{g rest}
{\sf g}_{\mu\nu}=g'_{\mu\nu} + \e h^{1'}_{\mu\nu}+\e^2 h^{2'}_{\mu\nu}+\O(\e^3), 
\eeq
where $g'_{\mu\nu}=\ord{0}{g}_{\mu\nu}$ is given by Eq.~\eqref{delta0g}, and the perturbations are
\begin{subequations}\label{h1 rest}
\begin{align}
h^{1'}_{tt} &= \frac{2m}{r} + \tfrac{5}{3} mr \mathcal{E}^{ai} \hat{n}_{ai} + r^2\Big(2m\dot{\mathcal{E}}^{ai} \hat{n}_{ai} 
			- \delta \mathcal{E}^{ai} \hat{n}_{ai}\nonumber\\
			&\quad + \tfrac{7}{12}m\mathcal{E}^{aij} \hat{n}_{aij}\Big)+\O(r^3)\label{rest-gauge_h1tt}\\
h^{1'}_{ta} &= -\tfrac{2}{3}mr \bigl(\mathcal{E}_{ab} \hat{n}^{b} + \mathcal{B}^{bi} \epsilon_{aij} \hat{n}_{b}{}^{j}\bigr) 
			+ r^2\Big(\tfrac{49}{30}m \dot{\mathcal{E}}_{ab} \hat{n}^{b} \nonumber\\
			&\quad + \tfrac{13}{9}m\dot{\mathcal{B}}^{bi} \epsilon_{aij} \hat{n}_{b}{}^{j} - \tfrac{2}{3}\delta \mathcal{B}^{bi} \epsilon_{aij} \hat{n}_{b}{}^{j}
			+ \tfrac{1}{12} m\mathcal{E}_{abi} \hat{n}^{bi} \nonumber\\
			&\quad + \tfrac{1}{18}m \dot{\mathcal{E}}^{bi} \hat{n}_{abi} 
			- \tfrac{2}{9}m \mathcal{B}^{bij} \epsilon_{ab}{}^{k} \hat{n}_{ijk}\Big)+\O(r^3) \label{rest-gauge_h1ta}\\
h^{1'}_{ab} &= \frac{2m\delta_{ab}}{r} + mr\Big(\tfrac{2}{45} \mathcal{E}_{ab} + \tfrac{88}{21} \mathcal{E}_{(a}{}^{c} \hat{n}_{b)c} 
			- \tfrac{9}{7} \mathcal{E}^{cd} \delta_{ab} \hat{n}_{cd}\nonumber\\
			&\quad + \tfrac{1}{3} \mathcal{E}^{cd} \hat{n}_{abcd}\Big) + \tfrac{1}{9}r^2\Big(16m\dot{\mathcal{E}}_{ab}-\delta \mathcal{E}_{ab}  \nonumber\\
			&\quad + \tfrac{141}{35} m\mathcal{E}_{abc} \hat{n}^{c} - \tfrac{74}{5} m\dot{\mathcal{B}}_{(a}{}^{d} \epsilon_{b)cd} \hat{n}^{c} 
			+ 6\delta\mathcal{E}_{(a}{}^{c} \hat{n}_{b)c} \nonumber\\
			&\quad - 3\delta \mathcal{E}^{cd} \delta_{ab} \hat{n}_{cd} 
			+ 8m\dot{\mathcal{E}}^{cd} \delta_{ab} \hat{n}_{cd} - 8m \mathcal{B}_{c(a}{}^{i} \epsilon_{b)di} \hat{n}^{cd} \nonumber\\
			&\quad + 14m\mathcal{E}_{(a}{}^{cd} \hat{n}_{b)cd} + m\dot{\mathcal{B}}^{cd} \epsilon_{c(a}{}^{i} \hat{n}_{b)di} \nonumber\\
			&\quad - \tfrac{53}{12}m \mathcal{E}^{cdi} \delta_{ab} \hat{n}_{cdi} + \tfrac{3}{4} m\mathcal{E}^{cdi} \hat{n}_{abcdi}\Big) \!\!+\!\O(r^3),\label{rest-gauge_h1ab}
\end{align}
\end{subequations}
and
\begin{subequations}\label{h2 rest}
\begin{align}
h^{2'}_{tt} &= - \frac{2 m^2}{r^2}-\tfrac{4}{3} m^2 \mathcal{E}^{ai} \hat{n}_{ai}  + mr\Big(\tfrac{5}{3} \delta \mathcal{E}^{ai} \hat{n}_{ai} \nonumber\\
			&\quad - \tfrac{20}{3} m \dot{\mathcal{E}}^{ai} \hat{n}_{ai} - \tfrac{1}{2} m \mathcal{E}^{aij} \hat{n}_{aij}\Big)+\O(r^2),\label{rest-gauge_h2tt}\\
h^{2'}_{ta} &= -\tfrac{2}{15} m^2 \big(6 \mathcal{E}_{ab} \hat{n}^{b} - 5 \mathcal{B}^{bi} \epsilon_{aij} \hat{n}_{b}{}^{j} 
			- 10 \mathcal{E}^{bi} \hat{n}_{abi}\big)\nonumber\\
			&\quad + mr\Big(\tfrac{1}{5} m \dot{\mathcal{E}}_{ab} \hat{n}^{b}-\tfrac{2}{3} \delta \mathcal{E}_{ab} \hat{n}^{b} 
			+ \tfrac{31}{9} m \dot{\mathcal{B}}^{bi} \epsilon_{aij} \hat{n}_{b}{}^{j} \nonumber\\
			&\quad - \tfrac{2}{3} \delta \mathcal{B}^{bi} \epsilon_{aij} \hat{n}_{b}{}^{j} - \tfrac{8}{21} m \mathcal{E}_{abi} \hat{n}^{bi} 
			- \tfrac{13}{9} m \dot{\mathcal{E}}^{bi}\hat{n}_{abi} \nonumber\\
			&\quad + \tfrac{4}{9} m \mathcal{B}^{bij} \epsilon_{ab}{}^{k} \hat{n}_{ijk} + \tfrac{1}{12} m \mathcal{E}^{bij} \hat{n}_{abij}\Big)\nonumber\\&\quad+\O(r^2),\label{rest-gauge_h2ta}\\
h^{2'}_{ab} &= \frac{m^2\Big(\tfrac{4}{3}\delta_{ab} + \hat{n}_{ab}\big)}{r^2}+m^2\Big(\tfrac{68}{45} \mathcal{E}_{ab}
			- \tfrac{16}{5} \mathcal{B}_{(a}{}^{d} \epsilon_{b)cd} \hat{n}^{c} \nonumber\\
			&\quad + \tfrac{136}{21} \mathcal{E}_{(a}{}^{c} \hat{n}_{b)c} - \tfrac{29}{21}\mathcal{E}^{cd} \delta_{ab} \hat{n}_{cd} 
			- \tfrac{1}{3}\mathcal{E}^{cd} \hat{n}_{abcd}\nonumber\\
			&\quad + 4 \mathcal{B}^{cd} \epsilon_{c(a}{}^{i} \hat{n}_{b)di}\Big)  
			+ mr \Big(\tfrac{2}{45} \delta \mathcal{E}_{ab} 
			+ \tfrac{256}{45} m\dot{\mathcal{E}}_{ab}  \nonumber\\
			&\quad + \tfrac{68}{35} m\mathcal{E}_{abc} \hat{n}^{c} - \tfrac{70}{9} m\dot{\mathcal{B}}_{(a}{}^{d} \epsilon_{b)cd} \hat{n}^{c} 
			+ \tfrac{88}{21} \delta \mathcal{E}_{(a}{}^{c} \hat{n}_{b)c} \nonumber\\
			&\quad - \tfrac{32}{7} m\dot{\mathcal{E}}_{(a}{}^{c} \hat{n}_{b)c} - \tfrac{9}{7}\delta \mathcal{E}^{cd} \delta_{ab} \hat{n}_{cd} 
			+ \tfrac{208}{63} m\dot{\mathcal{E}}^{cd} \delta_{ab} \hat{n}_{cd} \nonumber\\
			&\quad - \tfrac{208}{63}m \mathcal{B}_{c(a}{}^{i} \epsilon_{b)di} \hat{n}^{cd} + \tfrac{41}{18} m\mathcal{E}_{(a}{}^{cd} \hat{n}_{b)cd}  \nonumber\\
			&\quad - \tfrac{10}{27} m\mathcal{E}^{cdi} \delta_{ab} \hat{n}_{cdi} + \tfrac{1}{3}\delta \mathcal{E}^{cd} \hat{n}_{abcd} 
			- \tfrac{16}{9} m\dot{\mathcal{E}}^{cd} \hat{n}_{abcd} \nonumber\\
			&\quad+ \tfrac{8}{9} m\mathcal{B}^{cdi} \epsilon_{c(a}{}^{j} \hat{n}_{b)dij} - \tfrac{1}{3} m\mathcal{E}^{cdi} \hat{n}_{abcdi}\nonumber\\
			&\quad + \tfrac{20}{9} m\dot{\mathcal{B}}^{cd} \epsilon_{c(a}{}^{i} \hat{n}_{b)di}\Big)+\O(r^2).\label{rest-gauge_h2ab}
\end{align}
\end{subequations}
This is the final form of the metric in the rest gauge. It has several important properties, already mentioned in the previous subsection, but reiterated here for emphasis. First, there is no explicit appearance of the regular field; it has been entirely bundled into the tidal moments $\delta\E_{ij}$ and $\delta\B_{ij}$. Next, there is no mass-dipole-moment term $\sim M_in^i/r^2$ in $h^2_{\mu\nu}$, and although I do not display $h^3_{\mu\nu}$, there is no dipole-moment term $\delta M_in^i/r^2$ in it either, as such a moment could only come from the expansion of $H^1_{\mu\nu}$. Hence, the object is mass-centered on $\gamma$. Finally, there is no acceleration term $\sim a_ix^i$ in either the background or the perturbations. This tells us that the object is not only centered on $\gamma$, but also at rest there; since the expansion here is around $\ord{0}{g}_{\mu\nu}$ rather than $g_{\mu\nu}$, one can imagine that in an expansion around $g_{\mu\nu}$, the perturbations $h^1_{\mu\nu}$ and $h^2_{\mu\nu}$ in this gauge would contain terms $+2f^1_ix^i u_\mu u_\nu$ and $+2f^2_ix^i u_\mu u_\nu$ that exactly cancel the acceleration term $-2a_ix^iu_\mu u_\nu$ in $g_{\mu\nu}$. 

As we shall see in the next section, the transformation to the Lorenz gauge unspools the regular field throughout the metric, determines how it relates to the tidal moments $\delta\E_{ij}$ and $\delta\B_{ij}$, and most importantly, determines in which piece of the Lorenz-gauge metric the motion is geodesic.

\section{Transformation from rest gauge to Lorenz gauge}\label{matching}


With the metric determined in both the Lorenz gauge and the rest gauge, we are now in a position to find the transformation between them---and thereby determine the acceleration of $\gamma$ in the Lorenz gauge. The two metrics already agree at leading order, implying that the perturbations must be related by a gauge transformation. In the rest gauge, the first- and second-order perturbations $h^{1'}_{\mu\nu}$ and $h^{2'}_{\mu\nu}$ are given in Eqs.~\eqref{h1 rest} and \eqref{h2 rest}. In the Lorenz gauge, the first- and second-order perturbations $h^{1\dagger}_{\mu\nu}$ and $h^{2\dagger}_{\mu\nu}$  are given in Eqs.~\eqref{h1New} and \eqref{h2New}.

\subsection{Form of transformation}

Under a gauge transformation generated by a small vector $\xi^\mu$, a metric ${\sf g}_{\mu\nu}$ can be expanded along the flow lines of $\xi^\mu$ as  
\beq\label{gauge g}
{\sf g}_{\mu\nu}\to{\sf g}_{\mu\nu}+\Lie_\xi {\sf g}_{\mu\nu}+\frac{1}{2}\Lie^2_\xi{\sf g}_{\mu\nu}+\ldots
\eeq
Substituting $\xi^\mu=\e\xi^\mu_1+\e^2\xi^\mu_2+\ldots$ and ${\sf g}_{\mu\nu}=g_{\mu\nu}+\e h^1_{\mu\nu}+\e^2 h^2_{\mu\nu}+\ldots$ yields a transformation law for each of the perturbations,
\begin{equation}
h^{n,\rm new}_{\mu\nu} = h^{n,\rm old}_{\mu\nu}+\Delta h^{n}_{\mu\nu},
\end{equation}
where
\begin{align}
\Delta h^{1}_{\mu\nu} &= \Lie_{\xi_1}g_{\mu\nu},\label{Dh1}\\
\Delta h^{2}_{\mu\nu} &= \Lie_{\xi_2}g_{\mu\nu}+\frac{1}{2}\Lie^2_{\xi_1}g_{\mu\nu}+\Lie_{\xi_1}h^{1,\rm old}_{\mu\nu}.\label{Dh2}
\end{align}
In a chart, the gauge vectors $\xi_1^\mu$ and $\xi_2^\mu$ correspond to the coordinate transformation~\eqref{coord_transformation}~\cite{Bruni-etal:96}.

A transformation from the rest gauge to the Lorenz gauge must therefore satisfy
\begin{align}
h^{1\dagger}_{\mu\nu} &= h^{1'}_{\mu\nu} + \Lie_{\xi_1}\ord{0}{g}_{\mu\nu},\label{gauge_equation1}\\
h^{2\dagger}_{\mu\nu} &= h^{2'}_{\mu\nu} + \Lie_{\xi_2}\ord{0}{g}_{\mu\nu} + \frac{1}{2}\Lie^2_{\xi_1}\ord{0}{g}_{\mu\nu} + \Lie_{\xi_1}h^{1'}_{\mu\nu}.\label{gauge_equation2}
\end{align} 
I will solve these equations for $\xi^\mu_n$ order by order in $r$. However,  ensuring that the transformation is worldline preserving requires that we also consider the transformation of the third-order field; recall that a subleading dipole moment would appear as a term  of the form $\e^3 \delta M_in^i/r^2$ in the outer expansion. The transformation of the third-order perturbation is easily derived by adding the next term, $\frac{1}{6}\Lie^3_\xi{\sf g}_{\mu\nu}$, to Eq.~\eqref{gauge g}. The result is
\begin{align}
\Delta h^{3}_{\mu\nu} &= \Lie_{\xi_3}g_{\mu\nu}+\frac{1}{2}(\Lie_{\xi_1}\Lie_{\xi_2}+\Lie_{\xi_2}\Lie_{\xi_1})g_{\mu\nu}+\frac{1}{6}\Lie^3_{\xi_1}g_{\mu\nu}\nonumber\\
							&\quad +\frac{1}{2}\Lie^2_{\xi_1}h^{1, \rm old}_{\mu\nu}+\Lie_{\xi_2}h^{1, \rm old}_{\mu\nu}+\Lie_{\xi_1}h^{2, \rm old}_{\mu\nu}.\label{Dh3}
\end{align}
To keep the object centered on $\gamma$, I demand that the gauge transformation does not induce a dipole term of the form $\delta M_in^i/r^2$ in $\Delta h^3_{\mu\nu}$ (equivalently, if one converts the transformation to scaled coordinates $\tilde x^a=x^a/\e$,  I demand that no $\delta M_i n^i/\tilde r^2$ term appears in $\Delta H^1_{\mu\nu}$). But I do not otherwise seek to control the gauge of the third-order perturbation, leaving it in the form $h^{3'}_{\mu\nu}+\Delta h^3_{\mu\nu}$ with $\xi^\mu_3=0$. 

To solve Eqs.~\eqref{gauge_equation1} and \eqref{gauge_equation2} I assume an expansion
\begin{equation}\label{xi form}
\xi^n_\alpha = \sum_{p\geq-n+1}\sum_{q,l\geq0}r^p(\ln r)^q\xi^{(n,p,q)}_{\alpha L}\nhat^L,
\end{equation}
where $\xi^{(n,p,q)}_{\alpha L}$ is STF in the indices $L$, and I assume that for a given $p$, there exists a finite maximum $q$. As with the metric perturbations, I abbreviate  $\xi^{(n,p,0)}_{\alpha L}$ as  $\xi^{(n,p)}_{\alpha L}$. The expansion~\eqref{xi form} might not be the most general gauge vector possible, but it is likely the most general transformation that preserves the form \eqref{hn mode expansion} of the metric perturbations. In terms of this expansion, the worldline-preserving condition~\eqref{worldline-preserving} becomes
\beq\label{worldline preserving 2}
\xi^{(n,0)}_a = 0.
\eeq
The transformation~\eqref{xi form} will be found to be unique if and only if this condition is imposed, and this condition will suffice to preserve the center-of-mass condition $\delta M_i=0$.

Finding $\xi^n_\alpha$ now reduces to a straightforward procedure of substituting the expansion \eqref{xi form} into Eqs.~\eqref{gauge_equation1} and \eqref{gauge_equation2} and finding the coefficients $\xi^{(n,p,q)}_{\alpha L}$. At each order in $r$, $\xi^{(n,p,q)}_{\alpha L}$ is found by decomposing the equation into coefficients of the STF tensors $\nhat^L$. Because these tensors $\nhat^L$ form an orthogonal basis, from an equation of the form $\sum_l a_{P\langle L\rangle}\nhat^L=\sum_l b_{P\langle L\rangle}\nhat^L$, one can equate $a_{P\langle L\rangle}$ with $b_{P\langle L\rangle}$. Even after equating coefficients of $\nhat^L$, it is sometimes nontrivial to solve for the tensors $\xi^{(n,p,q)}_{\alpha L}$, since they can be contracted with other tensors. In those instances, it is necessary to take a final step of writing $a_{P\langle L\rangle}$ and $b_{P\langle L\rangle}$ in irreducible form, using Eqs.~\eqref{s=1_decomposition} and \eqref{s=2_decomposition}. Since the decomposition into irreducible pieces is again unique, one can equate the terms in the decomposition of $a_{P\langle L\rangle}$ with those in the decomposition of $b_{P\langle L\rangle}$. To facilitate this process, I use Eq.~\eqref{s=1_decomposition} to write $\xi^{(n,p,q)}_{aL}$ itself in the irreducible form
\begin{align}
\xi^{(n,p,q)}_{aL} &= \hat\Upsilon^{(n,p,q)}_{aL} +\epsilon_{a\langle i_i}{}^j\hat \Lambda_{L-1\rangle j}^{(n,p,q)} \nonumber\\
&\quad +\delta_{a\langle i_1}\hat\Psi_{L-1\rangle}^{(n,p,q)},\label{xi decomposition}
\end{align}
where $\hat\Upsilon^{(n,p,q)}_{L+1}$, $\hat \Lambda_{L}^{(n,p,q)}$, and $\hat\Psi_{L-1}^{(n,p,q)}$ are STF tensors.  In terms of this decomposition, the condition~\eqref{worldline preserving 2} becomes $\hat\Upsilon^{(n,0)}_a=0$.

The main result of the calculation is that the metrics in the two gauges are related by a worldline-preserving gauge transformation if and only if the forces $f^a_1$ and $f^a_2$ satisfy Eqs.~\eqref{a1_Fermi} and \eqref{a2-hR}. To understand how this comes about, consider the order-$r$, $l=1$ piece of the $tt$ component of Eq.~\eqref{gauge_equation1}. The left-hand side reads simply $\ord{0}{h}^{\R1}_{tt,i}x^i-2f^1_i x^i$, and when the worldline-preserving condition is imposed on the right-hand side, this piece of Eq.~\eqref{gauge_equation1} becomes  an equation for $f^1_i$. Analogously, the order-$r$, $l=1$ piece of the $tt$ component of Eq.~\eqref{gauge_equation2} becomes an equation for $f^2_i$. In a similar manner, the calculation yields formulas for the tidal moments $\delta\E_{ab}$ and $\delta\B_{ab}$; these expressions for $\delta\E_{ab}$ and $\delta\B_{ab}$, although derived in Ref.~\cite{Pound:12a}, appear here explicitly for the first time.

\subsection{Transformation at first order}
I first consider the solution to Eq.~\eqref{gauge_equation1}. The worldline-preserving order-$\e$ transformation satisfying this equation is given in Eq.~\eqref{xi1}. It has been simplified using the gauge conditions~\eqref{STF gauge relations 1}. Since each step of the calculation is straightforward (if lengthy), I will omit most of the details and instead describe, order by order in $r$, the effect of the transformation and its implications.

\subsubsection{Order $1/r$}
Because of the preliminary transformation in Sec.~\ref{preliminary transformation}, the $1/r$ terms in Eqs.~\eqref{h1 rest} and \eqref{hS1} agree. This determines that $\xi^{(1,p,q)}_{\mu L}=0$ for $p<0$, and $\xi^{(1,0,q)}_{\mu L}=0$ for all $l>0$ for any $q$; $\xi^{(1,0,q)}_{\mu L}$ are involved at this order because $\partial_a \ln r\sim \partial_a n^i\sim 1/r$. Hence, we have $\xi^1_\mu = \xi^{(1,0)}_{\mu}(t)+\O(r\ln r)$. $\xi^{(1,0)}_a$ will eventually be set to zero due to the condition~\eqref{worldline preserving 2}, but for the moment I leave it arbitrary to better illustrate its role.

\subsubsection{Order $r^0$}
The metric~\eqref{h1 rest} in the rest gauge contains no terms of order $r^0$, while the metric~\eqref{h1New} contains terms at this order comprising $h^{\R1}_{\mu\nu}\big|_{\gamma}$. The transformation hence serves to introduce the regular field on the worldline. Specifically, the order-$r^0$ term in Eq.~\eqref{xi1} introduces $h^{\R1}_{tt}\big|_{\gamma}$ ($:=\hat A^{(1,0)}$), and the non-integral order-$r$ terms introduce $h^{\R1}_{ta}\big|_{\gamma}$ ($:=\hat C^{(1,0)}_a$) and $h^{\R1}_{ab}\big|_{\gamma}$ ($:=\delta_{ab}\hat K^{(1,0)} + \hat H^{(1,0)}_{ab}$). Of these effects, the transformation  $\xi_t^1=\tfrac{1}{2}\int dt h^{\R1}_{tt}$ is especially significant: it is an adjustment of the proper time along the worldline, telling us that the proper time in the black hole's rest frame is the proper time in the metric $g_{\mu\nu}+\e h^{\R1}_{\mu\nu}$; the quantity $\tfrac{1}{2}h^{\R1}_{tt}$ in the integrand of the transformation is the Detweiler redshift~\cite{Detweiler:08}, which has played an important role in interfacing self-force with post-Newtonian theory~\cite{LeTiec:14} and has recently been computed for the first time in numerical relativity~\cite{Zimmerman-etal:16}.

At this stage, the antisymmetric piece of $\xi^{(1,1)}_{ai}$, encoded in the vector $\hat \Lambda^{(1,1)}_{i}$, is undetermined, and the vector $\xi^{(1,0)}_a$ remains arbitrary. Each of these quantities will carry a dynamical meaning.

\subsubsection{Order $r$}
At order $r$, the metric~\eqref{h1 rest} in the rest gauge contains terms of the form $m\E_{ab}$ and $m\B_{ab}$. The metric~\eqref{h1New} contains terms of this form, but in addition it contains terms $\sim\partial h^{\R1}_{\mu\nu}\big|_{\gamma}$ and an explicit appearance of the first-order acceleration $f^a_1$ (via $\ord{1}{g}_{\mu\nu}$). The order-$r^2$ piece of the transformation brings the $m\E_{ab}$ and $m\B_{ab}$ terms into agreement and introduces the $\sim\partial h^{\R1}_{\mu\nu}\big|_{\gamma}$ terms into the metric. Most significantly, at this order, the vectors $\hat \Lambda^{(1,1)}_{i}$  and $\xi^{(1,0)}_a$ are determined. 

The $\hat \Lambda^{(1,1)}_{i}$ term in the transformation appears as the integral in the order-$r$ piece of Eq.~\eqref{xi1a}, which can be written as $\int dt\,\partial_{[a}h^{\R1}_{b]t}x^b$. This indicates that the rest frame of the object rotates relative to the Fermi-Walker frame (which is parallel-propagated with respect to $g_{\mu\nu}$ along $\gamma$). 

Finally, one finds that the vector $\xi^{(1,0)}_a$ must satisfy
\beq\label{master equation1}
\ddot{\xi}^{(1,0)}_a = -\E_{a}{}^i\xi^{(1,0)}_i + f^1_a - \frac{1}{2}\partial_a \ord{0}{h}^{\R1}_{tt} + \partial_t \ord{0}{h}^{\R1}_{ta},
\eeq
where the derivatives of the regular field are evaluated at $r=0$. Given the worldline-preserving condition $\xi^{(1,0)}_a=0$, this equation yields the standard formula~\eqref{a1_Fermi} for the first-order self-force. It is worth mentioning that Eq.~\eqref{master equation1} can be derived in only a few lines of calculation. As stated in the opening of this section, we know in advance which piece of Eq.~\eqref{gauge_equation1}  determines $f_1^a$: the order-$r$, $l=1$ piece of the $tt$ component. That piece of  Eq.~\eqref{gauge_equation1} is easily found to be 
\beq\label{r,l=1}
(\partial_a \ord{0}{h}^{\R1}_{tt}-2f^1_a)x^a = - 2 (\dot \xi^{(1,1)}_{ta} +\xi^{(1,0)}_i\E^i{}_a)x^a.
\eeq
$\dot \xi^{(1,1)}_{ta}$ is determined from the order-$r^0$, $l=0$ piece of the $ta$ component of Eq.~\eqref{gauge_equation1}, which reads $\ord{0}{h}^{\R1}_{ta}=\dot\xi^{(1,0)}_a-\xi^{(1,1)}_{ta}$. Solving for $ \xi^{(1,1)}_{ta}$ and substituting this into Eq.~\eqref{r,l=1} returns Eq.~\eqref{master equation1}.

If we had not imposed the worldline-preserving condition, then the gauge vector $\xi^{(1,0)}_a$, via the $\Lie_{\xi_1}h^{1'}_{\mu\nu}$ term in Eq.~\eqref{gauge_equation2}, would produce a term $\Delta h^2_{tt}=-\frac{2m\xi^{(1,0)}_an^a}{r^2}$ in the second-order field, corresponding to a mass dipole moment $ M_a=-m\xi^{(1,0)}_a$. Equation~\eqref{master equation1} would then tell us how the object's Lorenz-gauge center of mass moves relative to a nearby worldline with arbitrary acceleration $f_1^a$ (refer to Ref.~\cite{Pound:10a} for a discussion). But since $M_a$ has been set to zero in both the rest gauge and the Lorenz gauge, Eq.~\eqref{gauge_equation2} can only be satisfied if $\xi^{(1,0)}_a=0$. That is, as was anticipated in the Introduction, even though the equation of motion~\eqref{master equation1} is a consequence of the second-order field equations in the outer expansion, knowing that $M_a=0$ in the second-order field allows us to obtain that equation of motion without performing any (nontrivial) second-order computations. Appendix~\ref{supertranslations} makes some additional comments on the ramifications of this fact.

\subsubsection{Order $r^2$}\label{tidal matching}
At order $r^2$, both the rest-gauge metric~\eqref{h1 rest} and the Lorenz-gauge metric~\eqref{h1New} contain $m\dot\E_{ab}$, $m\dot\B_{ab}$, $m\E_{abc}$, and $m\B_{abc}$ terms. The order-$r^3$ terms in $\xi^1_\mu$ bring these terms into agreement. More significantly, the rest-gauge metric contains $\delta\E_{ab}$ and $\delta\B_{ab}$ terms, while the Lorenz-gauge metric contains $\partial\partial h^{\R1}_{\mu\nu}$ terms. The transformation partially serves to introduce the $\partial\partial h^{\R1}_{\mu\nu}$ terms into the metric, but the transformation is only possible if, as anticipated in Sec.~\ref{inner_expansion}, $\delta\E_{ab}$ and $\delta\B_{ab}$ are closely related to the tidal moments of $h^{\R1}_{\mu\nu}$. Specifically, one finds that the metrics can be matched if and only if $\delta \E_{ab}$ and $\delta\B_{ab}$ are related to $\partial\partial h^R$ and $m\E$ and $m\B$ by
\begin{align}
\delta\E_{ab} &= -\frac{1}{2} h^{{\rm R}1}_{tt,\langle ab\rangle} + h^{{\rm R}1}_{tt}\E_{ab} + \frac{8}{3} m\dot{\E}_{ab} -\E_{\langle a}^i h^{{\rm R}1}_{b\rangle i} \nonumber\\
&\quad  + 2\mathop{\rm STF}_{ab} \E_a^i\int h^{{\rm R}1}_{t[b,i]}dt + h^{{\rm R}1}_{t\langle a,b\rangle t}\nonumber\\
&\quad + \frac{1}{2}\dot{\E}_{ab}\int h^{{\rm R}1}_{tt} dt -\frac{1}{2}h^{{\rm R}1}_{\langle ab\rangle,tt},\label{dE}
\end{align}
\begin{align}
\delta\B_{ab} &= \frac{13}{6} m \dot{\B}_{ab} + \mathop{\rm STF}_{ab}\epsilon_a{}^{ij}h^{{\rm R}1}_{i[t,b]j} -h^{{\rm R}1}_{ti}\E_{j(a}\epsilon_{b)}{}^{ij}\nonumber\\
&\quad -\frac{1}{2}\B_{ab}\delta^{ij}h^{{\rm R}1}_{ij}+2\mathop{\rm STF}_{ab}\B_a^i\int h^{{\rm R}1}_{t[b,i]}+\frac{1}{2}\B_{ab}h^{{\rm R}1}_{tt}\nonumber\\
&\quad +\frac{1}{2}\dot{\B}_{ab}\int h^{{\rm R}1}_{tt}dt.\label{dB}
\end{align}
I omit the implied left-superscript $0$ on all regular-field terms.

The meaning of these results is not especially transparent. However, we can make them clearer by noting two simplifications. First, all the $\partial\partial h^{\R1}_{\mu\nu}$ terms in Eq.~\eqref{dE} can be written as $\delta R_{atbt}[h^{\rm R1}]$, and all those in Eq.~\eqref{dB} as $\frac{1}{2}\epsilon^{pq}{}_{(a}\delta R_{b)tpq}[h^{\rm R1}]$, where 
\beq
\delta R_{\alpha\beta\gamma\delta}[h] = h_{\mu[\alpha}R^\mu{}_{\beta]\gamma\delta} - h_{\gamma[\alpha;\beta]\delta} + h_{\delta[\alpha;\beta]\gamma}
\eeq
is the linearized Riemann tensor associated with a perturbation $h_{\mu\nu}$. Next, all the terms involving zeroth or first derivatives of $h^{\R1}_{\mu\nu}$ in Eq.~\eqref{dE} can be written as $-(\Lie_{\xi_1}R)_{tatb}$, and those in Eq.~\eqref{dB} as $-\frac{1}{2}\epsilon^{pq}{}_{(a}(\Lie_{\xi_1} R)_{b)tpq}$. Noting that $\Lie_{\xi_1}R_{\alpha\beta\gamma\delta}=\delta R_{\alpha\beta\gamma\delta}[\Lie_{\xi_1}g]$, we obtain the much simpler formulas
\begin{align}
\delta\E_{ab} &= \delta R_{tatb}[h^{\rm R1}-\Lie_{\xi_1}g]  + \frac{8}{3} m\dot{\E}_{ab},\label{dE simple}\\
\delta\B_{ab} &=  \frac{1}{2}\epsilon^{pq}{}_{(a}\delta R_{b)tpq}[h^{\rm R1}-\Lie_{\xi_1}g] + \frac{13}{6} m \dot{\B}_{ab}. \label{dB simple}
\end{align}

Equations \eqref{dE simple} and \eqref{dB simple} almost have a simple interpretation: $\delta\E_{ab}$ and $\delta\B_{ab}$ are almost the tidal moments of the regular field $h^{\R1}_{\mu\nu}$, up to gauge. But due to the presence of the $m\dot\E_{ab}$ and $m\dot\B_{ab}$ terms, these interpretations are not quite correct. This tells us that we cannot always safely interpret the effective metric $g_{\mu\nu}+h^{\R}_{\mu\nu}$ as the ``external'' metric that the object feels. In particular,  the tidal quantities defined in Refs.~\cite{Dolan-etal:14,Bini-Damour:14}, which are computed from $h^{\R1}_{\mu\nu}$ alone, cannot always be associated with the tidal moments that appear in the metric of a tidally perturbed black hole or material body (see Refs.~\cite{Poisson:14,Landry-Poisson:15} and references therein for a review of such metrics). This likely stems from there being some degree of ambiguity in the split between $(\dot\E_{ab},\dot\B_{ab})$ and $(\delta\E_{ab},\delta\B_{ab})$; as one can see in Eq.~\eqref{h1 rest}, they ultimately appear in similar ways in the second-order metric perturbation. In fact, by making the redefinitions $\delta\E_{ab}\to \delta\E_{ab}-\frac{8}{3} m\dot{\E}_{ab}$ and $\delta\B_{ab}\to \delta\B_{ab}-\frac{13}{6} m\dot{\B}_{ab}$, we would ensure that $\delta\E_{ab}$ and $\delta\B_{ab}$ precisely correspond to the moments of $h^{\R1}_{\mu\nu}$ (up to gauge). Indeed, in Ref.~\cite{Gralla:12}, Gralla {\em defines} his regular field such that this is true. However, since he does not show that his first-order regular field agrees with the Detweiler-Whiting field (while the one I use here does at least through order $r^2$), it is not clear whether it is his regular field or his tidal moments that differ from the ones used here.

Besides this ambiguity in the definitions of tidal moments, the above results point to a limitation in the typical construction of metrics of tidally perturbed objects. Equations~\eqref{dE} and \eqref{dB} show that these metrics are generically nonuniform in time. For example, imagine that the small object moves on a quasicircular orbit around a much larger black hole. Then $h^{\R1}_{\mu\nu}$ and its derivatives are  approximately constant in time, and the tidal moments defined in the object's rest gauge  grow approximately linearly with time. Therefore a single inner expansion of this sort is unlikely to serve well on long timescales in a binary inspiral. Since the growth is a natural effect of the growing mismatch between the object's rest frame and the background Fermi-Walker frame, one should construct a new rest gauge every so often, effectively resetting the frame's clocks and gyroscopes.

\subsection{Transformation at second order}
Solving the second-order equation~\eqref{gauge_equation2} proceeds in the same way as at first order, and Eq.~\eqref{xi2} gives the final result for the worldline-preserving order-$\e^2$ transformation.

\subsubsection{Order $1/r^2$}
Unlike at first order, the leading terms in the second-order perturbations~\eqref{h2 rest} and \eqref{h2New} do not agree. The $m^2/r$ term in Eq.~\eqref{xi2a} brings them into agreement. Note that although the leading-order terms  superficially appear to lack spherical symmetry, this is an artifact of using Cartesian coordinates: in terms of tensor harmonics, they are purely $l=0$. This can be seen either from using the irreducible STF decomposition in Ref.~\cite{Blanchet-Damour:86} or by converting to polar coordinates $(t,r,\theta^A)$ using $h^2_{\alpha r}=h^2_{\alpha a}n^a$, $h^2_{\alpha A}=h^2_{\alpha a}\frac{\partial x^a}{\partial \theta^A}$, $\delta_{ab}\frac{\partial x^a}{\partial \theta^A}\frac{\partial x^b}{\partial \theta^B}=r^2\Omega_{AB}$ (where $\Omega_{AB}$ is the metric on the unit two-sphere), and  $n_a\frac{\partial x^a}{\partial \theta^A}=0$. At order $1/r^2$, the mixed terms $h^2_{tA}$ and $h^2_{rA}$ vanish, $h^2_{rr}$ is independent of $\theta^A$, and $h^2_{AB}$ is proportional to $\Omega_{AB}$. Similarly, the gauge vector corresponds to a monopolar radial transformation $r\to r-\xi^r$, with $\xi^r = \xi^an_a$.

\subsubsection{Order $1/r$}
The rest-gauge metric \eqref{h2 rest} contains no terms at order $1/r$. The Lorenz-gauge metric~\eqref{h2New}, on the other hand, contains terms of the form $mh^{\R1}_{\mu\nu}$ in both $h^{\S\R}_{\mu\nu}$ and $h^{\delta m}_{\mu\nu}$. Some terms of that form arise from the first-order transformation, by virtue of the $\Lie_{\xi_1}h^{1'}_{\mu\nu}$ term in Eq.~\eqref{gauge_equation1}, but additional terms are required in $\xi^2_\mu$. These appear as the order-$r^0$ terms in Eq.~\eqref{xi2}.

\subsubsection{Order $r^0$}
At order $r^0$, the rest-gauge metric \eqref{h2 rest} and Lorenz-gauge metric~\eqref{h2New} both contain terms of the form $m^2\E_{ab}$ and $m^2\B_{ab}$, while the Lorenz-gauge metric in addition contains $h^{\R2}_{\mu\nu}$ and $m\partial h^{\R1}_{\mu\nu}$, the latter through $h^{\S\R}_{\mu\nu}$ as well as $\ord{1}{h}^1_{\mu\nu}$. $\xi^2_\mu$ brings the  $m^2\E_{ab}$ and $m^2\B_{ab}$ terms into agreement, introduces $h^{\R2}_{\mu\nu}$ exactly as at first order, and further introduces the $m\partial h^{\R1}_{\mu\nu}$ terms. It also serves to remove terms of the form $(h^{\R1}_{\mu\nu})^2$ that appear via $\Lie^2_{\xi_1}\ord{0}{g}_{\mu\nu}$ in Eq.~\eqref{gauge_equation2}.

\subsubsection{Order $r$}
Finally, at order $r$, Eq.~\eqref{gauge_equation2} determines the second-order force $f_2^a$.  This result follows from the $tt$ component of Eq.~\eqref{gauge_equation2}, and I do not seek to solve the remaining components of the equation, leaving one piece of $\xi^{(2,1)}_a$ and all of $\xi^{(2,2)}_\mu$ undetermined.

The equation for $f_2^a$ comes in the same form as Eq.~\eqref{master equation1}: as a differential equation satisfied by the translation $\xi_a^{(2,0)}$. It reads
\begin{align}
\ddot{\xi}^{(2,0)}_a  &=   -\E_{a}{}^i\xi^{(2,0)}_i + f_a^2 -\tfrac{1}{2}\partial_a(\ord{0}{h}^{\R2}_{tt}+\ord{1}{h}^{\R1}_{tt})  \nonumber\\
			&\quad + \partial_t (\ord{0}{h}^{\R2}_{ta}+\ord{1}{h}^{\R1}_{ta})   + \tfrac{1}{2}\partial_t\ord{0}{h}^{\R1}_{tt}\,\ord{0}{h}^{\R1}_{ta},\label{a2 STF}
\end{align}
where all quantities are evaluated at $r=0$. As at first order, this equation is specifically the order-$r$, $l=1$ piece of the $tt$ component of Eq.~\eqref{gauge_equation2}; though unlike at first order, explicitly evaluating that piece to arrive at the above equation is nontrivial, complicated as it is by the $\Lie_{\xi_1}h^{1'}_{\mu\nu}$ term in Eq.~\eqref{gauge_equation2} and the presence of negative powers of $r$ in $\xi^2_a$. 

Just as at first order, I now impose the worldline-preserving condition $\xi^{(2,0)}_a=0$. This prevents a mass dipole moment $\delta M^i$ from appearing via the $\Lie_{\xi_2}h^{1,\rm old}_{\mu\nu}$ term in Eq.~\eqref{Dh3}. We can directly confirm that the entirety of the gauge transformation then leaves $\delta M^i=0$. One way of doing this is by substituting the explicit results for $\xi^\mu_1$ and $\xi^\mu_2$ into Eq.~\eqref{Dh3}, but a more efficient way is by converting to scaled coordinates $\tilde x^a$ and computing $\Delta H^1_{\mu\nu}$. Because $x^a$ derivatives lower the power of $\e$ by one, $\Delta H^1_{\mu\nu}$ gets contributions from both the linear- and quadratic-in-$\e$ pieces of the scaled transformation. The powers of $\e$ in the transformation $\tilde \xi^\mu(\tilde r)=\e\xi^\mu_1(\tilde r) +\e^2\xi^\mu_2(\tilde r)$ can be obtained by adding $p$ to $n$ in Eq.~\eqref{xi form}. Referring to Eqs.~\eqref{xi1} and \eqref{xi2}, we find $\tilde \xi^\mu_1=\tilde r^0\xi^\alpha_{(1,0)}+\frac{1}{\tilde r}\xi^{\mu i}_{(2,-1)}n_i$ and $\tilde\xi^\mu_2=\tilde r \xi^{\mu i}_{(1,1)}n_i+\tilde r^0(\xi^\mu_{(2,0)}+\xi^{\mu i}_{(2,0)}n_i)$, with $\xi^a_{(1,0)}=\xi^a_{(2,0)}=0$. The  coefficient of the order-$\e$ piece of $\Lie_{\tilde\xi}g^{\rm obj}_{\mu\nu}$ has components $\Delta H^1_{tt}=\tilde \xi^a_2\partial_{\tilde a}g^{\rm obj}_{tt}+2\tilde\xi^t_{1,t}g^{\rm obj}_{tt}$, $\Delta H^1_{ta}=\tilde \xi^t_{2,\tilde a}g^{\rm obj}_{tt}$, and $\Delta H^1_{ab}=\tilde \xi^c_2\partial_{\tilde c}g^{\rm obj}_{ab}+2\tilde\xi^c_{2,(\tilde a}g^{\rm obj}_{b)c}$, where I have omitted vanishing $t$ derivatives. It is now easy to see that $\Delta H^1_{\mu\nu}$ contains only even values of $l$, and so in particular, no terms with $l=1$. 

So, confidently imposing $\xi^{(2,0)}_a=0$, we find that Eq.~\eqref{a2 STF} becomes
\begin{align}
f^2_a &= \frac{1}{2}\partial_a \ord{0}{h}^{\R2}_{tt}-\partial_t \ord{0}{h}^{\R2}_{ta} +\frac{1}{2}\partial_a \ord{1}{h}^{\R1}_{tt}-\partial_t \ord{1}{h}^{\R1}_{ta} \nonumber\\
		&\quad - \frac{1}{2}\partial_t\ord{0}{h}^{\R1}_{tt}\,\ord{0}{h}^{\R1}_{ta}.\label{a2-hR}
\end{align}
One can verify that this is equivalent to Eq.~(16) of Ref.~\cite{Pound:12a} using Eqs.~\eqref{A10_to_tail}, \eqref{C10_to_tail}, and \eqref{A11_to_tail} (with $h^{\R1}_{tt}\big|_\gamma=\hat A^{(1,0)}$, $h^{\R1}_{ta}\big|_\gamma=\hat C^{(1,0)}_{a}$, and $h^{\R1}_{tt,a}\big|_\gamma=\hat A^{(1,0)}_{a}$).

With this, we have found the first two terms in the acceleration $a^\mu=\e f^\mu_1+\e^2 f^\mu_2+\O(\e^3)$. Summing the two terms, we find that the result can be written in the compact covariant form 
\begin{align}
a^\mu &= -\frac{1}{2}P^{\mu\nu}\left(g_\nu{}^\rho-h^{{\rm R}}_\nu{}^\rho\right)\!\left(2h^{\rm R}_{\rho\sigma;\lambda}-h^{\rm R}_{\sigma\lambda;\rho}\right)\!u^\sigma u^\lambda\nonumber\\
&\quad +\O(\e^3),\label{acceleration}
\end{align}
where $h^{\rm R}_{\mu\nu}=\e h^{{\rm R}1}_{\mu\nu}+\e^2h^{{\rm R}2}_{\mu\nu}$. If $g^{\rm obj}_{\mu\nu}$ had spin and quadrupole moments, then this equation would be expected to include the standard test-body-type quadrupole forces~\cite{Dixon:74,Thorne-Hartle:85,Harte:12} as well as a correction to the Papapetrou spin force. We may be able to correctly extract those forces from Harte's fully nonlinear equations~\cite{Harte:12}, though it is unclear whether the moments he defines would correspond to the ones defined from matched expansions.


\section{Geodesic motion in an effective spacetime}\label{geodesic_motion}
I opened this paper by promising that the second-order equation of motion was equivalent to the second-order geodesic equation in a meaningful effective metric. In this section, I show that Eq.~\eqref{acceleration} is in fact equivalent to the geodesic equations in $g_{\mu\nu}+\e h^{\R1}_{\mu\nu}+\e^2 h^{\R2}_{\mu\nu}$.

An appropriate expansion of the geodesic equation can be found in Sec.~IIIA.1 of Ref.~\cite{Pound:15b}, but I reproduce it here for the reader's convenience. In a metric ${\sf g}_{\mu\nu}=g_{\mu\nu}+h_{\mu\nu}$, the geodesic equation reads $\frac{d\dot z^\mu}{d\lambda}+{}^{\sf g}\Gamma^\mu_{\nu\rho}\dot z^\nu \dot z^\rho=\kappa \dot z^\mu$, where $\lambda$ is a potentially nonaffine parameter on the geodesic, $z^\mu(\lambda)$ are the geodesic's coordinates, $\dot z^\mu=\frac{dz^\mu}{d\lambda}$ is its tangent vector field, ${}^{\sf g}\Gamma^\mu_{\nu\rho}$ is the Christoffel symbol corresponding to ${\sf g}_{\mu\nu}$, and $\kappa=\frac{d}{d\lambda}\ln\sqrt{-{\sf g}_{\mu\nu}u^\mu u^\nu}$. If we take $\lambda=\tau$ ($=$Fermi-Walker coordinate $t$), the proper time on $\gamma$ in $g_{\mu\nu}$, then the geodesic equation becomes 
\begin{equation}
a^\mu = -C^\mu_{\nu\rho}u^\nu u^\rho+\kappa u^\mu,\label{exact_geodesic}
\end{equation}
where $a^\mu=\frac{Du^\mu}{d\tau}$, and
\begin{align}
C^\alpha_{\beta\gamma} &\equiv {}^{\sf g}\Gamma^\alpha_{\beta\gamma}-\Gamma^\alpha_{\beta\gamma}\\
				&= \frac{1}{2}{\sf g}^{\alpha\delta}\left(2h_{\delta(\beta;\gamma)}-h_{\beta\gamma;\delta}\right)
\end{align}
is the difference between the Christoffel symbols in the full metric ${\sf g}_{\mu\nu}$ and in the background $g_{\mu\nu}$. With $\tau$ as a parameter, $\kappa$ becomes
\begin{equation}
\kappa = \frac{\frac{d}{d\tau}\sqrt{1-h_{\mu\nu}u^\mu u^\nu}}{\sqrt{1-h_{\mu\nu}u^\mu u^\nu}}.
\end{equation}

So far Eq.~\eqref{exact_geodesic} is exact. I now expand $C^\mu_{\nu\rho}$ and $\kappa$ in powers of $h_{\mu\nu}$, yielding
\begin{align}
a^\alpha &= -\frac{1}{2}(g^{\alpha\delta}-h^{\alpha\delta})\!\left(2h_{\delta(\beta;\gamma)}-h_{\beta\gamma;\delta}\right)\!u^\beta u^\gamma\nonumber\\
		&\quad	-\frac{1}{2}h_{\beta\gamma;\delta}u^\alpha u^\beta u^\gamma u^\delta
			-\frac{1}{2}h_{\mu\nu}h_{\beta\gamma;\delta}u^\alpha u^\beta u^\gamma u^\delta u^\mu u^\nu \nonumber\\
		&\quad	-h_{\beta\gamma}u^\alpha a^\beta u^\gamma+\O(h^3). \label{expanded_geodesic}
\end{align}
This equation is complicated by the fact that the acceleration appears on both sides in a nontrivial way. To disentangle the acceleration from the perturbation, I assume that $a^\mu$, too, has an expansion in powers of $h_{\mu\nu}$,
\begin{equation}
a^\mu = a^\mu_{\rm lin}+a^\mu_{\rm quad}+\O(h^3),
\end{equation}
where $a^\mu_{\rm lin}$ is linear in $h_{\mu\nu}$ and $a^\mu_{\rm quad}$ is quadratic in it. Substituting this expansion into Eq.~\eqref{expanded_geodesic}, one finds
\begin{align}
a^\mu_{\rm lin} &= -\frac{1}{2}P^{\alpha\delta}\left(2h_{\delta(\beta;\gamma)}-h_{\beta\gamma;\delta}\right)\!u^\beta u^\gamma,\\
a^\mu_{\rm quad} &= -\frac{1}{2}P^{\alpha\mu}h^\delta{}_\mu\!\left(2h_{\delta(\beta;\gamma)}-h_{\beta\gamma;\delta}\right)\!u^\beta u^\gamma.
\end{align}
Summing these yields the compact form
\begin{align}
a^\mu &=  -\frac{1}{2}P^{\alpha\mu}(g^\delta{}_\mu-h^\delta{}_\mu)\!\left(2h_{\delta(\beta;\gamma)}-h_{\beta\gamma;\delta}\right)\!u^\beta u^\gamma\nonumber\\
		&\quad +\O(h^3).\label{2nd-geo}
\end{align}

Comparing Eq.~\eqref{2nd-geo} to Eq.~\eqref{acceleration}, we see that, up to terms cubic in $h^{\rm R}_{\mu\nu}$, the second-order self-forced motion is, as promised, identical to  geodesic motion in the  effective metric $g_{\mu\nu}+h^{\rm R}_{\mu\nu}$.


\section{Motion in alternative gauges}\label{gauge}

The equation of motion~\eqref{acceleration} is specific to the Lorenz gauge, but in practice we may wish to work in other gauges. In this section, I first describe how it applies in gauges smoothly related to Lorenz. I then show, more promisingly, how it applies in a highly regular gauge with a different singularity structure than Lorenz.

\subsection{Motion in gauges smoothly related to Lorenz}\label{smooth gauge}
In Ref.~\cite{Pound:15a}, I described how to transform between smoothly related gauges, given a specification, in the initial gauge, of a singular-regular split for which the motion is geodesic in $g_{\mu\nu}+h^{\R}_{\mu\nu}$. I only briefly reiterate that prescription here. 

Consider a  transformation away from Lorenz generated by  arbitrary smooth vectors $\xi^\mu_1$ and $\xi^\mu_2$. At first order we have $h^1_{\mu\nu}\to h^1_{\mu\nu}+\Lie_{\xi_1} g_{\mu\nu}$. Since the transformation is smooth, we can naturally assign its effect to the regular field, such that in the new gauge we have
\begin{align}
h^{\R1'}_{\mu\nu} &= h^{\R1}_{\mu\nu} + \Lie_{\xi_1} g_{\mu\nu},\label{DhR1}\\
h^{\S1'}_{\mu\nu} &= h^{\S1}_{\mu\nu}.
\end{align}
(Here primes denote perturbations in the new gauge, not perturbations in the rest gauge as they did in previous sections.) At second order we have  $h^2_{\mu\nu}\to h^2_{\mu\nu}+\Delta h^2_{\mu\nu}$, with $\Delta h^2_{\mu\nu}$ given by Eq.~\eqref{Dh2}. This transformation includes a singular term $\Lie_{\xi_1}h^{\S1}_{\mu\nu}$, but we can again assign the smooth remainder to the regular field:
\begin{align}
h^{\R2'}_{\mu\nu} &= \Lie_{\xi_2} g_{\mu\nu}+\frac{1}{2}\Lie^2_{\xi_1} g_{\mu\nu} + \Lie_{\xi_1} h^{\R1}_{\mu\nu},\label{DhR2}\\
h^{\S2'}_{\mu\nu} &= h^{\S2}_{\mu\nu} + \Lie_{\xi_1} h^{\S1}_{\mu\nu}.\label{DhS2}
\end{align}
With these transformation laws, the effective metric $g_{\mu\nu}+\e h^{\R1}_{\mu\nu}+\e^2 h^{\R2}_{\mu\nu}$ transforms as any ordinary smooth metric would, thus ensuring that it remains a vacuum metric in the new gauge. We also see that we can freely choose the gauge of the regular field, while the form of the singular field is dictated by (a) its form in the Lorenz gauge and (b) the gauge of $h^{\R1'}_{\mu\nu}$ (through $h^{\S2'}_{\mu\nu}$'s dependence on $\xi^\mu_1$). Because $\xi^\mu_1$ is associated with $h^{\R1}_{\mu\nu}$, we can think of $h^{\S\S}_{\mu\nu}$, like $h^{\S1}_{\mu\nu}$, as being invariant under smooth transformations, while the ``singular times regular'' pieces of the metric, $h^{\S\R}_{\mu\nu}$ and $h^{\delta m}_{\mu\nu}$, are altered by $\Lie_{\xi_1} h^{\S1}_{\mu\nu}$.

Here I consider a generic smooth transformation, not a worldline-preserving one, meaning it alters the worldline as well. Specifically, the coordinate description of the worldline changes as any coordinates do,\footnote{Because of this, the transformation laws~\eqref{DhR2}--\eqref{DhS2}, as written, will introduce a mass dipole moment into $h^{\S2'}_{\mu\nu}$, via the term  $\Lie_{\xi_1} h^{\S1}_{\mu\nu}$. However, in the self-consistent approximation, the metric perturbations are functionals of the worldline, and Eqs.~\eqref{DhR2}--\eqref{DhS2} as written leave the metric in the {\em new} gauge as functionals of the worldline in the {\em old} gauge. To make them functionals of the new worldline, the perturbations $h^{n}_{\mu\nu}[z]$ on the right-hand sides of the transformation laws should be expanded around $h^n_{\mu\nu}[z']$, shifting the worldline on which the singular field diverges and leading to additional terms in Eqs.~\eqref{DhR2}--\eqref{DhS2}. The additional term in Eq.~\eqref{DhS2} eliminates the dipole moment. I refer the reader to Sec. IVB of Ref.~\cite{Pound:15b} for a detailed discussion.}  
\beq\label{dz}
z^\mu\to z^\mu-\epsilon \xi^\mu_1-\epsilon^2\!\! \left(\xi^\mu_2-\frac{1}{2}\xi^\nu_1\partial_\nu\xi_1^\mu\right) +\O(\e^3).
\eeq
Now note that under a coordinate transformation, a geodesic of a given metric remains a geodesic of that metric, given that the metric transforms as an ordinary tensor under that transformation. Since the regular fields have been defined to transform as ordinary metric perturbations, it follows that the transformed worldline is a geodesic of the transformed effective metric. Hence, the equation of motion~\eqref{acceleration} applies in all gauges smoothly related to Lorenz, with the regular field defined to transform  according to Eqs.~\eqref{DhR1} and \eqref{DhR2}. 

\subsection{Motion in a highly regular gauge}
In the bulk of the paper, I have treated the Lorenz gauge as the ``practical gauge'', defining a regular field within it and transforming from the rest gauge to it. The preceding section extended the results to other gauges, but only those that share the same singularity structure as the Lorenz gauge (specifically, the same $h^{\S\S}_{\mu\nu}$). In this section I will show that there are more advantageous practical gauges, and I will derive the equation of motion in them.

To see why superior choices of gauge exist, recall that, as described in the Introduction, the most singular, $\sim m^n/r^n$ terms in the outer perturbations $h^n_{\mu\nu}$ correspond to terms in the inner background metric $g^{\rm obj}_{\mu\nu}$. For a spherical object, $g^{\rm obj}_{\mu\nu}$ in the buffer region is simply the Schwarzschild metric. Generically, it will contain all powers of $m/r$. But in light-cone coordinates, such as are used in Eq.~\eqref{gobj-lightcone}, the Schwarzschild metric is {\em linear} in $m/r$. This has dramatic consequences: if we simply take the inner expansion $g^{\rm obj}_{\mu\nu}+\sum \e^n H^n_{\mu\nu}$ from Eqs.~\eqref{gobj-lightcone}--\eqref{H3} and re-expand it in the buffer region, we obtain an outer expansion in a gauge that has {\em eliminated the most singular pieces of the metric}. I now show how to transform to a class of practical gauges that preserve this property. This has potentially significant benefits, described in Sec.~\ref{discussion} below.

\subsubsection{Outer expansion in the light-cone rest gauge}
Before proceeding to transform to the practical gauge, I first present the explicit form of the outer expansion in the original light-cone gauge. I begin with the inner metric of Eqs.~\eqref{gobj-lightcone}--\eqref{H3}, re-expand for small $\e$ at fixed $\sfr$, and then apply the transformation~\eqref{adv-to-Fermi} from advanced coordinates to Fermi-Walker coordinates. Unlike in Sec.~\eqref{preliminary transformation}, I {\em do not} combine this with the transformation~\eqref{adv-to-harmonic}. The result is a new, light-cone rest-gauge metric $\ord{0}{g}_{\mu\nu}+\e h^{1'}_{\mu\nu}+\e^2 h^{2'}_{\mu\nu}$, similar in form to (but less singular than) the one given in Eq.~\eqref{g rest}. At the first few  orders in $r$, the perturbations read
\begin{subequations}\label{h1 lightcone}
\begin{align}
h^{1'}_{tt} &= \frac{2m}{r}+\tfrac{11}{3}mr\E_{ij}\nhat^{ij}+\O(r^2\ln r),\\
h^{1'}_{ta} &= \frac{2m}{r}n_a+mr\left(\tfrac{22}{15}\E_{ai}n^i+\tfrac{4}{3}\B^{cd}\epsilon_{adi}\nhat_c{}^i+2\E_{ij}\nhat_a{}^{ij}\right)\nonumber\\
		&\quad +\O(r^2\ln r),\\
h^{1'}_{ab} &= \frac{2m}{r}n_{ab}+mr\left(\tfrac{22}{45} \mathcal{E}{}_{ab} -  \tfrac{8}{15} \mathcal{B}{}_{(a}{}^{d} \epsilon{}_{b)cd} \hat{n}{}^{c} \right.\nonumber\\
		&\quad  + \tfrac{32}{21} \mathcal{E}{}_{(a}{}^{c} \hat{n}{}_{b)c}- \tfrac{8}{3} \mathcal{B}{}^{cd}\epsilon{}_{c(a}{}^{i} \hat{n}{}_{b)di} + \tfrac{1}{21} \delta{}_{ab} \hat{n}{}_{cd}\mathcal{E}{}^{cd} \nonumber\\
		&\quad + \tfrac{1}{3} \mathcal{E}{}^{cd}\hat{n}{}_{abcd})+\O(r^2\ln r),
\end{align}
\end{subequations}
and 
\begin{subequations}\label{h2 lightcone}
\allowdisplaybreaks
\begin{align}
h^{2'}_{tt} &= -4m^2\E_{ij}\nhat^{ij}+\tfrac{11}{3}mr \delta \mathcal{E}_{ij} \hat{n}^{ij}-m^2r\bigl[ \tfrac{44}{3} \dot{\mathcal{E}}{}^{ci} \hat{n}{}_{ci} \nonumber\\
			&\quad - 8 \dot{\mathcal{E}}{}^{ci} \ln(2 m/r) \hat{n}{}_{ci} +\tfrac{8}{3} \mathcal{E}{}^{cij} \hat{n}{}_ {cij}\bigr]\nonumber\\
			&\quad +\O(r^2\ln r),\\
h^{2'}_{ta} &= - \tfrac{4}{15} (6 \mathcal{E}{}_{ac} \hat{n}{}^{c} + 5 \mathcal{B}{}^{cd} \epsilon{}_{adi} \hat{n}{}_{c}{}^{i} + 15 \mathcal{E}{}^{cd} \hat{n}{}_{acd})\nonumber\\
				&\quad +mr\left(\tfrac{22}{15}\delta\E_{ai}n^i+\tfrac{4}{3}\delta\B^{cd}\epsilon_{adi}\nhat_c{}^i+2\delta\E_{ij}\nhat_a{}^{ij}\right)\nonumber\\
				&\quad + m^2r\left(\tfrac{24}{5} \dot{\mathcal{E}}{}_ {ac} [\ln(2 m/r)-2] \hat{n}{}^{c} -  \tfrac{32}{21} \mathcal{E}{}_{acd} \hat{n}{}^{cd}\right. \nonumber\\
				&\quad + \tfrac{8}{9}\dot{\mathcal{B}}{}^{cd} \epsilon{}_ {adi} [3 \log(2m/r)-4] \hat{n}{}_{c}{}^{i} -  \tfrac{8}{9} \mathcal{B}{}^{cdi} \epsilon{}_{ac}{}^{j} \hat{n}{}_{dij} \nonumber\\
				&\quad  \left.+ \tfrac{4}{9} \dot{\mathcal{E}}{}^{cd} [12 \ln(2m/r)-19 ] \hat{n}{}_{acd}  - 2 \mathcal{E}{}^{cdi} \hat{n}{}_{acdi}\right)\nonumber\\
				&\quad+\O(r^2\ln r),\\
h^{2'}_{ab} &= - \tfrac{2}{21} (28 \mathcal{B}{}_{(a}{}^{d} \epsilon{}_{b)cd} \hat{n}{}^{c} + 48 \mathcal{E}{}_{(a}{}^{c} \hat{n}{}_{b)c} - 2 \mathcal{E}{}^{cd} \delta{}_{ab} \hat{n}{}_{cd}\nonumber\\
				&\quad - 70 \mathcal{B}{}^{cd} \epsilon{}_{c(a}{}^{i} \hat{n}{}_{b)di}  + 35 \mathcal{E}{}^{cd} \hat{n}{}_{abcd}) 
				+mr\Bigl(\tfrac{22}{45} \delta\mathcal{E}{}_{ab}  \nonumber\\
		&\quad  -  \tfrac{8}{15} \delta\mathcal{B}{}_{(a}{}^{d} \epsilon{}_{b)cd} \hat{n}{}^{c}+ \tfrac{32}{21}\delta \mathcal{E}{}_{(a}{}^{c} \hat{n}{}_{b)c}
		- \tfrac{8}{3}\delta \mathcal{B}{}^{cd}\epsilon{}_{c(a}{}^{i} \hat{n}{}_{b)di} 
		 \nonumber\\
		&\quad + \tfrac{1}{21} \delta{}_{ab} \hat{n}{}_{cd}\delta\mathcal{E}{}^{cd} + \tfrac{1}{3}\delta \mathcal{E}{}^{cd}\hat{n}{}_{abcd}\Bigr)\nonumber\\
		&\quad +m^2r\Bigl\{\tfrac{8}{45} \dot{\mathcal{E}}{}_{ab} [12 \ln(2m/r)-31] -  \tfrac{16}{21} \mathcal{E}{}_{abc} \hat{n}{}^{c} \nonumber\\
		&\quad  -  \tfrac{16}{45} \dot{\mathcal{B}}{}_{(a}{}^{d} \epsilon{}_{b)cd} [ 3 \ln(2m/r)-4] \hat{n}{}^{c} \nonumber\\
		&\quad  + \tfrac{16}{7} \dot{\mathcal{E}}{}_{(a}{}^{c} [3 \ln(2m/r)-4] \hat{n}{}_{b)c}  + \tfrac{32}{63} \mathcal{B}{}_{c(a}{}^{i} \epsilon{}_{b)di} \hat{n}{}^{cd}  \nonumber\\
		&\quad -  \tfrac{20}{9} \mathcal{E}{}_{(a}{}^{cd} \hat{n}{}_{b)cd}  - \tfrac{16}{9} \dot{\mathcal{B}}{}^{cd} [3 \ln(2m/r)-4] \epsilon{}_{c(a}{}^{i} \hat{n}{}_{b)di} \nonumber\\
		&\quad +\tfrac{4}{63} \dot{\mathcal{E}}{}^{cd} \bigl(\delta{}_ {ab} [6 \ln(2m/r)-29] \hat{n}{}_{cd} \nonumber\\
		&\quad + 14 [3 \ln(2m/r)-4] \hat{n}{}_ {abcd}\bigr) + \tfrac{16}{9} \mathcal{B}{}^{cdi} \epsilon{}_{c(a}{}^{j} \hat{n}{}_{b)dij} \nonumber\\
		&\quad  -  \tfrac{4}{27} \mathcal{E}{}^{cdi} (\delta{}_ {ab} \hat{n}{}_{cdi} + 9 \hat{n}{}_{abcdi}) \Bigr\}+\O(r^2\ln r).
\end{align}
\end{subequations}
Unlike the second-order field in the Lorenz gauge (and indeed, in all gauges previously considered in the literature), which diverges as $1/r^2$ at $r=0$, the second-order field in the light-cone gauge is actually {\em finite} at $r=0$.

\subsubsection{Singular and regular fields}\label{singular and regular in highly regular gauge}
We can naturally divide the perturbations in the light-cone gauge into singular and regular pieces, though this division will ultimately differ from the one in the Lorenz gauge. At order $r^2$, $h^{1'}_{\mu\nu}$ contains the  tidal moments $\delta\E_{ab}$ and $\delta\B_{ab}$ [this order is omitted for brevity in Eq.~\eqref{h1 lightcone}, but it looks schematically the same as in Eq.~\eqref{h1 rest}]; at order $r^3$ it would contain octupolar moments $\delta\E_{abc}$ and $\delta\B_{abc}$; and so on. In accordance with the fact that these terms take an identical form in $h^{1'}_{\mu\nu}$ as in the external background $\ord{0}{g}_{\mu\nu}$, and the idea that the effective metric $\ord{0}{g}_{\mu\nu}+h^{\R'}_{\mu\nu}$ is perceived as the {\em external} metric in the neighbourhood of the object, we can define $h^{\R1'}_{\mu\nu}$ to comprise everything in $h^{1'}_{\mu\nu}$ involving these moments. Explicitly, it is then given by
\begin{subequations}
\begin{align}
h^{\R1'}_{tt} &=  -r^2\delta \E_{ij}\nhat^{ij} +\O(r^3),\\
h^{\R1'}_{ta} &= -\tfrac{2}{3}r^2\epsilon_{adi}\delta\B^{cd}\nhat_c{}^i+\O(r^3),\\
h^{\R1'}_{ab} &= -\tfrac{1}{9}r^2 (\delta\mathcal{E}_{ab} 
				- 6\delta \mathcal{E}_{(a}{}^{c} \hat{n}_{b)c} 
				+ 3 \delta\mathcal{E}^{cd} \delta_{ab} \hat{n}_{cd})\nonumber\\
				&\quad +\O(r^3).
\end{align}
\end{subequations}
The singular field $h^{\S1'}_{\mu\nu}$ then consists of all terms with an explicit factor of $m$. Through order $r$, it is the whole of $h^{1'}_{\mu\nu}$, given by Eq.~\eqref{h1 lightcone}. 

Similarly, at second order, second-order moments $\delta^2\E_{ab}$, $\delta^2\B_{ab}$, etc., appear, along with terms quadratic in the first-order moments $\delta \E_{ab}$, $\delta\B_{ab}$, etc. I define the second-order regular field to comprise all such terms. This guarantees that it is a vacuum metric, satisfying the vacuum EFE 
\beq\label{vac EFE}
\delta R_{\mu\nu}[h^{\R1'}]+\delta R_{\mu\nu}[h^{\R2'}]+\delta^2R_{\mu\nu}[h^{\R1'}]=\O(\e^3)
\eeq
to all orders in $r$. And since the tidal terms begin at order $r^2$, we have
\beq
h^{\R2'}_{\mu\nu} = \O(r^2).
\eeq
Everything else in $h^{2'}_{\mu\nu}$ should go into $h^{\S2'}_{\mu\nu}$. In analogy with the $h^{\S\R}_{\mu\nu}$ terms in the Lorenz gauge, this should include the ``singular times regular'' terms such as $m\delta\E_{ab}$ and $m\delta\B_{ab}$, which appear at order $r$ in Eq.~\eqref{h2 lightcone}; if these were included in the regular field, then it would cease to satisfy Eq.~\eqref{vac EFE}. Hence,  through order $r$, $h^{\S2'}_{\mu\nu}$ is the whole of $h^{2'}_{\mu\nu}$, given by Eq.~\eqref{h2 lightcone}.

\subsubsection{Transformation to a smoothly related practical gauge}\label{smoothly related highly regular gauges}
At this stage, although we have a natural split into singular and regular fields, the metric in this rest gauge is not fit for practical use. It constrains the form of the regular field, forcing the regular field and its first derivative to vanish on the worldline and its second derivatives to take a particular form. It is not obvious how one would impose such a gauge condition in a global numerical scheme. Furthermore, the metric in this gauge is not uniform in time: as described in Sec.~\ref{tidal matching}, the tidal moments $\delta\E_{ab}$ and $\delta\B_{ab}$ grow large with time. 

In order to transform to a practical gauge without losing the advantages of the light-cone gauge, I transform the gauge of the regular field while, insofar as is possible, leaving the gauge of the singular field intact. This can be done as described in Sec.~\ref{smooth gauge}. However, as when I transformed to the Lorenz gauge, here I wish to ensure that the transformation does not alter the worldline. So I begin with smooth vectors $\xi^\mu_1$ and $\xi^\mu_2$ that satisfy the worldline-preserving condition~\eqref{worldline-preserving} but are otherwise arbitrary. For smooth vectors, the condition~\eqref{worldline-preserving} reduces to
\beq
\xi^a_n\big|_\gamma=0.
\eeq

Given that, as in Sec.~\ref{inner_expansion}, the metric $\ord{0}{g}_{\mu\nu}+\e h^{1'}_{\mu\nu}+\e^2 h^{2'}_{\mu\nu}$ includes an expansion of the acceleration, a transformation will bring it to a metric $\ord{0}{g}_{\mu\nu}+\e h^{1\dagger}_{\mu\nu}+\e^2 h^{2\dagger}_{\mu\nu}$, that is likewise expanded. The transformation  laws are then~\eqref{gauge_equation1} and \eqref{gauge_equation2}. Splitting them into laws for the singular and regular fields, as in Eqs.~\eqref{DhR1}--\eqref{DhS2}, we have
\begin{align}
h^{\R1\dagger}_{\mu\nu} &= h^{\R1'}_{\mu\nu} + \Lie_{\xi_1} \ord{0}{g}_{\mu\nu},\label{DhR1 dagger}\\
h^{\S1\dagger}_{\mu\nu} &= h^{\S1'}_{\mu\nu},\\
h^{\R2\dagger}_{\mu\nu} &= \Lie_{\xi_2} \ord{0}{g}_{\mu\nu}+\frac{1}{2}\Lie^2_{\xi_1} \ord{0}{g}_{\mu\nu} + \Lie_{\xi_1} h^{\R1'}_{\mu\nu},\label{DhR2 dagger}\\
h^{\S2\dagger}_{\mu\nu} &= h^{\S2'}_{\mu\nu} + \Lie_{\xi_1} h^{\S1'}_{\mu\nu}.\label{DhS2 dagger}
\end{align}
If $h^{\R n}_{\mu\nu}$ are the ``exact'' regular fields in the new practical gauge, with no expansion of the acceleration, then the daggered fields are 
\begin{align}
h^{\R1\dagger}_{\mu\nu}&:=\ord{0}{h}^{\R1}_{\mu\nu}+\ord{0}{g}_{\mu\nu},\label{hR1 dagger}\\
h^{\R2\dagger}_{\mu\nu}&:=\ord{0}{h}^{\R2}_{\mu\nu}+\ord{1}{h}^{\R1}_{\mu\nu}+\ord{2}{g}_{\mu\nu}.
\end{align}
Similarly,
\begin{align}
h^{\S1\dagger}_{\mu\nu}&:=\ord{0}{h}^{\S1}_{\mu\nu},\\
h^{\S2\dagger}_{\mu\nu}&:=\ord{0}{h}^{\S2}_{\mu\nu}+\ord{1}{h}^{\S1}_{\mu\nu}.
\end{align}
Note that I group the background terms $\ord{n}{g}_{\mu\nu}$ with the regular fields $h^{\R n\dagger}_{\mu\nu}$;  this again corresponds to the idea that the background plus regular field together form the ``external metric''.

I now adopt Gralla's approach from Ref.~\cite{Gralla:12}. The key realization in his approach is that we do not need to explicitly impose any given gauge condition in the target gauge. Instead, we can take the regular fields $h^{\R n}_{\mu\nu}$ as given, to be determined in any desired gauge by a puncture scheme, and express $\xi^\mu_n$ (and the ``singular times regular'' piece of $h^{\S2}_{\mu\nu}$) in terms of them. As in the derivation in the Lorenz gauge, the worldline-preserving condition $\xi^a_n|_\gamma=0$ will suffice to determine an equation of motion; here my approach differs from Gralla's, corresponding to my self-consistent ({i.e.}, unexpanded) treatment of the worldline.

On the worldline, given that $\xi^a_1\big|_\gamma=0$, that $\ord{1}{g}_{\mu\nu}=\O(r)$, and that $h^{\R1'}_{\mu\nu}=O(r^2)$, Eq.~\eqref{DhR1 dagger} reads 
\begin{align}
\frac{d}{dt}\xi^1_t\big|_{\gamma} &= \frac{1}{2} \ord{0}{h}^{\rm R1}_{tt}\big|_{\gamma},\label{ddtxit}\\
\partial_a\xi^1_t\big|_{\gamma}\big|_{\gamma} &=  \ord{0}{h}^{\rm R1}_{ta}\big|_{\gamma},\\
\partial_{(a}\xi^1_{b)}\big|_{\gamma} &= \frac{1}{2} \ord{0}{h}^{\rm R1}_{ab}\big|_{\gamma}.\label{dxi symmetric}
\end{align}
These, together with $\xi^a_1\big|_\gamma=0$, determine all components of $\xi^1_\mu$ and $\partial_{a}\xi^1_{\mu}$ on the worldline, with the exception of $\partial_{[a}\xi^1_{b]}$. That remaining piece can be obtained from the formula
\beq\label{ddxi}
\ord{0}{\nabla}_\alpha\ord{0}{\nabla}_\beta \xi^1_\gamma +\ord{0}{R}_{\beta\gamma\alpha}{}^\delta\xi^1_\delta = \delta\Gamma_{\gamma\alpha\beta}[\Delta h^{\R1}],
\eeq
where $\Delta h^{\R1}_{\mu\nu}:=h^{\R1\dagger}_{\mu\nu} - h^{\R1'}_{\mu\nu}$, $\delta\Gamma_{\gamma\alpha\beta}[h]=\frac{1}{2}(2h_{\gamma(\alpha;\beta)}-h_{\alpha\beta;\gamma})$  is the standard correction to the Christoffel symbol (with its first index down), and all covariant derivatives and the Riemann tensor are compatible with $\ord{0}{g}_{\mu\nu}$. (Note that on $\gamma$, $\ord{0}{R}_{\beta\gamma\alpha\delta}=R_{\beta\gamma\alpha\delta}$.) Equation~\eqref{ddxi} can be derived by writing $2\ord{0}{\nabla}_{[\alpha}\ord{0}{\nabla}_{\beta]} \xi^1_\gamma=\ord{0}{R}_{\alpha\beta\gamma}{}^\delta\xi^1_\delta$, using Eq.~\eqref{DhR1 dagger} to replace $\ord{0}{\nabla}_\alpha\xi^1_\gamma$ with $\Delta h^{\R1}_{\gamma\alpha}-\ord{0}{\nabla}_\gamma\xi^1_\gamma$, and then adding the resulting equation to its cyclic permutations $\alpha\beta\gamma\to\gamma\alpha\beta$ and $\alpha\beta\gamma\to\beta\gamma\alpha$. 

On $\gamma$, the $tab$ component of Eq.~\eqref{ddxi} reads $\frac{d}{dt}\xi^1_{b,a}=\partial_{[a}h^{\R1\dagger}_{b]t}+\frac{1}{2}\partial_th^{\R1\dagger}_{ab}$. Substituting Eqs.~\eqref{hR1 dagger} and~\eqref{dxi symmetric}, we obtain a formula for the remaining piece of $\partial_a \xi^1_\mu$,
\begin{align}\label{dxi antisymmetric}
\frac{d}{dt}\xi^1_{[a,b]}\big|_{\gamma} &= - \partial_{[a}\ord{0}{h}^{\R1}_{b]t}\big|_{\gamma}.
\end{align}
The $tta$ component of Eq.~\eqref{ddxi} will be discussed below.

Using these results for $\xi^1_\mu$, we can now write the $\Lie_{\xi_1} h^{\S1'}_{\mu\nu}$ term in the second-order singular field~\eqref{DhS2 dagger} in terms of the regular field. Equation~\eqref{h1 lightcone} with Eqs.~\eqref{ddtxit}--\eqref{dxi symmetric} and \eqref{dxi antisymmetric} together yield
\begin{subequations}\label{Lie hS1}
\begin{align}
\Lie_{\xi_1} h^{\S1'}_{tt} &= -\frac{2m}{r}\left(\ord{0}{h}^{\R1}_{tt}+\tfrac{1}{2}\ord{0}{h}^{\R1}_{ij}n^{ij}\right)+\O(r^0),\\
\Lie_{\xi_1} h^{\S1'}_{ta} &= -\frac{2m}{r}\left(\ord{0}{h}^{\R1}_{ta} +\tfrac{1}{2}\ord{0}{h}^{\R1}_{tt}n_a-\ord{0}{h}^{\R1}_{ai}n^i+\ord{0}{h}^{\R1}_{ij}n_a{}^{ij}\right) \nonumber\\
								&\quad +\O(r^0),\\
\Lie_{\xi_1} h^{\S1'}_{ab} &= -\frac{2m}{r}\left(\tfrac{3}{2}\,\ord{0}{h}^{\R1}_{ab}+2\,\ord{0}{h}^{\R1}_{t(a}n_{b)}-2\,\ord{0}{h}^{\R1}_{i(a}n_{b)}{}^i\right)\nonumber\\
							&\quad +\O(r^0).
\end{align}
\end{subequations}
Note that this introduces a divergent, $1/r$ term into the singular field, making it less regular than in the original light-cone gauge. However, it remains less singular than the Lorenz gauge, and since the Lorenz gauge itself is often considered ``regular'' in comparison to highly singular gauges like the radiation gauge~\cite{Pound-Merlin-Barack:14}, the title ``highly regular'' remains apt. I will discuss practical implications of this in Sec.~\ref{discussion}. 

Terms higher order in $r$ in $\Lie_{\xi_1} h^{\S1'}_{\mu\nu}$ can be obtained from higher derivatives of $\xi_\mu^1$, which can be found by expanding Eq.~\eqref{ddxi} in powers of $r$. Alternatively, if we impose the Lorenz-gauge condition on $h^{\R1}_{\mu\nu}$, we can simply set $m=0$ in the gauge vector~\eqref{xi1} and straightforwardly compute $\Lie_{\xi_1} h^{\S1'}_{\mu\nu}$. 

This still leaves $h^{\S2}_{\mu\nu}$ dependent on the regular field in the old gauge, through the $\delta\E_{ab}$ and $\delta\B_{ab}$ terms in Eq.~\eqref{h2 lightcone}. We can express the moments in terms of the regular field in the new gauge by writing them as $\delta\E_{ab} = \delta R_{tatb}[h^{\R1'}]\big|_\gamma$ and $\delta\B_{ab} = \frac{1}{2}\epsilon^{pq}{}_{(a}\delta R_{b)tpq}[h^{\R1'}]\big|_\gamma$. Using Eq.~\eqref{DhR1 dagger}, $\delta R_{\alpha\beta\gamma\delta}[\Lie_{\xi_1} \ord{0}{g}] = \Lie_{\xi_1}\ord{0}{R}_{\alpha\beta\gamma\delta}$, and $R_{\alpha\beta\gamma\delta}[\ord{1}{g}]\big|_\gamma=0$, we arrive at 
\begin{align}
\delta\E_{ab} &= \delta R_{tatb}[\ord{0}{h}^{\R1}]-(\Lie_{\xi_1}R)_{tatb},\label{dE no Edot}\\
\delta\B_{ab} &= \frac{1}{2}\epsilon^{pq}{}_{(a}\left\{\delta R_{b)tpq}[\ord{0}{h}^{\R1}]-(\Lie_{\xi_1}R)_{b)tpq}\right\}, \label{dB no Bdot} 
\end{align}
where all quantities are evaluated on $\gamma$. These are given more explicitly by Eqs.~\eqref{dE} and \eqref{dB} with $\dot \E_{ab}$ and $\dot \B_{ab}$ set to zero; since Eqs.~\eqref{dE} and \eqref{dB} differ from Eqs.~\eqref{dE no Edot} and \eqref{dB no Bdot} by those $\dot \E_{ab}$ and $\dot \B_{ab}$ terms, this tells us that beginning at order $r^2$, the regular fields in the Lorenz gauge and highly regular gauge differ by more than a gauge transformation. To avoid this disagreement, we could make the redefinitions $\delta\E_{ab}\to \delta\E_{ab}-\frac{8}{3} m\dot{\E}_{ab}$ and $\delta\B_{ab}\to \delta\B_{ab}-\frac{13}{6} m\dot{\B}_{ab}$ (with a corresponding change to the singular field).

We now have a practical formulation with a convenient split into singular and regular fields. Through order $r$, the singular field is given by 
\begin{align}
h^{\S1}_{\mu\nu} &= h^{1'}_{\mu\nu}+\O(r^2),\label{hS1 highly regular}\\
h^{\S2}_{\mu\nu} &= h^{2'}_{\mu\nu}+\Lie_{\xi_1} h^{\S1'}_{\mu\nu}+\O(r^2)\label{hS2 highly regular}
\end{align}
with Eqs.~\eqref{h1 lightcone}, \eqref{h2 lightcone}, \eqref{Lie hS1}, \eqref{dE no Edot}, and \eqref{dB no Bdot}. Here I have cavalierly discarded the $\dagger$ notation, with the understanding that in these singular fields, acceleration terms have been implicitly moved from $h^{\S 1}_{\mu\nu}$ into $h^{\S 2}_{\mu\nu}$. In analogy with the notation in the Lorenz gauge, we can write
\beq\label{hSS and hSR}
h^{\S2}_{\mu\nu} = h^{\S\S}_{\mu\nu}+h^{\S\R}_{\mu\nu},
\eeq
where $h^{\S\S}_{\mu\nu}$ comprises all terms in $h^{2'}_{\mu\nu}$ explicitly proportional to $m^2$, and $h^{\S\R}_{\mu\nu}=h^{\S\R'}_{\mu\nu}+\Lie_{\xi_1} h^{\S1'}_{\mu\nu}$, with $h^{\S\R'}_{\mu\nu}$ comprising all terms in $h^{2'}_{\mu\nu}$ proportional to $m\delta \E_{ab}$, $m\delta \B_{ab}$, $m\delta \E_{abc}$, $m\delta \B_{abc}$, etc. These fields behave as $h^{\S\S}_{\mu\nu}\sim m^2r^0$ and $h^{\S\R}_{\mu\nu}\sim m h^{\R1}_{\mu\nu}/r$. 

With the singular fields defined, the regular fields can  be written implicitly, as the difference 
\beq\label{hR highly regular}
h^{\R n}_{\mu\nu}=h^{n}_{\mu\nu}-h^{\S n}_{\mu\nu}. 
\eeq
This regular field, like the one defined in the Lorenz gauge, is a ``physical'' field, causal on the worldline and satisfying the vacuum EFE. Its causality on the worldline follows from the same argument given for the Lorenz-gauge field in Ref.~\cite{Pound:15a}. It satisfying the vacuum equation follows immediately from it being a gauge transformation of the regular field defined in Sec.~\ref{singular and regular in highly regular gauge}.

In the above, I have made no mention of finding the second-order gauge vector $\xi^2_\mu$. We can express $\xi^2_\mu$ in terms of $h^{\R1}_{\mu\nu}$ and $h^{\R2}_{\mu\nu}$ in a similar way as we did for $\xi^1_\mu$, but doing so is not necessary: with the regular field defined implicitly through Eq.~\eqref{hR highly regular}, all we require explicitly is an expression for the singular field, and for that, $\xi^1_\mu$ suffices. However, one may need to consider $\xi^2_\mu$ if one needs to refine the form of $h^{\S2}_{\mu\nu}$. That might be necessary if, for example, higher-order terms in Eq.~\eqref{Lie hS1} are found to grow large with time. Such growth is highly possible, given that Eqs.~\eqref{ddtxit} and \eqref{dxi antisymmetric} dictate that $\xi^\mu_1$ will generically grow large on long time scales. Indeed, Gralla's singular field in Ref.~\cite{Gralla:12} appears to contain numerous terms that grow with time, stemming from the growth of his gauge vector (in addition to, and distinct from, the growth associated with his use of an expanded worldline). If growing terms arise in the highly regular gauge, they will have to be eliminated with a second-order gauge refinement. 

\subsubsection{Equation of motion}
All that remains to be determined in the new practical gauge is the equation of motion governing $\gamma$. 

At first order, the equation can be obtained from Eq.~\eqref{ddxi}. Given the worldline-preserving condition $\xi^1_a\big|_\gamma=0$, on $\gamma$ the $tta$ component of Eq.~\eqref{ddxi} reads simply $0=\partial_t \Delta h^{\R1}_{ta}-\tfrac{1}{2}\partial_a\Delta h^{\R1}_{tt}$. Noting that $ \Delta h^{\R1}_{\mu\nu} = h^{\R1}_{\mu\nu}-2f^1_a x^a \delta^t_\mu\delta^t_\nu+\O(r^2)$, we find that this is the standard formula for the first-order self-force:
\beq
f^1_a = \frac{1}{2}\partial_a h^{\R1}_{tt}-\partial_t h^{\R1}_{ta}.
\eeq

At second order, the analog of  Eq.~\eqref{ddxi} is
\beq
\ord{0}{\nabla}_\alpha\ord{0}{\nabla}_\beta \xi^2_\gamma +\ord{0}{R}_{\beta\gamma\alpha}{}^\delta\xi^2_\delta = \delta\Gamma_{\gamma\alpha\beta}[k],
\eeq
where $k_{\alpha\beta} = h^{\R2\dagger}_{\alpha\beta}-h^{\R2'}_{\alpha\beta}-\tfrac{1}{2}\Lie^2_{\xi_1}g_{\alpha\beta}-\Lie_{\xi_1}h^{\R1'}_{\alpha\beta}$. Given the worldline-preserving condition $\xi^2_a\big|_\gamma=0$, on $\gamma$ the $tta$ component reads $0=\partial_t k_{ta}-\tfrac{1}{2}\partial_ak_{tt}$. Noting that $h^{\R2'}_{\mu\nu}=\O(r^2)$ and evaluating the Lie derivatives, we find the simple formula $0=\partial_t h^{\R2\dagger}_{ta}-\tfrac{1}{2}\partial_ah^{\R2\dagger}_{tt}+\frac{1}{2}\partial_t\ord{0}{h}^{\R1}_{tt}\ord{0}{h}^{\R1}_{ta}$. Since $h^{\R2\dagger}_{\mu\nu} = \ord{0}{h}^{\R2}_{\mu\nu}+\ord{1}{h}^{\R1}_{\mu\nu}-2f^1_a x^a \delta^t_\mu\delta^t_\nu+\O(r^2)$, we recover Eq.~\eqref{a2-hR} for the second-order self-force:
\begin{align}
f^2_a &= \frac{1}{2}\partial_a \ord{0}{h}^{\R2}_{tt}-\partial_t \ord{0}{h}^{\R2}_{ta} +\frac{1}{2}\partial_a \ord{1}{h}^{\R1}_{tt}-\partial_t \ord{1}{h}^{\R1}_{ta} \nonumber\\
		&\quad - \frac{1}{2}\partial_t\ord{0}{h}^{\R1}_{tt}\,\ord{0}{h}^{\R1}_{ta}.
\end{align}

Therefore the expanded geodesic equation~\eqref{acceleration} holds true in the highly regular gauge,  with the regular field given by Eqs.~\eqref{hS1 highly regular}--\eqref{hR highly regular}. Since the gauge of the regular field is unspecified, this formulation in fact applies to a class of smoothly related highly regular gauges.

Before concluding, I make two remarks. First, we could have established in advance, without performing any calculations, that Eq.~\eqref{acceleration} would hold true, as it follows from the same argument given in Sec.~\eqref{smooth gauge}: the motion is trivially geodesic in $\ord{0}{g}_{\mu\nu}+h^{\R'}_{\mu\nu}$ because $u^\mu=\delta^\mu_t$ and the Christoffel symbols of this metric vanish on $\gamma$; and the geodesic equation is preserved under the transformation laws~\eqref{DhR1 dagger} and \eqref{DhR2 dagger}. The second remark relates to Gralla's results. Although I utilized key aspects of his methods in this section, I seem to have arrived at a different conclusion. He uses an expanded form of the worldline, $z^\mu_0+\e z^\mu_1+\e^2z^\mu_2+\O(\e^3)$ in coordinate form, and begins in a rest gauge centered on the background geodesic $\gamma_0=\{z_0^\mu\}$, such that $z^\mu_1=z^\mu_2=0$. In that context, when transforming to a practical gauge, $\xi^\mu_1$ and $\xi^\mu_2$ are allowed to take arbitrary values on $\gamma_0$, introducing deviation vectors $z^\mu_1=-\xi^\mu_1\big|_{\gamma_0}$ and $z^\mu_2=(-\xi^\mu_2+\tfrac{1}{2}\xi_1^\nu\partial_\nu \xi^\mu_1)\big|_{\gamma_0}$ that point toward the accelerated worldline in the new gauge.\footnote{Compare to Eq.~\eqref{dz}. These formulas differ from Gralla's due to our differing conventions.} He then finds evolution equations for $z^\mu_1$ and $z^\mu_2$ that are {\em not} equivalent to the geodesic equation in his effective metric, seemingly contrary to my results. However, this outward discrepancy stems from his definition of his  singular and regular fields. Instead of Eqs.~\eqref{DhR2 dagger} and \eqref{DhS2 dagger}, he effectively makes the choice $h^{\R2}_{\mu\nu} = h^{\R2'}_{\mu\nu} + \Lie_{\xi_2} g_{\mu\nu}+\frac{1}{2}\Lie^2_{\xi_1} g_{\mu\nu}$ and $h^{\S2}_{\mu\nu} = h^{\S2'}_{\mu\nu} + \Lie_{\xi_1} h^{1'}_{\mu\nu}$ (since the rest gauge is centered on a background geodesic, there is no need for daggers or left-superscripts here). By including the term $\Lie_{\xi_1} h^{\R1'}_{\mu\nu}$ in the singular field instead of the regular field, he arrives at an effective metric $g_{\mu\nu}+h^{\R}_{\mu\nu}$ that is not a vacuum metric and in which the motion is not geodesic. 

%


\section{Summary and discussion}\label{discussion}

The primary result of this paper is the second-order equation of motion~\eqref{acceleration} for a small, compact, approximately spherical and nonspinning object, whether a black hole, a  neutron star, or something more exotic. It is equivalent to the geodesic equation in the effective metric $g_{\mu\nu}+h^{\rm R}_{\mu\nu}$ defined in Sec.~\ref{outer_expansion}. This  metric satisfies ``physical'' properties: it is a vacuum solution, and if the full metric satisfies retarded boundary conditions, then the effective metric and its derivatives on the worldline depend only on the causal past. Therefore Eq.~\eqref{acceleration} represents a generalized equivalence principle of the sort described in the Introduction.

Equation~\eqref{acceleration} also has more pragmatic consequences. As discussed in Refs.~\cite{Pound:12a,Pound:12b,Pound-Miller:14}, the equation of motion can be combined with the field equations in a puncture scheme. Suppose we truncate the local expansion of the singular fields~\eqref{hS1} and \eqref{hS2} at a some order in $r$ and then attenuate them in some appropriate way away from the worldline. This defines puncture fields $h^{\P n}_{\mu\nu}$, which locally agree with $h^{\S n}_{\mu\nu}$, and  residual fields $h^{\res n}:=h^n_{\mu\nu}-h^{\P n}_{\mu\nu}$, which locally agree with $h^{\R n}_{\mu\nu}$. If the truncation of $h^{\S n}_{\mu\nu}$ is of sufficiently high order in $r$, then we can replace $h^{\R n}_{\mu\nu}$ with $h^{\res n}_{\mu\nu}$ in the equation of motion without introducing any error. We can also rewrite the field equations~\eqref{EFE1} and \eqref{EFE2} as equations for the first- and second-order residual fields,\footnote{Typically a point-particle stress-energy tensor would be included on the right-hand side of Eq.~\eqref{EFE1eff}, but here I follow Ref.~\cite{Gralla:12} by defining the right-hand sides of these equations pointwise off $\gamma$. They can then be defined on $\gamma$ by continuity.}
\begin{align}
E_{\mu\nu}[h^{\res1}] &= - E_{\mu\nu}[h^{\P1}],\label{EFE1eff}\\
E_{\mu\nu}[h^{\res2}] &= 2\delta^2R_{\mu\nu}[h^1]- E_{\mu\nu}[h^{\P2}].\label{EFE2eff}
\end{align}
The coupled system of equations~\eqref{acceleration}, \eqref{EFE1eff}, and \eqref{EFE2eff}, with the punctures moving on the worldline determined by~\eqref{acceleration},  provides a way of finding the physical fields $h^n_{\mu\nu}=h^{\res n}_{\mu\nu}+h^{\P n}_{\mu\nu}$ globally and the effective fields and their derivatives $\partial_{\alpha_1\cdots\alpha_p} h^{\R n}_{\mu\nu}\big|_\gamma=\partial_{\alpha_1\cdots\alpha_p} h^{\res n}_{\mu\nu}\big|_\gamma$ on the worldline (up to a maximum $p$ corresponding to the power of $r$ at which the singular field was truncated). By construction, the metric $g_{\mu\nu}+\e h^1_{\mu\nu}+\e^2 h^2_{\mu\nu}$ obtained in this way is guaranteed to agree locally, near $\gamma$, with the physical metric outside a compact object. Practical, covariant forms of the singular fields $h^{\S n}_{\mu\nu}$ are available in Ref.~\cite{Pound-Miller:14} for use in such a scheme. 

\subsection{Self-force computations in a highly regular gauge}

The above results were previously derived in the Lorenz gauge~\cite{Pound:12a} and smoothly related gauges~\cite{Pound:15b}. In the present paper, I derived a  promising extension to a class of highly regular gauges, in which the singular field is given by Eqs.~\eqref{hS1 highly regular} and \eqref{hS2 highly regular}, and the gauge of the regular field is freely specified. We can formulate a puncture scheme in these gauges by truncating and attenuating the singular fields, as described above, and then, rather than imposing a gauge condition on the exact regular field, imposing it on the residual field. There are some subtleties in imposing gauge conditions in the self-consistent context~\cite{Pound:15a}, but there should be no obstacle to imposing the Lorenz gauge condition, for example, on the total residual field $\sum_n \e^n h^{\res n}_{\mu\nu}$. We can then write the field equations as 
\begin{align}
E_{\mu\nu}[h^{\res1}] &= - \delta R_{\mu\nu}[h^{\P1}],\\
E_{\mu\nu}[h^{\res2}] &= 2\delta^2R_{\mu\nu}[h^1]- \delta R_{\mu\nu}[h^{\P2}],\label{EFE2 highly regular}
\end{align}
coupled, as above, to the equation of motion~\eqref{acceleration}.

Directly specifying the gauge of the residual field in this way, while leaving the puncture in any convenient gauge, was previously advocated by Gralla~\cite{Gralla:12}. Along the same lines, we could use the Lorenz-gauge puncture to compute a residual field that satisfies a gauge condition more convenient for black-hole perturbation theory, such as the radiation gauge condition that has been critical for self-force computations in a Kerr background~\cite{vandeMeent:16}. However,  the highly regular puncture in  Eqs.~\eqref{hS1 highly regular} and \eqref{hS2 highly regular} should provide significant advantages over the puncture in the Lorenz gauge (or in any gauge with a generic, $1/r^2$ divergence in $h^2_{\mu\nu}$, including Gralla's). 

The most obvious benefit of this gauge is simply that with its weaker divergences, the numerical cancellations between the two source terms in Eq.~\eqref{EFE2 highly regular} will be less delicate. But it has many other merits. To see this, consider the source $\delta^2R_{\mu\nu}[h^1]=\delta^2R_{\mu\nu}[h^{\S1}]+\delta^2R_{\mu\nu}[h^{\S1},h^{\R1}]+\delta^2R_{\mu\nu}[h^{\R1},h^{\S1}]+\delta^2R_{\mu\nu}[h^{\R1}]$ in Eq.~\eqref{EFE2 highly regular}. With a first-order singular field given by Eq.~\eqref{hS1 highly regular}, the source $\delta^2R_{\mu\nu}[h^{\S1}]$ diverges as $1/r^2$. We can see this from the fact that $\delta^2R_{\mu\nu}[h^{\S1}]$ is the source for the $h^{\S\S}_{\mu\nu}$ term in $h^2_{\mu\nu}$, described below Eq.~\eqref{hSS and hSR}: since $h^{\S\S}_{\mu\nu}\sim r^0$, we have $\delta^2R_{\mu\nu}[h^{\S1}] = - \delta R_{\mu\nu}[h^{\S\S}]\sim 1/r^2$. In a generic gauge, $\delta^2R_{\mu\nu}[h^{\S1}]$ is far more singular, behaving as $1/r^4$, and worse, it is not well defined as a distribution.\footnote{Though once $h^{\S\S}_{\mu\nu}$ is known in a given gauge, we might be able to define $\delta^2R_{\mu\nu}[h^{\S1}]$ distributionally as $-\delta R_{\mu\nu}[h^{\S\S}]$, which as a linear operator acting on an integrable function, is well defined as a distribution.} But $1/r^2$  is integrable, meaning $\delta^2R_{\mu\nu}[h^{\S1}]$ in the highly regular gauge {\em is} a well-defined distribution. Also, although the ``singular times regular'' source $\delta^2R_{\mu\nu}[h^{\S1},h^{\R1}]+\delta^2R_{\mu\nu}[h^{\R1},h^{\S1}]$ diverges as $1/r^3$, it too is a well-defined distribution because it is a linear operator acting on the integrable function $h^{\S1}_{\mu\nu}$. (This is true in a generic gauge, not only the highly regular gauge.) Therefore in the highly regular gauge we can write down a distributional equation for the second-order field $h^2_{\mu\nu}$. The equation will likely contain a $\delta$ function source in addition to $\delta^2R_{\mu\nu}[h^1]$; the correct source should be found by analyzing $\delta R_{\mu\nu}[h^{\S2}]$ as a distribution, in the same manner that the point-particle stress-energy tensor is obtained from $\delta R_{\mu\nu}[h^{\S1}]$~\cite{Gralla-Wald:08}. Once the correct source is found, we can develop numerical schemes to solve for $h^2_{\mu\nu}$ directly, rather than via the puncture scheme~\eqref{EFE2 highly regular}. In that case, the regular field could be extracted after the fact, by subtracting $h^{\S2}_{\mu\nu}$ from $h^{2}_{\mu\nu}$ using, for example, mode-sum regularization~\cite{Barack:09,Wardell:15}. Having a distributionally well-defined equation for $h^2_{\mu\nu}$ would also allow us to straightforwardly write down solutions in terms of Green's functions. From them, we will be able to define quasilocal singular and regular fields analogous to the Detweiler-Whiting definitions at first order (while ensuring, of course, that these definitions reduce to the purely local ones used here). These Green's function representations would provide yet another way of computing both the full field $h^2_{\mu\nu}$ and the regular field~\cite{Wardell-etal:14}. Working in the highly regular gauge should also reduce a computational difficulty that arises in generic gauges. If one uses a spherical-harmonic decomposition to solve the field equations, then computing any given mode of $\delta^2R_{\mu\nu}[h^1]$ near the worldline becomes laborious~\cite{Miller-Wardell-Pound:16}. The diminished singularity in the highly regular gauge should ameliorate the problem.

Considerable effort will be required to bring the highly regular gauge to the same state of development as the Lorenz gauge. Since a puncture scheme capable of computing the self-force requires a puncture through order $r$ (such that first derivatives of $h^{\R2}_{\mu\nu}$ can be computed), a first concrete step would be to calculate the singular field~\eqref{hS2 highly regular} to order $r$. This would require continuing the expansion of Eq.~\eqref{Lie hS1} to that order. As discussed in Sec.~\ref{smoothly related highly regular gauges}, additional gauge refinements may also be necessary if secularly growing terms arise in the puncture. Once that step is complete, the puncture can be written in covariant form using the methods of Ref.~\cite{Pound-Miller:14} and then expanded in harmonic modes for use in a mode decomposition of the field equations~\eqref{EFE2 highly regular}~\cite{Wardell-Warburton:15}.


\subsection{Rest gauges and  effective metrics}
Regardless of which of the two classes of gauges one uses, the underlying method of derivation was the same. It begins with the construction of a local metric in a gauge in which (a) the object is centered on some worldline $\gamma$ and (b) the regular field and its first derivatives vanish on that worldline. If the object is nonspinning and spherical, then in this gauge it appears to be manifestly at rest on $\gamma$, perturbed only by tidal fields. The existence of this gauge implies that for the nonspinning, spherical object, the worldline is a geodesic in {\em some} effective metric, and the heart of the derivation then becomes a matter of transforming to a more practical gauge and determining {\em which} piece of the full metric, in the practical gauge, constitutes that effective metric. 

As alluded to in the Introduction, this method is closely related to many others, both at first and second order. In particular, the basic form of the rest-gauge metric recurs throughout the literature on equations of motion. It was used in derivations of the first-order self-forced equation of motion~\cite{Mino-Sasaki-Tanaka:97, Detweiler:01, Poisson:03, Detweiler:05}. At second order, it has appeared as Gralla's ``P gauge''~\cite{Gralla:12}, Rosenthal's ``Fermi gauge''~\cite{Rosenthal:06a}, and the gauge that Detweiler uses in his Eq.~(21) to define his singular field~\cite{Detweiler:12}. Even before any derivations of the gravitational self-force, a rest-gauge metric was used by Thorne and Hartle~\cite{Thorne-Hartle:85} in their derivation of equations of motion for compact objects immersed in some external gravitational field. Indeed, the self-force problem of ``determining which piece of the full metric constitutes the effective metric'' could be phrased as ``finding the `external' metric in which Thorne and Hartle's equations of motion are valid,'' a point discussed at length in Ref.~\cite{Pound:15a} (and in a different way by Detweiler~\cite{Detweiler:01,Detweiler:05}). 

However, any description of finding ``the'' effective metric is only heuristic. In fact, there is no one unique effective metric. Various choices of  $g_{\mu\nu}+h^{\R}_{\mu\nu}$ would lead to the same generalized equivalence principle and could be used in an equally practical puncture scheme.  An illustration of this is provided by my derivations in the Lorenz gauge and in the highly regular gauge: I utilized two different regular fields, which differ at order $r^2$, but which possess all the same  essential ``physical'' properties. We might think that one cannot choose an alternative regular field at order $r^0$ and $r$, since the self-force involves those orders. But the equation of motion~\eqref{acceleration} only involves specific components of $h^{\R}_{\mu\nu}$ and its derivatives on the worldline, and one could easily move portions of $h^\R_{\mu\nu}$ into $h^\S_{\mu\nu}$, thereby defining new singular and regular fields, while leaving Eq.~\eqref{acceleration} intact. Hence, there is always a potential danger of ascribing too much physical meaning (or too specific an interpretation) to any one choice of effective metric.




\begin{acknowledgments}
The possibility of finding a highly regular gauge was suggested to me by Soichiro Isoyama. I also thank Leor Barack  for helpful discussions. This work was supported by the Natural Sciences and Engineering Research Council of Canada and by the European Research Council under the European Union's Seventh Framework Programme (FP7/2007-2013)/ERC Grant No. 304978.
\end{acknowledgments}

\appendix

\section{STF decompositions}\label{STF decomposition}
This appendix reproduces standard formulas from Ref.~\cite{Blanchet-Damour:86}.

Any Cartesian tensor $T^S(t,r,\theta,\phi)$ can be expanded as
\begin{equation}
T^S(t,r,\theta,\phi)=\sum_{\ell\geq0}T^{S\langle L\rangle}(t,r)\nhat_L,\label{nhat_decomposition}
\end{equation}
with coefficients given by 
\begin{equation}
T^{S\langle L\rangle}(t,r) = \frac{(2\ell+1)!!}{4\pi\ell!}\int T^S(t,r,\theta,\phi)\nhat^Ld\Omega,\label{lth_coeff}
\end{equation}
where $x!!=x(x-2)\cdots1$.

For $s=1$ and 2 the coefficients can be put in irreducible form using
\begin{equation}\label{s=1_decomposition}
T^{a\langle L\rangle} = \hat T_{(+)}^{aL}+\epsilon^{ja\langle i_\ell}\hat T_{(0)}^{L-1\rangle}{}_j+\delta^{a\langle i_\ell}\hat T_{(-)}^{L-1\rangle},
\end{equation}
where
\begin{subequations}\label{T1}
\begin{align}
\hat T_{(+)}^{L+1} & \equiv T^{\langle L+1\rangle}, \\
\hat T_{(0)}^{L} & \equiv \frac{\ell}{\ell+1}T^{pq\langle L-1}\epsilon^{i_\ell\rangle}{}_{pq}, \\
\hat T_{(-)}^{L-1} & \equiv \frac{2\ell-1}{2\ell+1}T_j{}^{jL-1}.
\end{align} 
\end{subequations}
and
\begin{align}\label{s=2_decomposition}
T_{ab\langle L\rangle} & = \delta_{ab}\hat K_L+\hat T^{(+2)}_{abL}\nonumber\\
						&\quad +\mathop{\STF}_L\mathop{\STF}_{ab}\Big(\epsilon^p{}_{ai_\ell}\hat T^{(+1)}_{bpL-1}+\delta_{ai_\ell}\hat T^{(0)}_{b L-1}\nonumber\\
						&\quad +\delta_{a i_\ell}\epsilon^p{}_{bi_{\ell-1}}\hat T^{(-1)}_{pL-2} +\delta_{ai_\ell}\delta_{bi_{\ell-1}}\hat T^{(-2)}_{L-2}\Big),
\end{align}
where
\begin{subequations}\label{T2}
\allowdisplaybreaks
\begin{align}
\hat T^{(+2)}_{L+2} & \equiv T_{\langle L+2\rangle}, \\
\hat T^{(+1)}_{L+1} & \equiv \frac{2\ell}{\ell+2}\mathop{\STF}_{L+1}(T_{\langle pi_\ell\rangle qL-1}\epsilon_{i_{\ell+1}}{}^{pq}), \\
\hat T^{(0)}_L & \equiv \frac{6\ell(2\ell-1)}{(\ell+1)(2\ell+3)}\mathop{\STF}_L(T_{\langle ji_\ell\rangle}{}^j{}_{L-1}), \\
\hat T^{(-1)}_{L-1} & \equiv \frac{2(\ell-1)(2\ell-1)}{(\ell+1)(2\ell+1)}\mathop{\STF}_{L-1}(T_{\langle jp\rangle q}{}^j{}_{L-2}\epsilon_{i_{\ell-1}}{}^{pq}), \\
\hat T^{(-2)}_{L-2} & \equiv \frac{2\ell-3}{2\ell+1}T_{\langle jk\rangle}{}^{jk}{}_{L-2} \\
\hat K_L & \equiv \tfrac{1}{3}T^j{}_{jL}.
\end{align}
\end{subequations}

\section{Decomposition of the regular field}\label{hR relations}
\subsection{STF decomposition}
In Sec.~\ref{outer_expansion}, I decompose the regular field into irreducible STF pieces. Specifically, according to Eq.~\eqref{hR decomposition}, the functions $\frac{1}{p!}h^{{\rm R}n}_{\mu\nu,\langle L\rangle}=h^{(n,l)}_{\mu\nu L}$ have the following irreducible decompositions:
\begin{subequations}\label{hR0 decomposition}
\begin{align}
h^{(n,0)}_{tt} &= \hat A^{(n,0)},\\
h^{(n,0)}_{ta} &= \hat C^{(n,0)}_{a},\\
h^{(n,0)}_{ab} &= \delta_{ab}\hat K^{(n,0)} + \hat H^{(n,0)}_{ab},
\end{align}
\end{subequations}
\begin{subequations}\label{hR1 decomposition}
\begin{align}
h^{(n,1)}_{tti} &= \hat A^{(n,1)}_i,\\
h^{(n,1)}_{tai} &= \hat C^{(n,1)}_{ai} + \epsilon^b{}_{ai}\hat D^{(n,1)}_{b} + \delta_{ai}\hat B^{(n,1)},\\
h^{(n,1)}_{abi} &= \delta_{ab}\hat K^{(n,1)}_i + \hat H^{(n,1)}_{abi} +\epsilon_i{}^c{}_{(a}\hat I^{(n,1)}_{b)c} \nonumber\\
						&\quad +\delta_{i\langle a}\hat F^{(n,1)}_{b\rangle},
\end{align}
\end{subequations}
and
\begin{subequations}\label{hR2 decomposition}
\begin{align}
h^{(n,2)}_{ttij} &= \hat A^{(n,2)}_{ij},\\
h^{(n,2)}_{taij} &= \hat C^{(n,2)}_{aij} + \epsilon^b{}_{a\langle i}\hat D^{(n,2)}_{j\rangle b}+ \delta_{a\langle i}\hat B^{(n,2)}_{j\rangle},\\
h^{(n,2)}_{abij} &= \delta_{ab}\hat K^{(n,2)}_{ij} + \hat H^{(n,2)}_{abij}
					+\mathop{\STF}_{ij}\Big(\epsilon_i{}^c{}_{\langle a}\hat I^{(n,2)}_{b\rangle cj}\nonumber\\
					&\quad +\delta_{i\langle a}\hat F^{(n,2)}_{b\rangle j}+\delta_{i\langle a}\epsilon^c{}_{b\rangle j}\hat G^{(n,2)}_c\nonumber\\
					&\quad +\delta_{i\langle a}\delta_{b\rangle j}\hat E^{(n,2)}\Big).
\end{align}
\end{subequations}

We can invert these relationships using Eqs.~\eqref{T1} and \eqref{T2} to express the STF tensors in terms of $h_{\mu\nu}^{\R1}$:
\begin{subequations}\label{inverse hR0 decomposition}
\begin{align}
\hat A^{(1,0)} &= h^{\R1}_{tt},\\
\hat C^{(1,0)}_a &= h^{\R1}_{ta},\label{C10_to_hR}\\
\hat H^{(1,0)}_{ab} &= h^{\R1}_{\langle ab\rangle},\\ 
\hat K^{(1,0)} &= \frac{1}{3}\delta^{ij} h^{\R1}_{ij},
\end{align}
\end{subequations}
and
\begin{subequations}\label{inverse hR1 decomposition}
\begin{align}
\hat A^{(1,1)}_a &= h^{\R1}_{tt,a},\label{A11_to_hR}\\
\hat B^{(1,1)} &= \frac{1}{3} h^{\R1}_{t}{}^{a}{}_{,a} ,\\
\hat C^{(1,1)}_{ab} &= h^{\R1}_{t\langle a,b\rangle} ,\\
\hat D^{(1,1)}_a &= \frac{1}{2}\epsilon{}_{a}{}^{bc} h^{\R1}_{tb,c}, \\
\hat F^{(1,1)}_a &= \frac{3}{5}\delta^{bc}h^{\R1}_{\langle ab\rangle,c},\\
\hat H^{(1,1)}_{abc} &= h^{\R1}_{\langle ab,c\rangle},\\
\hat I^{(1,1)}_{ab} &= \frac{2}{3}  h^{\R1}_{(a}{}^{c,d} \epsilon{}_{b)cd},\\
\hat K^{(1,1)}_{a} &=  \frac{1}{6} h^{\R1}_{b}{}^b{}_{,a},
\end{align}
\end{subequations}
where $h^{\R1}_{\mu\nu}$ and $\partial_\rho h^{\R1}_{\mu\nu}$ are evaluated on $\gamma$. I forgo writing the similar but lengthier relationships for the pieces of $h^{(1,2)}_{\mu\nu ij}$.

At second order, the decompositions \eqref{hR0 decomposition}--\eqref{hR2 decomposition} can be inverted to express the STF tensors in terms of $h_{\mu\nu}^{\R2}$ in precise analogy with Eqs.~\eqref{inverse hR0 decomposition}--\eqref{inverse hR1 decomposition}. 

As mentioned in Sec.~\ref{outer_expansion}, the regular field used here differs from the regular field $\bar h^{\R \mu\nu}_{2}$ defined in Ref.~\cite{Pound:12b}; the two are not simply the trace reversal of one another. This difference can be determined by taking the trace reverse of $\bar h^{2}_{\mu\nu}$, decomposing the result into coefficients $h^{(2,p)}_{\mu\nu L}$, and picking out the particular coefficients $h^{(2,p)}_{\mu\nu P}$ that determine the regular field. After further decomposing those coefficients into irreducible pieces, one finds
\begin{subequations}\label{hR hbarR relation1}%
\allowdisplaybreaks
\begin{align}
\hat A^{(2,0)} &= -\frac{59}{6} m^2 a{}_{a} a{}^{a} + \frac{1}{2} \bar h^{{\rm R}a}_{2}{}_{a} + \frac{1}{2} \bar h^{{\rm R}tt}_{2},\\
\hat C^{(2,0)}_a &= - \bar h^{{\rm R}t}_{2}{}_{a},\\
\hat H^{(2,0)}_{ab} &= -\frac{5}{9} m^2 \mathcal{E}{}_{ab} + \bar h^{{\rm R}}_{2\langle ab\rangle} ,\\ 
\hat K^{(2,0)} &= -\frac{31}{6} m^2 a{}_{a} a{}^{a} - \frac{1}{6} \bar h^{{\rm R}a}_{2}{}_{a} + \frac{1}{2} \bar h^{{\rm R}tt}_{2},
\end{align}
\end{subequations}%
and
\begin{subequations}\label{hR hbarR relation2}%
\allowdisplaybreaks
\begin{align}
\hat A^{(2,1)}_a &= \frac{1}{2} \bar h^{{\rm R}b}_{2}{}_{b,a} + \frac{1}{2} \bar h^{{\rm R}tt}_{2}{}_{,a} 
					- \frac{317}{45} m^2 \mathcal{E}_{ab} a{}^{b} \nonumber\\
					&\quad - \frac{601}{90} m^2 a{}_{a} a{}_{b} a{}^{b} + a{}_{a} \bar h^{{\rm R}b}_{2}{}_{b} + 2 a{}_{a} \bar h^{{\rm R}tt}_{2},\\
\hat B^{(2,1)} &= -\frac{1}{3} \bar h^{{\rm R}ta}_{2}{}_{,a} - \frac{2}{3} a{}^{a} \bar h^{{\rm R}t}_{2}{}_{a} + \frac{2}{9} m^2 a{}^{a} \dot a{}_{a},\\
\hat C^{(2,1)}_{ab} &= -\bar h^{{\rm R}t}_{2}{}_{\langle a,b\rangle} + \frac{1}{10} m^2 \dot{\mathcal{E}}{}_{ab} 
					+ \frac{68}{45} m^2 \mathcal{B}_{(a}{}^{d} \epsilon_{b)cd} a{}^{c} \nonumber\\
					&\quad- 2a_{\langle a} \bar h^{{\rm R}t}_{2}{}_{b\rangle} + \frac{1}{15} m^2 a_{\langle a} \dot a_{b\rangle},\\
\hat D^{(2,1)}_a &= -\frac{1}{2} \bar h^{{\rm R}tb,c}_{2} \epsilon{}_{abc} + \frac{47}{15} m^2 \mathcal{B}{}_{ab} a{}^{b} 
					+ \epsilon{}_{abc} a{}^{b} \bar h^{{\rm R}tc}_{2} \nonumber\\
					&\quad + \frac{1}{6} m^2 \epsilon{}_{ab}{}^{c} a^{b} \dot a_{c}, \\
\hat F^{(2,1)}_a &= \frac{3}{5} \bar h^{{\rm R}b}_{2}{}_{a,b} - \frac{1}{5}\bar h^{{\rm R}b}_{2}{}_{b,a} - \frac{67}{90} m^2 \mathcal{E}_{ab} a{}^{b},\\
\hat H^{(2,1)}_{abc} &= \bar h^{{\rm R}}_{2\langle ab,c\rangle}- \frac{1}{6} m^2 \mathcal{E}{}_{abc}+ \frac{7}{9}m^2a_{\langle a}\mathcal{E}_{bc\rangle},\\
\hat I^{(2,1)}_{ab} &= -\frac{4}{9} m^2 \dot{\mathcal{B}}{}_{ab} + \frac{2}{3} \bar h^{{\rm R}}_{2(a}{}^{c,d} \epsilon_{b)cd} \nonumber\\
					&\quad - \frac{319}{135} m^2 \mathcal{E}_{(a}{}^{d} \epsilon{}_{b)cd} a{}^{c},\\
\hat K^{(2,1)}_{a} &= -\frac{1}{6} \bar h^{{\rm R}b}_{2}{}_{b,a} + \frac{1}{2} \bar h^{{\rm R}tt}_{2}{}_{,a} 
					- \frac{437}{135} m^2 \mathcal{E}_{ab} a{}^{b} \nonumber\\
					&\quad + \frac{89}{18} m^2 a{}_{a} a{}_{b} a{}^{b} + a{}_{a} \bar h^{{\rm R}tt}_{2}.
\end{align}
\end{subequations}
The fact that the two regular fields are not simple trace reversals of one another is manifested by the explicit presence of $m$ in these relationships.

\subsection{Acceleration terms}
Equation~\eqref{d1h1} involves the terms in the first-order regular field that are linear in the acceleration. If we assume that the full metric satisfies retarded boundary conditions, then these acceleration terms can be obtained from the analytical form of the retarded field~\cite{Poisson-Pound-Vega:11},
\beq
h^1_{\mu\nu}= 4m \int_\gamma\!\bar G_{\mu\nu\mu'\nu'}u^{\mu'}u^{\nu'}dt',
\eeq
where $G_{\mu\nu\mu'\nu'}$ is the retarded Green's function, and primed indices refer to the point $\gamma(t')$. By expanding this integral near the worldline using standard methods reviewed in Ref.~\cite{Poisson-Pound-Vega:11}, one can  read off the various STF tensors appearing in the regular field. 

The results, taken from Ref.~\cite{Pound:10b} (and reproduced in Ref.~\cite{Poisson-Pound-Vega:11}), are\footnote{Table I in Ref.~\cite{Pound:10b} and Table II in Ref.~\cite{Poisson-Pound-Vega:11} are missing a factor of 4 from the $ma_a$ term in $\hat C_a^{(1,0)}$. The factor appears correctly in Eq.~(E.9) of the former reference and (23.10) of the latter.}
\begin{subequations}\label{hR explicit}
\begin{align}
\hat A^{(1,0)} &= \tail_{tt},\label{A10_to_tail}\\
\hat C_a^{(1,0)} &= \tail_{ta}+4ma_a,\label{C10_to_tail}\\
\hat K^{(1,0)} &= \tfrac{1}{3}\delta^{ab}\tail_{ab},\\
\hat H_{ab}^{(1,0)} &= \tail_{\langle ab\rangle},
\end{align}
\end{subequations}
and
\begin{subequations}\label{dhR explicit}
\allowdisplaybreaks
\begin{align}
\hat A_{a}^{(1,1)} &= \tail_{tta}+2\tail_{tt}a_a+\tfrac{2}{3}m\dot a_a,\label{A11_to_tail}\\
\hat B^{(1,1)} &= \tfrac{1}{3}\tail_{tij}\delta^{ij}+\tfrac{1}{3}\tail_{ti}a^i,\\
\hat C_{ab}^{(1,1)} &= \tail_{t\langle ab\rangle}+2m\E_{ab}+\tail_{t\langle a}a_{b\rangle}, \\
\hat D_{a}^{(1,1)} &= \tfrac{1}{2}\epsilon_a{}^{bc}(\tail_{tbc}+\tail_{tb}a_c),\\
\hat H_{abc}^{(1,1)} &= \tail_{\langle abc\rangle},\\
\hat F_{a}^{(1,1)} &= \tfrac{3}{5}\delta^{ij}\tail_{\langle ia\rangle j},\\
\hat I_{ab}^{(1,1)} &= \tfrac{2}{3}\displaystyle{\mathop{\STF}_{ab}} \left(\epsilon_b{}^{ij}\tail_{\langle ai\rangle j}\right),\\
\hat K_{a}^{(1,1)} &= \frac{1}{3}\delta^{bc}\tail_{bca}+\tfrac{2}{3}m\dot a_a.
\end{align}
\end{subequations}
Here I have defined the tail integrals
\begin{align}
\tail_{\mu\nu}(t) &= 4m \int_{-\infty}^{t^-}\!\!\!\!\!\!\bar G_{\mu\nu\mu'\nu'}u^{\mu'}u^{\nu'}dt',\label{tail}\\
\tail_{\mu\nu\rho}(t) &= 4m \int_{-\infty}^{t^-}\!\!\!\!\!\!\nabla_{\!\rho}\bar G_{\mu\nu\mu'\nu'}u^{\mu'}u^{\nu'}dt',\label{dtail}
\end{align}
which are tensors on the worldline.  $t^-=t-0^+$ indicates that the integral covers the past history $t'<t$ but excludes $t'=t$. 

The terms linear in $a_i$ in Eq.~\eqref{hR explicit} constitute the term $\ord{1}{h}^{(1,0)}_{\mu\nu}$ in Eq.~\eqref{d1h1}, and those in Eq.~\eqref{dhR explicit} constitute the term $\ord{1}{h}^{(1,1)}_{\mu\nu i}$.

\section{Radial functions in tidally perturbed black hole metric}\label{radial_functions}
In this appendix I list the functions $e_k$ and $b_k$ appearing in Eqs.~\eqref{H2} and \eqref{H3}. With $\zeta:= 2m/\tilde\sfr$ and $f=1-\zeta$,
\begin{subequations}
\allowdisplaybreaks
\begin{align}
e_1 &= f^2 \\ 
e_2 &= f\Big[ 1 + \frac{1}{4}\zeta \left(5 - 12\ln\zeta\right)-\frac{3}{4}\zeta^2 \left(9-4\ln \zeta\right) \nonumber\\
&\quad + \frac{7}{4}\zeta^3 + \frac{3}{4}\zeta^4\Big], \\ 
e_3 &= f^2\left(1 - \frac{1}{2}\zeta\right), \\ 
e_4 &= f, \\
e_5 &= f\!\left[ 1 + \frac{1}{6}\zeta\left(13 - 12\ln\zeta\right) - \frac{5}{2}\zeta^2 - \frac{3}{2}\zeta^3 - \frac{1}{2}\zeta^4\right]\!,\!\! \\ 
e_6 &= f\left(1 - \frac{2}{3}\zeta\right), \\ 
e_7 &= 1 - \frac{1}{2}\zeta^2, \\ 
e_8 &= 1 + \frac{2}{5}\zeta\left(4 - 3\ln\zeta\right) - \frac{9}{5}\zeta^2 - \frac{1}{5}\zeta^3\left(7 - 3\ln \zeta\right) \nonumber\\
&\quad + \frac{3}{5}\zeta^4, \\ 
e_9 &= f + \frac{1}{10}\zeta^3,
\end{align}
\end{subequations}
and
\begin{subequations}
\allowdisplaybreaks
\begin{align}
b_4 &= f, \\
b_5 &= f\!\left[ 1 + \frac{1}{6}\zeta\left(7 - 12\ln \zeta\right) - \frac{3}{2}\zeta^2 - \frac{1}{2}\zeta^3 - \frac{1}{6}\zeta^4\right]\!,\!\! \\ 
b_6 &= f\left(1 - \frac{2}{3}\zeta\right), \\ 
b_7 &= 1 - \frac{3}{2}\zeta^2, \\ 
b_8 &= 1 + \frac{1}{5}\zeta\left(5 - 6\ln \zeta\right) - \frac{9}{5}\zeta^2 - \frac{1}{5}\zeta^3\left(2 - 3\ln\zeta\right) \nonumber\\
&\quad+ \frac{1}{5}\zeta^4, \\  
b_9 &= f - \frac{1}{10}\zeta^3.
\end{align}%
\end{subequations}

\begin{widetext}
\section{Gauge transformation}
The transformation from the rest gauge to the Lorenz gauge, described in Sec.~\ref{matching}, is given by the following expansion for small $r$:
\allowdisplaybreaks
\begin{subequations}\label{xi1}
\begin{align}
\xi_t^{1} &= \tfrac{1}{2}\int \!\!dt\hat{A}^{(1,0)} + r\hat{C}^{(1,0)}_{a} \hat{n}^{a} - r^2 \bigg[-\tfrac{1}{12} \partial_t \hat{A}^{(1,0)} 
			 + \tfrac{1}{2}\Big(\tfrac{1}{2}\partial_t\hat{H}^{(1,0)}_{ab}-\tfrac{2}{3}m\mathcal{E}_{ab}
			-\mathcal{E}_{ab}\int\!\!dt\hat{A}^{(1,0)} - \hat{C}^{(1,1)}_{ab}\Big) \hat{n}^{ab}\bigg] \nonumber\\
			&\quad - r^3 \bigg[\tfrac{5}{18} m \dot{\mathcal{E}}_{bc} \hat{n}^{bc} 
			- \Big( \tfrac{1}{3} \hat{C}^{(1,2)}_{ijk}
			-\tfrac{2}{3} \mathcal{B}_{ij} \int\!\! dt\hat{D}^{(1,1)}_{k} +\tfrac{2}{3} \hat{C}^{(1,0)}_{k} \mathcal{E}_{ij} 
			+ \tfrac{1}{3} \mathcal{B}^{q}{}_{j} \epsilon_{q}{}^{p}{}_{i} \hat{H}^{(1,0)}_{kp}\Big)\hat{n}^{ijk} \bigg]+\O(r^4),\label{xi1t}\\
\xi_a^{1} &= r\Big(\tfrac{1}{2} \hat{K}^{(1,0)} \hat{n}_{a} + \epsilon_{ab}{}^c\int\!\! dt\hat{D}^{(1,1)}_c \hat{n}^{b} 
			+ \tfrac{1}{2} \hat{H}^{(1,0)}_{ad} \hat{n}^{d}\Big) - r^2\Big(\tfrac{1}{12} \hat{K}^{(1,1)}_{a} -\tfrac{5}{18} \hat{F}^{(1,1)}_{a} 
			+ \tfrac{8}{5} m \mathcal{E}_{ad} \hat{n}^{d} - \tfrac{1}{4} \hat{H}^{(1,1)}_{adi} \hat{n}^{di}\nonumber\\
			&\quad + \tfrac{1}{6} \hat{F}^{(1,1)}_{d} \hat{n}_{a}{}^{d} 
			- \tfrac{1}{2} \hat{K}^{(1,1)}_{d} \hat{n}_{a}{}^{d} - \tfrac{1}{2} \epsilon_{ab}{}^c \hat{I}^{(1,1)}_{cd} \hat{n}^{bd} 
			 - \tfrac{1}{6} m \mathcal{E}_{di} \hat{n}_{a}{}^{di} 
			+ \tfrac{1}{3} \mathcal{B}^{cd} \epsilon_{abc} \int \!\!dt\hat{A}^{(1,0)}\hat{n}{}^{b}{}_{d}\Big)\nonumber\\
			&\quad -r^3\bigg[\Big(\tfrac{1}{10} \mathcal{B}_{b}{}^{d} \hat{C}^{(1,0)}{}^{b} \epsilon{}_{acd} 
			- \tfrac{1}{15} \mathcal{B}{}_{(a}{}^{d} \epsilon{}_{c)bd}\hat{C}^{(1,0)}{}^{b} 
			- \tfrac{7}{30} \mathcal{E}^{bd} \epsilon{}_{acd} \int\!\! dt\hat{D}^{(1,1)}_{b}\Big) \hat{n}^{c} 
			+ \tfrac{1}{18} m \dot{\mathcal{B}}{}^{bc} \epsilon{}_{acd} \hat{n}{}_{b}{}^{d} 
			+ \tfrac{11}{21} m \mathcal{E}{}_{abc} \hat{n}{}^{bc} \nonumber\\
			&\quad -\tfrac{1}{9}\Big(\mathcal{E}^{cd}\int\!\! dt\hat{D}^{(1,1)}_{b}-4\mathcal{B}^{cd}\hat{C}^{(1,0)}_{b}\Big) \epsilon^b{}_{c}{}^{i}\hat{n}_{adi} 
			+ \tfrac{1}{3} \mathcal{E}_{b[c} \hat{H}^{(1,0)}_{a]d}\hat{n}^{bcd} 
			 - \tfrac{2}{9}\Big(\mathcal{E}^{cd}\int\!\! dt\hat{D}^{(1,1)b}
			+\mathcal{B}{}^{cd} \hat{C}^{(1,0)}{}^{b}\Big) \epsilon{}_{ac}{}^{i} \hat{n}{}_{bdi} \nonumber\\
			&\quad - \tfrac{1}{6} \hat{H}^{(1,2)}_{abcd} \hat{n}{}^{bcd} 
			+ \Big(\tfrac{2}{9} m \mathcal{B}{}^{bcd} \epsilon{}_{ab}{}^{i} - \tfrac{1}{9} \mathcal{B}{}^{cd} \hat{C}^{(1,0)}{}^{b} \epsilon{}_{ab}{}^{i} 
			+ \tfrac{1}{4} \epsilon{}_{a}{}^{bc} \hat{I}^{(1,2)di}_{b} - \tfrac{1}{9} \mathcal{E}{}^{cd} \epsilon_{a}{}^{bi} \int\!\! dt\hat{D}^{(1,1)}_{b} 
			 - \tfrac{1}{6} \mathcal{B}{}^{bcd} \epsilon{}_{ab}{}^{i}\int \!\!dt\hat{A}^{(1,0)}\Big)\hat{n}{}_{cdi}\nonumber\\
			&\quad- \tfrac{1}{24} m \mathcal{E}{}^{bcd} \hat{n}{}_{abcd}   + \Big(\tfrac{1}{9} \mathcal{E}{}^{bc} \hat{K}^{(1,0)}
			-\tfrac{1}{3} \hat{A}^{(1,2)}{}^{bc} - \tfrac{5}{12} \hat{F}^{(1,2)}{}^{bc} 
			 - \tfrac{1}{24} \dot{\mathcal{E}}{}^{bc} \int \!\!dt\hat{A}^{(1,0)}
			 + \tfrac{1}{6} \mathcal{E}{}^{bc} \hat{A}^{(1,0)}\Big)\hat{n}{}_{abc} - \tfrac{5}{9} \mathcal{E}^{bc}\hat{H}^{(1,0)}_{bd}\hat{n}_{ac}{}^d\nonumber\\ 
			&\quad + \Big(\tfrac{1}{10} \hat{A}^{(1,0)} \mathcal{E}{}_{ab}-\tfrac{1}{30} \delta \mathcal{E}_{ab} 
			+ \tfrac{8}{15} m \dot{\mathcal{E}}{}_{ab} - \tfrac{1}{30} \mathcal{E}{}_{(b}{}^{c} \hat{H}^{(1,0)}_{a)c} 
			+ \tfrac{1}{15} \mathcal{E}_{ab} \hat{K}^{(1,0)} + \tfrac{1}{24} \dot{\mathcal{E}}_{ab}\!\!\int \!\!dt\hat{A}^{(1,0)}
			- \tfrac{1}{10} \epsilon{}_{ab}{}^{c} \partial_t \hat{D}^{(1,1)}_{c} \nonumber\\
			&\quad
			- \tfrac{1}{20} \partial_t^{2} \hat{H}^{(1,0)}_{ab}\Big)\hat{n}^{b}  + \Big(\tfrac{1}{3} \partial_t \hat{C}^{(1,1)}_{bc} - \tfrac{1}{12} \partial_t^{2} \hat{H}^{(1,0)}_{bc}\Big)\hat{n}{}_{a}{}^{bc}
			 + \Big(\tfrac{1}{30} \mathcal{E}{}^{bc} \hat{H}^{(1,0)}_{bc} - \tfrac{1}{20} \partial_t^{2} \hat{K}^{(1,0)}\Big)\hat{n}{}_{a} \bigg]+\O(r^4),\label{xi1a}
\end{align}
\end{subequations}
and
\begin{subequations}\label{xi2}
\begin{align}
\xi^{2}_t &= \tfrac{1}{2}\int\!\!\left(\hat{A}^{(2,0)} + \tfrac{1}{4}\hat{A}^{(1,0)}\hat{A}^{(1,0)}\right)dt 
			+\tfrac{1}{8} \hat{A}^{(1,0)}\!\!\int\!\! \hat{A}^{(1,0)}dt +m \hat{C}^{(1,0)}_{a} \hat{n}^{a}+ r \bigg[ 
			\Big(\tfrac{1}{2} m \hat{C}^{(1,1)}{}^{bc} + \tfrac{4}{3} m^2 \mathcal{E}{}^{bc}\Big) \hat{n}{}_{bc} \nonumber\\
			&\quad - \Big(\tfrac{1}{4} \hat{C}^{(1,0)b} \hat{H}^{(1,0)}_{ab}-\tfrac{1}{4}\partial_t \hat{C}^{(1,0)}{}_{a}\!\!\int \!\!\hat{A}^{(1,0)}dt
			-\tfrac{3}{4} \hat{A}^{(1,0)} \hat{C}^{(1,0)}_{a} -  \hat{C}^{(2,0)}_{a} 
			+ \tfrac{1}{4} \hat{C}^{(1,0)}{}_{a} \hat{K}^{(1,0)} \nonumber\\
			&\quad
			- \tfrac{1}{2} \hat{C}^{(1,0)}{}^{b} \epsilon_{ab}{}^{c} \!\!\int\!\! \hat{D}^{(1,1)}_c dt\Big)\hat{n}{}^{a}  -\tfrac{1}{6} m \partial_t \hat{A}^{(1,0)} - \tfrac{1}{2} m \partial_t \hat{K}^{(1,0)}\bigg] 
			 + \O(r^2\ln r),\label{xi2t} \\
\xi^{2}_a &=  \frac{2 m^2 \hat{n}{}_{a}}{r} + \tfrac{1}{2} m \hat{A}^{(1,0)} \hat{n}{}_{a} + \tfrac{1}{2} m \hat{K}^{(1,0)} \hat{n}{}_{a} 
			- m \hat{H}^{(1,0)}_{ab} \hat{n}{}^{b} - \frac{8}{15}m^2r\ln r\, \E_{ai}n^i - r\bigg[\tfrac{1}{3} m \hat{A}^{(1,1)}_{a} 
			- \tfrac{1}{2} \hat{C}^{(1,0)}_{a} \hat{C}^{(1,0)}_{b} \hat{n}^{b}   \nonumber\\
			&\quad + \tfrac{1}{2} m \hat{H}^{(1,1)}_{abc} \hat{n}{}^{bc} + \Big(\tfrac{1}{2}\epsilon^{bd}{}_{(c}\hat{H}^{(1,0)}_{a)d}\!\!\int\!\! \hat{D}^{(1,1)}_{b}dt
			+\epsilon{}_{a}{}^b{}_c \Lambda^{(2,1)}_{b}\Big) \hat{n}{}^{c} 
			-\tfrac{1}{40} \Big(11 m \hat{A}^{(1,1)}_{b} + 9 m \hat{K}^{(1,1)}_{b}\Big) \hat{n}^b{}_{a} + \tfrac{3}{5} m^2 \mathcal{E}^{bc} \hat{n}{}_{abc} 
			\nonumber\\
			&\quad + \Big(2 m^2 \mathcal{B}_{bc}  - \tfrac{1}{4} m \hat{I}^{(1,1)}_{bc}\Big)\epsilon_{a}{}^{c}{}_d \hat{n}^{bd} 
			 + \tfrac{1}{20} m \partial_t \hat{C}^{(1,0)}{}_{b}\hat{n}{}_{a}{}^{b}+ \Big(\tfrac{1}{4} \hat{K}^{(1,0)}\hat{K}^{(1,0)} 
			- \tfrac{1}{2} \hat{K}^{(2,0)} - \tfrac{1}{8} \partial_t \hat{K}^{(1,0)}\!\!\int \!\!\hat{A}^{(1,0)} dt\Big)\hat{n}_{a}\nonumber\\
			&\quad + \Big(\tfrac{1}{4} \hat{H}^{(1,0)c}_{a} \hat{H}^{(1,0)}_{bc}-\tfrac{178}{45} m^2 \mathcal{E}_{ab} 
			- \tfrac{1}{2} \hat{H}^{(2,0)}_{ab} + \tfrac{1}{2} \hat{H}^{(1,0)}_{ab}\hat{K}^{(1,0)} 
			- \tfrac{1}{8} \partial_t \hat{H}^{(1,0)}_{ab}\!\!\int \!\!\hat{A}^{(1,0)} dt\Big)\hat{n}{}^{b}- \tfrac{1}{3} m \partial_t \hat{C}^{(1,0)}_{a}\bigg]\nonumber\\
			&\quad+\O(r^2\ln r).\label{xi2a}
\end{align}
\end{subequations}
Here $\Lambda^{(2,1)}_{b}$ is an unknown function of time that would be fixed by the $ta$ component of the order-$\e^2 r$ matching condition. For the sake of visual clarity, I have omitted superscript 0's and $\dagger$'s. To accord with the notation of Sec.~\ref{acceleration expansion}, in the above expressions all uppercase Latin tensors with $n=1$ ({e.g.}, $\hat{A}^{(1,0)}$) should have a left-superscript 0, and all those with $n=2$ ({e.g.}, $\hat{A}^{(2,0)}$) should have a superscript $\dagger$. 

\end{widetext}

\section{Supertranslations}\label{supertranslations}
An interesting consequence of the calculation in Sec.~\ref{matching} is that I {\em only} have to impose the worldline-preserving condition~\eqref{worldline-preserving}. This is a restriction on ordinary translations $x^a\to x^a-\hat\Upsilon^a(t)$. Yet ``supertranslations'' of the form $x^a\to x^a-\hat\Psi^b(t)\nhat^a{}_b$ also generate mass dipole moments, given by $-\tfrac{2}{3}m\hat\Psi^a$ (as compared to $-m\hat\Upsilon^a$). We may have surmised that first-order supertranslations, like ordinary translations, would only be ruled out once we impose the condition $M^a=0$. But in fact, they are found to be zero simply from the transformation equation~\eqref{gauge_equation1}. In the same way, second-order supertranslations are ruled out by Eq.~\eqref{gauge_equation2} alone. 

Consider the first-order case for simplicity. A nonzero supertranslation would be required in one particular scenario: if the rest gauge were parity regular in the sense of Gralla~\cite{Gralla:11}, and the target gauge were parity irregular, or vice versa. For example, this would be the case if $h^1_{ab}=\frac{2m\delta_{ab}}{r}+\O(r^0)$ in the rest gauge and $h^1_{ab}=\frac{2m\delta_{ab}+c_{abi}n^i+d_{abijk}\nhat^{ijk}}{r}+\O(r^0)$, for some $c_{abi}$ and $d_{abijk}$, in the target gauge. But even in that situation, the first-order metric would dictate the supertranslation. No worldline-preserving condition would need to be imposed to constrain it.

On the other hand, if one were to transform away from the rest gauge with no specified target gauge in mind, then the supertranslation would be arbitrary. Consider starting from the field~\eqref{h1 rest} and performing a transformation generated by $\xi^a_1=\Psi^i\nhat_i{}^a$ plus an arbitrary smooth vector. In the new gauge, the field's leading behavior is 
\begin{align}
h^1_{ab} &= \frac{(2m+\tfrac{6}{5}\Psi_in^i)\delta_{ab}+\tfrac{2}{5}\Psi_{(a}n_{b)}-4\Psi^{i}\nhat_{abi}}{r}\nonumber\\&\quad+\O(r^0), 
\end{align}
with the $tt$ and $ta$ components unchanged at this order. If we continue to define $h^{\R1}_{\mu\nu}$ according to Eq.~\eqref{hR1} in this new gauge, then a short calculation shows that Eq.~\eqref{master equation1} is unchanged except for the addition of a term $-\tfrac{1}{15}\E_{ai}\Psi^i$ to the right-hand side. If we also impose the condition $M^i=0$ in this new gauge, then instead of $\hat\Upsilon^{(1,0)}_a=0$, we have the relation $\hat\Upsilon^{(1,0)}_a=-\frac{2}{3}\hat\Psi_a$. Rearranging the new version of Eq.~\eqref{master equation1} to solve for $f^a_1$ yields 
\beq
f^1_a = \frac{1}{2}\partial_a h^{\R1}_{tt} + \partial_t h^{\R1}_{ta} - \tfrac{2}{3}\ddot{\hat\Psi}_a-\tfrac{3}{5}\E_a{}^i\hat\Psi_{i},
\eeq
a self-force that depends on the supertranslation in addition to the regular field. Since $\Psi_a$ forms a part of the singular field in this gauge, there is a sense in which the self-force depends on both the singular and regular fields. One might still be able to preserve the generalized equivalence principle, but to do so, one would  have to adopt a less natural definition of $h^{\R1}_{\mu\nu}$ in this gauge [by adding appropriate terms to $\partial_a h^{\R1}_{tt}\big|_\gamma$, for example, and subtracting them from $h^{\S1}_{\mu\nu}$, meaning that in the expansion~\eqref{hn mode expansion}, $h^{\S1}_{\mu\nu}$ would include some of the smooth term $rh^{(1,1)}_{\mu\nu i}n^i$]. 

\bibliography{../bibfile}
\end{document}